\documentclass{article} 

\usepackage[square, numbers]{natbib}
\usepackage{graphicx}
\usepackage{booktabs} 
\usepackage{amsmath} 
\usepackage{prodint} 
\usepackage{rotating} 
\usepackage{makecell} 
\usepackage{subcaption} 
\usepackage{hyperref} 
\usepackage{enumitem} 
\usepackage[table]{xcolor} 
\usepackage{float} 
\usepackage{authblk}
\usepackage{caption}
\usepackage{multirow}
\usepackage{bbold}

\title{Multi-state Models For Disease Histories Based On Longitudinal Data} 

\author[1,2]{Simon Wiegrebe\thanks{These authors contributed equally to this work.}}
\author[1,3]{Johannes Piller\protect\footnotemark[1]}
\author[2]{Mathias Gorski} 
\author[4]{Merle Behr}
\author[1]{Helmut Küchenhoff} 
\author[2]{Iris M. Heid} 
\author[3,5]{Andreas Bender} 

\affil[1]{Statistical Consulting Unit StaBLab, Department of Statistics, LMU Munich, Munich, Germany.} 
\affil[2]{Department of Genetic Epidemiology, University of Regensburg, Regensburg, Germany.} 
\affil[3]{Munich Center for Machine Learning, LMU Munich, Munich, Germany.} 
\affil[4]{Faculty of Informatics and Data Science, University of Regensburg, Regensburg, Germany.}
\affil[5]{Department of Statistics, LMU Munich, Munich, Germany.} 
\date{}                     
\begin{document} 

\maketitle 

\begin{abstract}
\noindent
Multi-stage disease histories derived from longitudinal data are becoming increasingly available as registry data and biobanks expand.
Multi-state models are suitable to investigate transitions between different disease stages in presence of competing risks. 
In this context, however, their estimation is complicated by 
dependent left-truncation, multiple time scales, index event bias, and interval-censoring.
In this work, we investigate the extension of piecewise exponential additive models (PAMs) to this setting and their applicability given the above challenges.
In simulation studies we show that PAMs can handle dependent left-truncation and accommodate multiple time scales. Compared to a stratified single time scale model, a multiple time scales model is found to be less robust to the data generating process.
We also quantify the extent of index event bias in multiple settings, demonstrating its dependence on the completeness of covariate adjustment.
In general, PAMs recover baseline and fixed effects well in most settings, except for baseline hazards in interval-censored data.
Finally, we apply our framework to estimate multi-state transition hazards and probabilities of chronic kidney disease (CKD) onset and progression in a UK Biobank dataset (n=$142,667$).
We observe CKD progression risk to be highest for individuals with early CKD onset and to further increase over age.
In addition, the well-known genetic variant rs77924615 in the \textit{UMOD} locus is found to be associated with CKD onset hazards, but not with risk of further CKD progression.

\vspace{1em}
\end{abstract}


\section{Introduction}
\label{sec:introduction}


As the adoption of registry data expands \citep{slawomirski2023ehr, rau2024implementation} and biobanks gain importance \citep{sudlow2015uk, all2019all, greiser2023german}, multi-stage disease histories are increasingly becoming available for use in both clinical routine and epidemiological research.
Examples include Chronic Kidney Disease (CKD), where disease stages are defined by thresholds of serum creatinine-based estimated glomerular filtration rate (eGFR) measurements \citep{levin2013kidney};
diabetes, defined based on fasting plasma glucose levels \citep{american20252};
or age-related macular degeneration (AMD), characterized by the accumulation of subretinal drusenoid deposits (early/intermediate AMD) and cell atrophy (late AMD).

\begin{figure}[!ht]
\begin{center}
\includegraphics[width=0.8\linewidth]{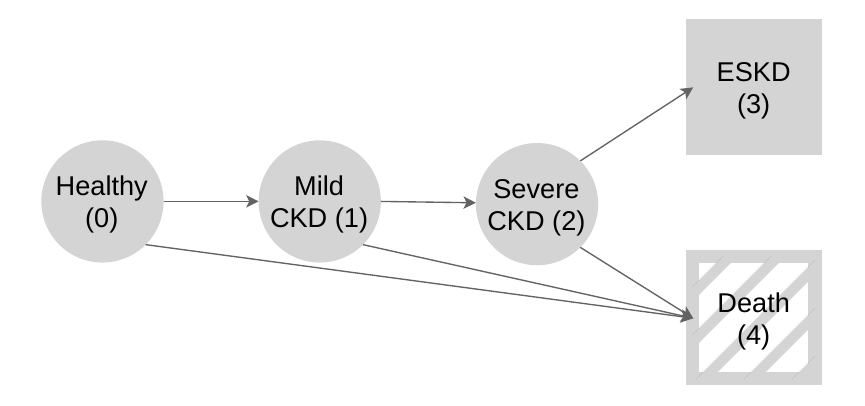}
\end{center}
\vspace{-0.7cm}
\caption{State diagram for Chronic Kidney Disease. This state diagram illustrates states and transitions within a chronic kidney disease (CKD) disease history. 
Circles represent transient states (Healthy (initial state), Mild CKD, Severe CKD).
Squares represent absorbing states.
Here, end-stage kidney disease (ESKD) is the terminal endpoint of interest (solid gray shading), while death before ESKD is the terminal competing event (hatched gray shading); this is why both ESKD and Death are considered as absorbing states. 
(For ESKD, this is an analysis-specific convention; see Section \ref{ssec:ukb-data}.)
The arrows between states indicate possible/allowed transitions:
0$\rightarrow$1 (CKD onset), 1$\rightarrow$2 (progression to Severe CKD), 2$\rightarrow$3 (progression to ESKD), as well as transitions into Death out of all transient states.
}
\label{fig:state-diagram-ckd}
\end{figure}

In this context, it is usually of interest to study transitions and risk factors conditional on current disease stage: 
For example, given onset of Mild CKD at 50 years of age and progression to Severe CKD 10 years later, what is the probability of developing end-stage kidney disease (ESKD) within the next $t$ years? 
And is a genetic variant a risk factor for CKD onset, for CKD progression, or for both? 
It is appealing to employ multi-state models \citep[cf.][]{andersen2012statistical, cook2018multistate} to derive transition probabilities and risk factor associations of multi-stage disease histories, such as CKD stages (Figure \ref{fig:state-diagram-ckd}).\\

\noindent
Multi-stage disease histories can be formalized as a stochastic process $(X_t)_{t \geq 0}$, where $X_t$ denotes the occupied state at time $t$ from a finite state space $\mathcal{S} := \{0, 1, \ldots, K\}$. As the process evolves over time, it induces a filtration $\mathcal{F}_t := \sigma(X_s, \mathbf{x}(s): 0 \leq s < t)$, representing the information available up to time $t$. Here, $\sigma(\cdot)$ denotes the $\sigma$-algebra generated by the enclosed random variables, i.e., the smallest $\sigma$-algebra with respect to which every $X_s$ and $\mathbf{x}(s)$, $0 \leq s < t$, is measurable. This filtration comprises the observed trajectory of the process on $0 \leq s < t$ as well as the covariates $\mathbf{x}(s)$, which may be time-dependent but are assumed to be observed prior to time $t$. Crucially, we allow $\mathbf{x}(s)$ to itself encode features of the process history up to $s$ (e.g., time since entry into the current state, number of prior transitions, or previously visited states).

Within this framework \citep[cf.][]{andersen2012statistical, beyersmann2011competing, andersen2002msm, putter2007competing}, for states $o, \ell \in \mathcal{S}$ with $o \neq \ell$ and times $s, t \in [0, \tau)$ such that $s < t$, multi-stage disease histories are characterized by transition probabilities
\begin{align}
    p_{o\ell}(s,t) = \mathbb{P}(X_t = \ell \mid X_s = o, \mathcal{F}_s),
\end{align}
or, equivalently, by transition intensities
\begin{align}
    \alpha_{o\ell}(t) = \lim_{\Delta t \to 0} \frac{1}{\Delta t} p_{o\ell}(t, t+\Delta t).
\end{align}

Assuming $\Delta t$ to be infinitesimally small and defining the probability to stay in state $o$ as 
$\alpha_{oo}(t)=-\sum_{\ell \neq o} \alpha_{o\ell}(t)$, 
the empirical transition probability matrix 
$\mathbf{P}(s,t) = \left( p_{o \ell}(s,t) \right)_{o, \ell \in \mathcal{S}}$ 
follows from the cumulative transition intensity 
$\text{d}\boldsymbol{\Lambda} = \left(\alpha_{o\ell}(t) \Delta t\right)_{o, \ell \in \mathcal{S}}$, 
and can be calculated via the product integral over time interval $[s, t)$ \citep{aalenEmpiricalTransitionMatrix1978, beyersmann2011competing}, i.e., 
\begin{equation}
    \label{eq:trans-prob}
    \mathbf{P}(s, t) = \Prodi_{u \in [s, t)} \left( \textbf{I} + \text{d}\boldsymbol{\Lambda}(u)\right).
\end{equation}

Oftentimes, $(X_t)_{t \geq 0}$ is assumed to be Markov, i.e., $\mathbb{P}(X_t = \ell \mid X_s = o, \mathcal{F}_s) = \mathbb{P}(X_t = \ell \mid X_s = o)$; 
in other words, the transition probability is assumed to solely depend on the past via the current time $s$ and the occupied state.
Traditional approaches to multi-state modeling, such as the non-parametric Aalen-Johansen (AJ) estimator \citep{aalenEmpiricalTransitionMatrix1978} or cause-specific Cox Proportional Hazards (PH) models \citep{cox1972regression}, have limited support for handling non-Markov scenarios and non-linear effect modeling.
Within a Markov framework, recent Bayesian and continuous-time approaches have advanced the modeling of time-inhomogeneous disease progression \citep{williams2020bayesian, luo2021bayesian, kendall2025beyond}. 
However, these methods remain fundamentally limited by the Markov assumption, which precludes the incorporation of subject-level history-dependent information. 
Alternative semi-Markov approaches, such as the Bayesian accelerated failure time (AFT) model \citep{klausch2023bayesian}, have been developed to handle interval-censored data while relaxing restrictive Markov assumptions.
However, these methods typically rely on parametric distribution families for transition times, which can entail severe bias in baseline hazard estimation if the underlying distribution is misspecified.
Unbiased estimation of the baseline hazards is relevant because inferential targets typically include not only relative covariate effects on transition-specific hazards, but also absolute, subject-specific transition probabilities.
The latter are frequently of direct interest in disease-history analyses, including for prognosis and risk prediction.
As Equation \eqref{eq:trans-prob} shows, these probabilities depend on the full cumulative transition-intensity matrix and thus on the time-varying baseline hazards \citep{putter2007competing}.

Furthermore, the application of multi-state models in the context of multi-stage disease histories comes with additional statistical challenges:

\begin{itemize}
\item[(a)] Independence between (history-dependent) state-entry times and events may be violated \citep{mackenzie2012survival, cheng2015causal, wang2024doubly}.
\item[(b)] There exist multiple time scales, which are not all jointly identifiable in subpopulations due to the age-period-cohort problem \citep{iacobelli2013multiple}.
\item[(c)] Negative correlation between otherwise uncorrelated risk factors may be induced within subsets of diseased individuals, causing index event bias \citep{dahabreh2011index}.
\item[(d)] Disease histories derived from longitudinal data tend to be interval-censored as traits are only assessed at (ir-)regular follow-up times \citep{zhang2010interval}.
\end{itemize}

\noindent
Here, we propose to use a novel multi-state modeling framework based on piecewise exponential additive models (PAMs) \citep{piller2026pammtools, piller2024flexible}: 
Within this framework, multi-state data encodes transitions via tuples $(y^{entry}_{i,k}, y^{exit}_{i,k}, d_{i,k}, \mathbf{x}_{i,k})$, where  $i \in \{1, \ldots, n\}$ is the subject index, $k$ refers to the transition from state $o$ to $\ell$ (also noted as $o \rightarrow \ell$), $y^{entry}_{i,k}$ and $y^{exit}_{i,k}$ are risk set entry and exit times, due to either transitioning or censoring, $d_{i,k}$ is the transition-specific status indicator, and $\mathbf{x}_{i,k}$ is the subject- and transition-specific covariate vector \citep{piller2025reduction}. 
Transition intensities $\alpha_{o\ell}(t)$ are then computed through transition-specific hazards $h_k(t \mid \mathbf{x}_{i,k})$ (see Equations \eqref{eq:baseline-hazard-ssts}, \eqref{eq:baseline-hazard-mts}, and \eqref{eq:baseline-hazard-general}).
This permits flexible specification of covariate-dependent hazards, thereby inducing transition probabilities that vary with the underlying covariate structure \citep{piller2024flexible}.
By further including history-dependent information -- such as state-entry times or multiple time scales -- into the estimation of $h_k(t \mid \mathbf{x}_{i,k})$ through covariate adjustment, this PAM-based framework can explicitly handle non-Markov scenarios.
Moreover, baseline hazards are estimated much more flexibly, without relying on distributional assumptions, yet still in a fully parametric fashion.
Finally, the proposed framework allows for seamless incorporation of time-varying covariates as well as flexible estimation of their (time-varying) effects, for instance via penalized splines.

In this work, we demonstrate the scope of the above challenges and how the PAM-based framework can cope with them.
Section \ref{sec:challenges} introduces terminology and describes the statistical challenges in detail.
Section \ref{sec:pam} introduces the modeling framework. 
Section \ref{sec:simulations} provides the setup and results of simulation studies investigating the properties of this modeling framework under the various challenges.
In Section \ref{sec:application}, we apply the framework to a UK Biobank dataset on CKD, estimating baseline hazards and transition probabilities as well as risk factor associations for CKD onset and progression.
Section \ref{sec:discussion} concludes, discusses limitations and provides directions for future research.

\section{Challenges in Disease History Analysis}
\label{sec:challenges}

We discuss four challenges that arise when applying multi-state modeling techniques to multi-stage disease histories: dependent left-truncation, choice of time scales, index event bias, and interval-censoring. 
The second and third challenges are not limited to time-to-event approaches.
Only the last one is specific to disease histories derived from interval-structured longitudinal data.

\subsection{Dependent left-truncation}
\label{ssec:challenges-left-truncation}

The multi-state process induces left-truncation because subjects enter the risk set for transition $k$ at different time points, i.e., at subject-specific state-entry times. 
For the above CKD example, a subject with CKD \textit{onset} at age 50 years is only at risk of CKD \textit{progression} after CKD onset and is thus left-truncated for CKD \textit{progression} prior to the age of 50 (see subject B in Figure \ref{fig:sample-trajectories}).
An estimator for the transition-specific hazards $h_{k}(t \mid \mathbf{x}_{i,k})$ therefore needs to be able to take this into account.
In general, left-truncation can be accounted for by non- and semi-parametric methods, provided that state-entry times (which represent the left-truncation times) can be assumed to be independent of event times.
Such methods include the AJ estimator and the Cox PH model, which adjust the risk sets accordingly \citep{andersen2012statistical, zhang2009mass}.

However, state-entry times may not necessarily be independent of event times \citep{mackenzie2012survival, cheng2015causal, wang2024doubly}: For instance, subjects with disease onset at a young age are possibly at higher risk of disease progression than subjects with late disease onset, implying dependence between truncation times and event occurrence. 
This dependence might not be linear on the log-hazard scale.

\subsection{Choice of time scales}
\label{ssec:challenges-time-scales}

Hazards for all transitions of a multi-state model are commonly modeled on a single time scale, typically the natural time scale for transitions out of the initial state \citep{iacobelli2013multiple}.
We call such a model with transition-specific stratification of the baseline hazard a \emph{stratified single time scale} model.
Risk sets for all transitions are defined along that single time scale $t$.
Since the popular AJ estimator and Cox PH model can, by design, only incorporate a single time scale, the use of the \emph{stratified single time scale} approach is further reinforced.

\begin{figure}[!ht]
\begin{center}
\includegraphics[width=1.0\linewidth]{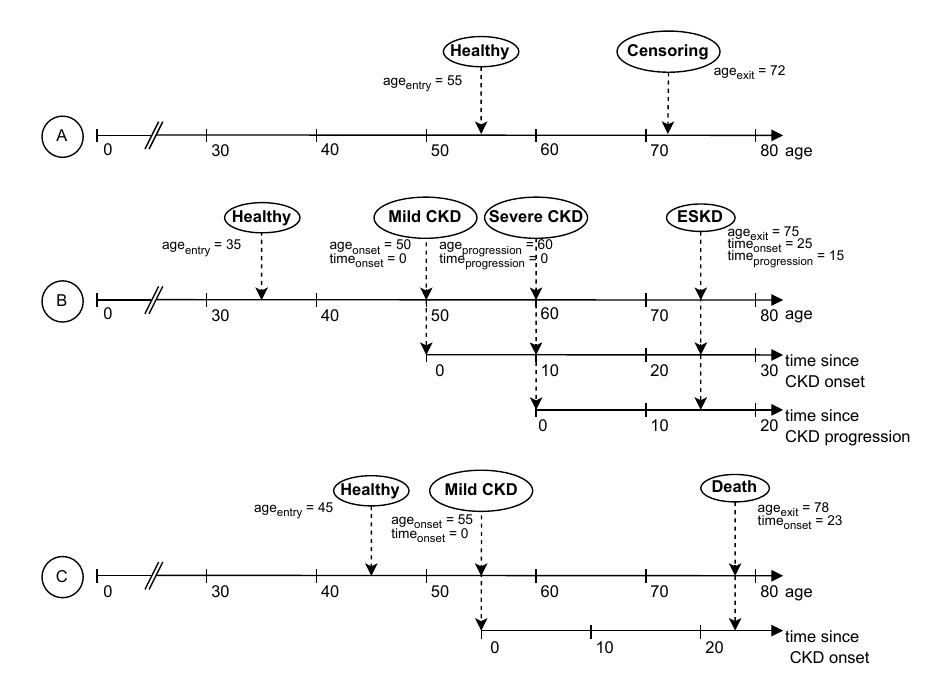}
\end{center}
\vspace{-0.8cm}
\caption{
Examples of disease histories regarding CKD. Based on the state diagram in Figure \ref{fig:state-diagram-ckd}, this figure illustrates CKD onset and progression histories of three hypothetical subjects.
Commencing at study entry (Healthy), the disease histories end at the absorbing states ESKD or Death (before ESKD) or cease to be observable following right-censoring. 
ESKD is the analysis-specific terminal endpoint of interest, whereas death before ESKD is the terminal competing event. Subsequent post-ESKD trajectories are outside the scope of the present model.
Horizontal arrows represent time scales.
The precise definitions of time scales and state-entry times, as well as the practical implementation via transition-specific helper variables, are described in Section \ref{sec:pam}.
}
\label{fig:sample-trajectories}
\end{figure}

In general registry data, entry into the dataset is random and not linked to the start of disease or intervention. 
Then, the baseline time scale is typically chronological age \citep{korn1997time, thiebaut2004choice} and age at disease onset is the time-to-event for the transition from "healthy" into "diseased". 
For transitions into higher disease states or for clinical studies in patients (disease progression), the time since entry into the intermediate state is the relevant time scale \citep{iacobelli2013multiple, wiegrebe2024analyzing}. 
In clinical studies, time to recovery or relapse can be alternative time scales.
This multiplicity of relevant time scales is illustrated for CKD onset and progression in Figure \ref{fig:sample-trajectories}:
For example, the hazard of progressing to Severe CKD for subject B at 57 years of age might depend not only on age, but also on time since CKD onset (i.e., 7 years).
As a consequence, it may be of interest to allow transition intensities of a multi-state model to vary along multiple time scales;
in the CKD example, time scales would be chronological age, time since CKD onset, and time since CKD progression.
This way, baseline hazards can be captured by (a combination of) distinct time scales which commence at the time of entry into the different states \citep{iacobelli2013multiple, copenhagen2024and}.
We call this the \emph{multiple time scales} approach.

It is possible to jointly estimate the effects of multiple time scales and state-entry times by fitting one model on the entirety of disease trajectories, thus circumventing the identifiability problem of age-period-cohort models \citep{iacobelli2013multiple}. 
However, this requires fully parametric models with respect to parameter estimation, such as the piecewise exponential model \citep{friedman1982piecewise} or discrete-time models \citep{tutz2016modeling}.
The specific choice of (multiple) time scales can then be considered as an empirical -- and thus, dataset-specific -- question \citep{iacobelli2013multiple}.


\subsection{Index event bias}
\label{ssec:challenges-ieb}

Diseases are typically multi-factorial, where risk factors can depend on each other \citep{DUARTE20131}.
Even if risk factors are independent in the healthy population, a negative correlation may be induced between them if analyses are restricted to subjects who have previously experienced a so-called index event whose occurrence is influenced by these risk factors.
As an example, consider the index event CKD onset and two independent risk factors: the genetic variant rs77924615 from the \textit{UMOD} locus (known to be associated with kidney function decline \citep{gorski2022genetic, wiegrebe2024analyzing}) and diabetes.
When studying risk factor associations with CKD progression and thus restricting to CKD patients, the genetic variant will likely be associated with absence of diabetes (and vice versa) because people with the genetic variant are less likely to require an additional risk factor in order to be in the CKD subgroup.

As a consequence, failing to include other (confounding) risk factors into an analysis restricted to such a subgroup causes biased estimation of the effect of the risk factor of interest, in the form of attenuation or even effect reversal.
This particular form of selection bias is referred to as index event bias (IEB) \citep{dahabreh2011index}.

Conditioning on index events is inherent to multi-state models, due to their conditional view: For transitions out of all non-initial states $\tilde{\ell} \neq 0$, we condition on having previously transitioned into $\tilde{\ell}$. 
As a consequence, we expect all risk factors to be affected by IEB for progression analysis.
In addition, as omitted variable bias (OVB) is known to affect time-to-event approaches by biasing effect size estimates towards zero \citep{bretagnolle1988effects}, we additionally expect OVB-driven attenuation for all states. 


\subsection{Interval-censoring}
\label{ssec:challenges-ic}

Longitudinal datasets are routinely obtained either from cohort studies with mostly prespecified examination times or from registry data with irregular and subject-specific visit numbers and times \citep{fitzmaurice2012applied, diggle2002analysis, lokku2020summarizing, hyun2017flexible, cheung2017mixture}.
In both cases, we have a natural interval structure for each subject, implying that event times from longitudinal data are typically interval-censored \citep{zhang2010interval}. 

The degree of interval-censoring (IC) inherently depends on interval lengths.
For registry data, interval lengths tend to be smaller for less healthy individuals, who in turn are more prone to have events ("informative observation") \citep{sperrin2017informative, bible2024accounting}.
This, as well as the fact that the common endpoint Death (cf. Figure \ref{fig:state-diagram-ckd}) is usually recorded precisely \citep["dual censoring";][]{boruvka2016sieve, schober2018survival}, limits the potential extent of IC.


\section{Flexible Modeling of Disease Histories Using Piecewise Exponential Additive Models}
\label{sec:pam}

We now introduce the piecewise exponential model (PEM) framework, a flexible approach for modeling disease histories, particularly in light of the challenges presented in Section \ref{sec:challenges}. 

PEMs \citep{friedman1982piecewise} reduce survival tasks (based on right-censored, potentially left-truncated data) to standard Poisson regression tasks (of uncensored data) via transformation of the original survival data into a piecewise exponential data (PED) format.
After partitioning follow-up into intervals $I_j$, and with $d_{ij}$ and $y_{ij}$ denoting the event indicator and time at risk, respectively, for subject $i$ in interval $j$, a standard single-event PEM can be represented as
\begin{equation*}
d_{ij}\sim\operatorname{Poisson}(\mu_{ij}), 
\qquad
\log(\mu_{ij})
=
\log(y_{ij})+\beta_{0j}+g(\mathbf{x}_{i}),
\label{eq:pem-generic}
\end{equation*}
where $\log(y_{ij})$ is an offset, $\beta_{0j}$ is the interval-specific log-baseline hazard, and $g(\mathbf{x}_{i})$ represents the covariate effects. 
(See Appendix \ref{app-ssec:ped} for details on data transformation and precise definitions of auxiliary variables.)
Thus, for $t\in I_j$, the hazard is
\[
h_i(t\mid\mathbf{x}_{i})
=
\exp\!\left\{\beta_{0j}+g(\mathbf{x}_{i})\right\}.
\]

The piecewise exponential additive model (PAM) \citep{bender2018generalized} is an extension of the PEM that replaces the interval-specific log-hazard intercepts $\beta_{0j}$ by a smooth function over time, typically estimated via penalized splines.
A standard single-event PAM thus specifies the hazard as
\begin{equation*}
h_i(t\mid\mathbf{x}_i)
=
\exp\!\left\{\beta_0+f_0(t)+g(\mathbf{x}_i)\right\},
\label{eq:pam-generic}
\end{equation*}
where $\beta_0$ is the global intercept and $f_0(t)$ is the smooth log-baseline hazard.
Note that penalized splines still require specifying a basis dimension (i.e., the number of knots) and a smoothing parameter.

The PAM approach is not entirely free of tuning choices as it requires choosing both a grid of cut points and a spline basis dimension.
However, in PAMs, the exact choice of cut points is less important: as long as the grid is sufficiently fine (e.g., at each observed event time, cf. transformation example in Appendix~A.1.2) and the resulting basis dimension is large enough to avoid underfitting, the associated roughness penalty controls the effective flexibility of $f_0(t)$ in a data-driven way, preventing overfitting \citep{eilers1996flexible, wood2017generalized, bender2018generalized}.

Related approaches have been discussed in the literature \citep{rodriguez2013model, argyropoulos2015analysis, sennhenn2016structured}.
Given the PED format, model estimation of a PAM is thus analogous to that of a penalized additive Poisson regression, typically via Restricted Maximum Likelihood (REML) estimation \citep{harville1977maximum, fahrmeir2022regression} and using standard software for the estimation of generalized additive models such as the \textsf{R} package \texttt{mgcv} \citep{wood2015package}.

For large datasets (e.g., the UK Biobank CKD dataset used in Section \ref{sec:application}), significant gains in memory efficiency and computational speed can be achieved by using \texttt{mgcv::bam()} with fast Restricted Maximum Likelihood (fREML) and covariate discretization.
The estimation techniques from single-event PEMs/PAMs can be directly applied to multi-state modeling, with the key difference that subjects can now enter and switch between different states. This, in turn, requires the PED to be further augmented (Appendix \ref{app-ssec:ped-msm}). Point estimation proceeds via the standard Poisson likelihood based on the hazard, as in \citep{bender2018generalized, piller2025reduction} (see \citep{holford1980, lairdoliver1981, friedman1982piecewise} for a more detailed derivation).

The PAM framework offers multiple approaches to derive standard errors. For arbitrary non-linear functionals, e.g., transition probabilities, confidence intervals can be obtained via an empirical Bayes approach \citep{piller2026pammtools}. By drawing a large number $N$ of coefficient vectors from their empirical Bayesian posterior, one can propagate each draw $\boldsymbol{\beta}^{(n)}, n=1,\ldots,N$ through the functional of interest and take the $(1-\alpha)$ confidence interval as the empirical $\alpha/2$ and $(1-\alpha/2)$ quantiles of the resulting distribution. Credible intervals derived from this posterior have been shown to attain close-to-nominal frequentist coverage for the smooth components of the GAM \citep{Wahba1983, Nychka1988, Marra2012}.\\

\noindent
In the following, we introduce two baseline multi-state PAMs: 
the \emph{stratified single time scale} (\emph{SSTS}) PAM, which estimates transition-specific baseline hazards along a single time scale, $t$;
and the \emph{multiple time scales} (\emph{MTS}) PAM, which expresses baseline hazards in terms of (combinations of) multiple time scales.
These baseline models are subsequently extended to account for dependent left-truncation and to include covariates.


\subsection*{Single Stratified Time Scale PAM}

\noindent
The transition-specific baseline hazard of the \emph{SSTS} PAM is
\begin{equation}
    h_{0,k}^{ssts}(t) = \exp \Big( \beta_{0,k} + f_k^{ssts}(t) \Big),
\label{eq:baseline-hazard-ssts}
\end{equation}
where $\beta_{0,k}$ is a log-hazard baseline level and $f_{k}^{ssts}(t)$ is a smooth, possibly non-linear function of $t$, estimated via penalized splines.
(See Equation \eqref{eq:pam-ssts-ukb} for an application of the transition-specific baseline hazard of an \emph{SSTS} PAM to the CKD data, using chronological age as time scale.). 
For estimation, we parameterize $f_{k}^{ssts}(t) = \sum_{m=1}^{M_k}\gamma_{m,k}B_{m,k}(t)$ with basis coefficients $\gamma_{m,k}$ and basis functions $B_{m,k}$ (e.g., B-splines or thin-plate splines). 
As the parameters of each function $f_{k}^{ssts}(t)$ are estimated independently of each other (with separate penalty terms), model \eqref{eq:baseline-hazard-ssts} can be viewed as a smooth version of the AJ estimator. \\

\subsection*{Multiple Time Scales PAM}

\noindent
For a general definition of the baseline hazard of an \emph{MTS} PAM, we consider progressive disease states $d \in \{0, \ldots, D\}$ ($0$ being the initial (healthy) state and $D$ the final, absorbing one).
We further consider $n_R$ (terminal) competing risks $R \in \{D+1, \ldots, D+n_R\}$.
For instance, in Figure \ref{fig:state-diagram-ckd} we have $D=3$ and a single competing risk $R=4$ ($n_R=1$).
In addition, we define:
\begin{itemize}
    \item the set of all disease onset and progression transitions out of state $d$ or higher states, $\mathcal{S}_{d D} := \{d \rightarrow d+1, ..., D-1\rightarrow D\}$ (in Figure \ref{fig:state-diagram-ckd} for $d=0$: $\mathcal{S}_{03} = \{0 \rightarrow 1, 1 \rightarrow 2, 2 \rightarrow 3 \}$);
    \item the set of all transitions into $R$ out of state $d$ or higher states, $\mathcal{S}_{dR} := \{d \rightarrow R, ..., D-1\rightarrow R\}$ (in Figure \ref{fig:state-diagram-ckd} for $d=0$: $\mathcal{S}_{04} = \{0 \rightarrow 4, 1 \rightarrow 4, 2 \rightarrow 4 \}$); 
    \item time scales $t_d$, representing the initial time scale for $d=0$ (identical to $t$ in Equation \eqref{eq:baseline-hazard-ssts}) or time since entry into state $d$ for $d>0$ (set to $0$ for transitions out of states $\tilde{d}<d$; cf. Figure \ref{fig:sample-trajectories});
    \item the effect of time scale $t_d$ on transitions in $\mathcal{S}_{d D}$, $f_{d D}(t_d)$; 
    \item the effect of time scale $t_d$ on transitions in $\mathcal{S}_{d R}$, $\tilde{f}_{d R}(t_d)$.
\end{itemize}
$\mathcal{S}_{0 D} \cup \left(\bigcup_{R=D+1}^{n_R} \mathcal{S}_{0 R}\right)$ thus represents the set of all possible transitions.
We distinguish between transitions in $\mathcal{S}_{d D} $ versus $\mathcal{S}_{d R}$, because the former ones characterize (consecutive) disease onset and progression transitions, whereas the latter ones refer to direct transitions into a terminal (competing risk) state. 

The baseline hazard of an \emph{MTS} PAM for transition $k$ out of state $d_k$ is then
\begin{equation}
h_{0,k}^{mts}(\mathbf{t}_k) =
\exp \Bigg(
   \beta_{0,k} 
   + \sum_{d=0}^{D-1} \Big\{ 
       \mathbb{1}_{\mathcal{S}_{d D}}(k) \cdot f_{d D}(t_d) 
       + \sum_{R=D+1}^{D+n_R} \mathbb{1}_{\mathcal{S}_{d R}}(k) \cdot \tilde{f}_{dR}(t_d)
     \Big\}
   \Bigg),
\label{eq:baseline-hazard-mts}
\end{equation}
where $\mathbf{t}_k = (t_0, \ldots, t_{d_k})$ is the vector of relevant time scales for transition $k$.
$\mathbb{1}_{A}(x)$ is the indicator function taking on the value 1 if $x \in A$ and 0 otherwise (cf. Equation \eqref{eq:pam-mts-ukb}).

\subsection*{Accounting for dependent left-truncation and inclusion of covariates}

\noindent
As discussed in Section \ref{ssec:challenges-left-truncation}, left-truncation may not be independent of event occurrence.
It is possible to account for this by incorporating state-entry times into the model \citep{beyersmann2011competing, emura2016semiparametric, rennert2022cox}.
While the exact state-entry time (e.g., the time of CKD onset) is interval-censored, its \textit{detection} time point is known and can be considered an actual new state because treatment, follow-up frequency, etc. change upon \textit{detection}.
Since PAMs do not rely on the Markov assumption and can flexibly estimate covariate effects, state-entry times (as well as other history-dependent variables) can be directly included into the model, typically as spline-based smooth effects.
The inclusion of state-entry times $\mathbf{x}_k^{entry} = (t_{\text{entry}\_1}, \ldots, t_{\text{entry}\_{d_k}}$) into a multi-state PAM requires the exact same stratification procedure as required for additional time scales $t_d$ of the \emph{MTS} PAM, including setting $t_{\text{entry}\_d} := 0$ for transitions out of states $\tilde{d}<d$.

When incorporating relevant state-entry times for transition $k$, $\mathbf{x}_k^{entry}$, in order to account for dependent left-truncation, the hazard becomes
\begin{align}
\label{eq:baseline-hazard-left-truncation} 
& h_{entry,k}^{m}(\mathbf{t}_k, \mathbf{x}_k^{entry})
  = h_{0,k}^{m}(\mathbf{t}_k) \cdot \\
& \quad \exp \Bigg(
     \underbrace{
     \sum_{d=1}^{D-1} \Big\{ \mathbb{1}_{\mathcal{S}_{d D}}(k) \cdot f_{\text{entry}\_d}(t_{\text{entry}\_d})
     + \sum_{R=D+1}^{D+n_R} \mathbb{1}_{\mathcal{S}_{d R}}(k) \cdot \tilde{f}_{\text{entry}\_d}(t_{\text{entry}\_d})
     \Big\}
     }_{=:\eta_k^{entry}}
     \Bigg), \nonumber 
\end{align}
for $\text{m} \in \{ssts, mts\}$.
Here, $\eta_k^{\mathrm{entry}}$ denotes the transition-specific contribution of the state-entry-time effects to the linear predictor on the log-hazard scale.
(Note that for the \emph{SSTS} PAM, $\mathbf{t}_k = t$, $\forall k$.)
Importantly, state-entry times are defined as the transition time (or age) \textit{into} another state, so they do not apply to the initial state. 
Consequently, the summation is restricted to non-initial transient states $d = 1, \ldots, D-1$.
In Section \ref{sssec:ukb-results-time-scales} (in particular, in Figure \ref{fig:ukb-left-truncation}), we show the relevance of modeling such state-entry time effects flexibly.

The hazards of multi-state PAMs can be further expanded to include any other (time-varying or time-constant) covariates $\mathbf{x}_k$, yielding the general form
\begin{equation}
h_{k}^{m}(\mathbf{t}_k, \mathbf{x}_k^{entry}, \mathbf{x}_k)
  = h_{0,k}^{m}(\mathbf{t}_k) \cdot 
  \exp \Bigg(
  \eta_k^{entry} + g(\mathbf{x}_k,\mathbf{x}_k^{entry},\mathbf{t_k},k)
  \Bigg).
\label{eq:baseline-hazard-general}
\end{equation}
Here, $g(\mathbf{x}_k,\mathbf{x}_k^{entry},\mathbf{t_k},k)$ is a potentially non-linear function representing the effects of covariates that may also interact with time scales, state-entry times, and transitions \citep{piller2026pammtools}.

\subsection*{Identifiability of multi-state PAMs}

\noindent
In order to jointly estimate multiple time scales and state-entry times, it is necessary to fit a single model to the entirety of disease trajectories \citep{iacobelli2013multiple} (see Section \ref{ssec:challenges-time-scales}).
In this case, however, the implementation of multi-state models so as to ensure model identifiability is not straightforward. 
For the design matrix to be of full rank, effects of $t_d$ and $t_{\text{entry}\_d}$ must only be estimated (and, e.g., spline bases set up) for those transitions where $t_d$ and $t_{\text{entry}\_d}$ actually vary.
Estimating such effects for transitions $\tilde{d}<d$ (where $t_d = t_{\text{entry}\_d} = 0$) introduces redundant (perfectly collinear) columns into the design matrix.
This, in turn, results in rank-deficiency and loss of identifiability, in particular for transition main effects and the effects of $t_d$ and $t_{\text{entry}\_d}$.

In Equations \eqref{eq:baseline-hazard-mts} and \eqref{eq:baseline-hazard-left-truncation}, transition-specific effects of multiple time scales and state-entry times are specified by means of the indicator functions $\mathbb{1}_{\mathcal{S}_{dD}}(k)$ and $\mathbb{1}_{\mathcal{S}_{dR}}(k)$.
For practical implementation, we need to specify helper variables for all transient states $d \in \{0, \ldots, D-1\}$ in order to stratify the effects of $t_d$ and $t_{\text{entry}\_d}$.

Due to the presence of multiple time scales, \emph{MTS} PAMs always require the construction of such helper variables in order to stratify the effect of each $t_d$ (and, potentially, $t_{\text{entry}\_d}$ and other history-dependent covariates) by all possible subsequent transitions, $d \rightarrow d+1$ and $d \rightarrow R$, $R \in \{D+1, \ldots, D+n_R\}$.
To this end, we define categorical variables $trans_{\text{after}\_d}$, which take on the value \textit{progression} for all progression transitions out of states $d, \ldots, D-1$ and $d \rightarrow R$ for transition into terminal state $R$.
In addition, for each non-initial transient state $d \in \{1, \ldots, D-1\}$, we set $trans_{\text{after}\_d}$ to the value \textit{none} for transitions out of states $\tilde{d}<d$.
This handles subjects who have not yet or never will transition into state $d$.
We can then estimate the effects of all terms $t_d$ and $t_{\text{entry}\_d}$, stratified by the subsequent transition, simultaneously, typically via spline-based smooth effects.

For \emph{SSTS} PAMs, such helper variables are necessary in case state-entry times (or any other history-dependent covariates) are included into the model.
In this case, the required helper variables $trans_{\text{after}\_d\_\text{exact}}$ are identical to the transition variable $k$ for transitions out of states $\tilde{d} \geq d$, but take on the value \textit{none} for transitions out of states $\tilde{d} < d$.\\


\noindent
We present two approaches to resolve the issue of non-identifiability.
The first approach consists in setting \textit{none} as the reference level of the helper variables, so that no separate effect is estimated for the corresponding transitions.
(For instance, in Figure \ref{fig:state-diagram-ckd}, we do not wish to estimate the effect of $t_1$ (time since CKD onset) for transitions out of state 0.)
For the \texttt{pammtools} implementation, which uses \texttt{mgcv::gam()} \citep{wood2015package} in the background, this can be achieved by coding $trans_{\text{after}\_d}$ and $trans_{\text{after}\_d\_\text{exact}}$ as \textit{ordered} factor variables with reference level \textit{none} and estimating $t_d$ and $t_{\text{entry}\_d}$ as penalized splines.
Then, \texttt{mgcv::gam()} only sets up basis functions for the non-reference levels of the helper variables, as desired.

The second approach removes the perfectly collinear design matrix columns (i.e., those where the helper variables are equal to \textit{none}) through penalization.
For the \texttt{pammtools} implementation, we can achieve this by using factor smooths (instead of penalized splines) to estimate the stratified $t_d$ and $t_{\text{entry}\_d}$ effects (\texttt{mgcv} version 1.9-3) \citep{wood2015package}. 
The factor smooth, which is actually a random effects-style smoother \citep{hagemann2025capturing}, only estimates a single, shared smoothing parameter $\lambda$ across all levels of the helper variable, while penalizing all null space components and thus "penalizing away" the non-informative and perfectly collinear spline basis coefficients of those subjects where the helper variable is equal to \textit{none}.

In the simulation study in Section \ref{ssec:sim-time-scales}, we test and contrast both of these two approaches to achieving identifiability.

\section{Simulation studies}
\label{sec:simulations}

In this section, we present the setup and results of simulation studies investigating the impact of the challenges in Sections \ref{ssec:challenges-time-scales}, \ref{ssec:challenges-ieb}, and \ref{ssec:challenges-ic}.
For this purpose, we design challenge-specific data generating processes (DGPs) and use different specifications of the PAM approach (Section \ref{sec:pam}) for estimation.

For all simulations, we employ a study time window of $t \in (0,10]$ with administrative censoring at $t=10$ and grid points $0.1, 0.2, \dots, 9.9, 10$ for evaluation.
Note that for transitions $1 \rightarrow 2$ and $1 \rightarrow 3$ of the multi-state DGPs (Sections \ref{ssec:sim-time-scales} and \ref{ssec:sim-ieb}), the evaluation grid is two-dimensional: $t \in \{0, 0.1, \dots, 9.9, 10\}$ and $t_{\text{entry}\_1} \in \{0, 0.1, \dots, t\}$, as the time at entry into state $1$ must not exceed the total time since study entry.
Penalized splines are fitted with $k=20$ basis functions to allow for sufficient flexibility.

We are interested, throughout, in the estimation of baseline (log-)hazards, cumulative hazards, and transition (or survival) probabilities (cf. Equations \eqref{eq:baseline-hazard-ssts}, \eqref{eq:baseline-hazard-mts}, and \eqref{eq:trans-prob}); as well as of associations of covariates (e.g., risk factors) with (log-)hazards (cf. Equation \eqref{eq:baseline-hazard-general}).
We judge estimation quality through coverage.
Pointwise coverage is computed for all grid points $t$ (or $t$-by-$t_{\text{entry}\_1}$) by evaluating whether the true quantity of interest is covered by the corresponding estimated 95\% confidence interval.
Averaging across all simulation runs yields pointwise mean coverage and 95\% confidence intervals, using an exact binomial test.
For overall coverage, we subsequently average over all evaluation grid points. 

We note that, by definition, results for log-hazards and hazards are identical in terms of coverage and directly proportional (up to exponentiation) in terms of scale.
This is why we only report log-hazards in the figures and tables of this section.

\subsection{Choice of time scales}
\label{ssec:sim-time-scales}

We start by looking into how well an \emph{SSTS} and an \emph{MTS} PAM (Equations \eqref{eq:baseline-hazard-ssts} and \eqref{eq:baseline-hazard-mts}) recover both baseline hazards and fixed effect hazards (e.g., of a risk factor) for the various transitions of a multi-state model.
We consider the case of the estimated model being correctly specified -- i.e., congruent with the DGP -- and the case of misspecification.

\subsubsection{Setup}
\label{sssec:sim-time-scales-setup}

\begin{figure}[!ht]
\begin{center}
\includegraphics[width=0.65\linewidth]{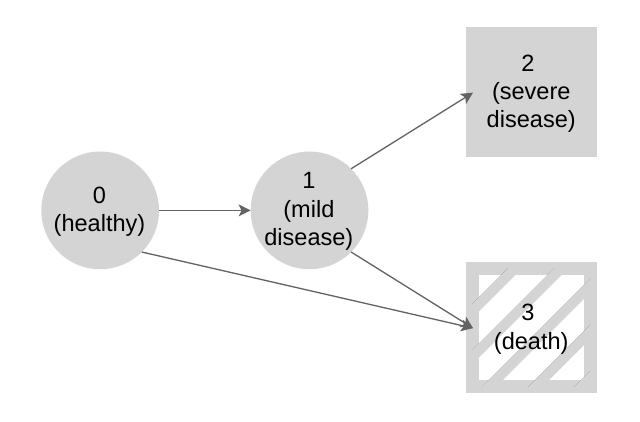}
\end{center}
\vspace{-0.7cm}
\caption{State diagram for simulation studies. This state diagram has initial state $0$, interim state $1$, and absorbing states $2$ and $3$.
Arrows depict possible transitions: $0 \rightarrow 1$ (\textit{onset}), $0 \rightarrow 3$ (\textit{death without disease}),
$1 \rightarrow 2$ (\textit{progression}), and $1 \rightarrow 3$ (\textit{death with disease}).
}
\label{fig:state-diagram-sim}
\end{figure}

We assume a state diagram with initial state $0$, interim state $1$, and absorbing states $2$ and $3$ ($k \in \{0\rightarrow1, 0\rightarrow3, 1\rightarrow2, 1\rightarrow3\}$); see Figure \ref{fig:state-diagram-sim}.
In terms of complexity, this state diagram ranks between a standard Illness-Death model and the model depicted in Figure \ref{fig:state-diagram-ckd}, allowing us to study onset and progression dynamics (transitions $0 \rightarrow 1$ and $1 \rightarrow 2$, respectively) in the presence of a competing risk ($3$).\\

\noindent
We consider two DGPs:
First, the \emph{SSTS} DGP assumes a single time scale $t$ (cf. Equation \eqref{eq:baseline-hazard-ssts}) and is characterized by the baseline hazard
\begin{equation}
    h_{0,k}^{ssts}(t) = \exp \Big( \beta_{0,k} + f_k^{ssts}(t) \Big).
\label{eq:baseline-hazard-ssts-dgp}
\end{equation}

Second, the \emph{MTS} DGP assumes the existence of one baseline hazard over time $t$ (i.e., the first time scale) for subjects with a disease transition ($0 \rightarrow 1$ or $1 \rightarrow 2$) and a distinct baseline hazard over $t$ for subjects with a death transition ($0 \rightarrow 3$ or $1 \rightarrow 3$). 
Additionally, the \emph{MTS} DGP assumes the existence of another baseline hazard over $t_1$ (time since entry into state $1$, the second time scale) for those individuals with disease onset.
This $t_1$ effect again differs between subjects with disease progression ($1 \rightarrow 2$) and subjects with death transition ($1 \rightarrow 3$).
The baseline hazard of the \emph{MTS} DGP is thus
\begin{equation}
\begin{split}
h_{0,k}^{mts}(\mathbf{t}_k) &= 
\exp \Bigg(
   \beta_{0,k} 
   + \sum_{d=0}^{D-1} \Big\{ 
       \mathbb{1}_{\mathcal{S}_{d D}}(k) \cdot f_{d D}(t_d) 
       + \sum_{R=D+1}^{D+n_R} \mathbb{1}_{\mathcal{S}_{d R}}(k) \cdot \tilde{f}_{dR}(t_d)
     \Big\}
   \Bigg) \\
   &=
\exp\Big(
  \beta_{0,k}
  + \mathbb{1}_{\{0\rightarrow1,1\rightarrow2\}}(k) \cdot f_{02}^{mts}(t) +
  \mathbb{1}_{\{0\rightarrow3,1\rightarrow3\}}(k) \cdot \tilde{f}_{03}^{mts}(t) \\
  &\quad + \mathbb{1}_{\{1\rightarrow2\}}(k) \cdot f_{12}^{mts}(t_1) 
  + \mathbb{1}_{\{1\rightarrow3\}}(k) \cdot \tilde{f}_{13}^{mts}(t_1)
\Big),
\raisetag{15pt}
\end{split}
\label{eq:baseline-hazard-mts-dgp}
\end{equation}
%
with possible from-states $d_k \in \{0,1\}$ and corresponding vectors of relevant time scales $\mathbf{t}_k = (t_0)$ or $(t_0, t_1)$ for $d_k=0$ or $1$, respectively;
$D=2$ and $R=3$;
$\mathcal{S}_{d D} := \{0 \rightarrow 1, 1 \rightarrow 2\}$ and $\mathcal{S}_{dR} := \{0 \rightarrow 3, 1 \rightarrow 3\}$ for $d=0$;
$\mathcal{S}_{d D} := \{1 \rightarrow 2\}$ and $\mathcal{S}_{dR} := \{1 \rightarrow 3\}$ for $d=1$.

For both DGPs, we assume the hazards of transitions after disease onset to be affected by the time of entry into state 1, $t_{\text{entry}\_1}$, again separately for subjects with disease progression and death transition.
Whenever we wish to investigate the coverage of fixed effects, we additionally include a Bernoulli($0.5$)-distributed covariate $x_1$ with transition-specific fixed effects $\beta_{x_1,k}$ on the log-hazard into the respective DGP.
The DGP hazard then becomes
\begin{align}
h_{k}^{m}(\mathbf{t}_k, t_{\text{entry}\_1}) 
  &= h_{0k}^{m}(\mathbf{t}_k) \cdot \exp\Big(
    \beta_{x_1,k} + \raisetag{-0.6\baselineskip} \label{eq:hazard-covariate-dgp} \\
  &\quad \mathbb{1}_{\{1\rightarrow2\}}(k) \cdot f_{\text{entry}\_1}(t_{\text{entry}\_1})
    + \mathbb{1}_{\{1\rightarrow3\}}(k) \cdot \tilde{f}_{\text{entry}\_1}(t_{\text{entry}\_1})
  \Big). \nonumber
\end{align}
%
for $\text{m} \in \{ssts, mts\}$.
Detailed specifications of the transition-specific log-hazard functions and fixed effect parameters can be found in Table \ref{tab:ts-dgp-details}.\\

\noindent
For a given DGP, we simulate datasets of size $n=5,000$ for a total of $500$ simulation runs.
For each dataset, we then draw transition times for each subject from a piecewise exponential distribution using the transition-specific log-hazards. We add random right-censoring to the simulated data by drawing censoring times $c_i$ from a Weibull distribution with shape parameter $\sigma = 1.5$ and scale parameter $\lambda = 10$.
We round event and censoring times to two decimals.\\

\noindent
We fit both an \emph{SSTS} PAM and an \emph{MTS} PAM, specified using Equations \eqref{eq:baseline-hazard-ssts} and \eqref{eq:baseline-hazard-mts}, to the data generated from the two DGPs.
In line with Sections \ref{ssec:challenges-left-truncation} and \ref{sec:pam}, we adjust both models for dependent left-truncation by including state-entry time $t_{\text{entry}\_1}$ in the model (cf. Equation \eqref{eq:baseline-hazard-left-truncation}).
We estimate both $t_1$ and $t_{\text{entry}\_1}$ either via penalized splines or factor smooths.
Whenever the DGP also contains a fixed effect, we include $x_1$ and its interaction with transition into the model (cf. Equation \eqref{eq:baseline-hazard-general}).\\

\noindent
We are interested in the estimation of baseline (log-)hazards, cumulative hazards, and transition probabilities, as well as of fixed effect hazards, for the different transitions and DGP-by-model-by-smooth scenarios.

\subsubsection{Results}
\label{sssec:sim-time-scales-results}

We find that baseline hazards for all transitions are recovered very well by the \emph{SSTS} PAM, regardless of DGP (Table \ref{tab:sim-ts-bh-coverage}).
For the \emph{MTS} PAM, coverage is nominal (i.e., 95\%) in case of correct model specification; in case of misspecification, however, mean coverage of (log-)hazards drops as low as $69\%$ (Table \ref{tab:sim-ts-bh-coverage}). 
For transitions $0 \rightarrow 1$ and $0 \rightarrow 3$, this is due to overestimation of the hazard especially at later time points (Figure \ref{fig:sim-ts-bh-line-plots}), leading to a sharp increase in bias and root mean squared error (RMSE; Figure \ref{fig:sim-ts-bh-bias_rmse}).
The choice of the smooth (penalized splines versus factor smooth) has little effect on baseline hazard coverage.
The choice of the quantity of interest, on the other hand, is relevant:
Coverage is generally higher for cumulative hazards and transition probabilities than for (log-)hazards.
This is because, as hazards are aggregated over time, up- and downward biases cancel each other out (Table \ref{tab:sim-ts-bh-coverage}).

\begin{figure}[!ht]
\begin{center}
\includegraphics[width=1.0\linewidth]{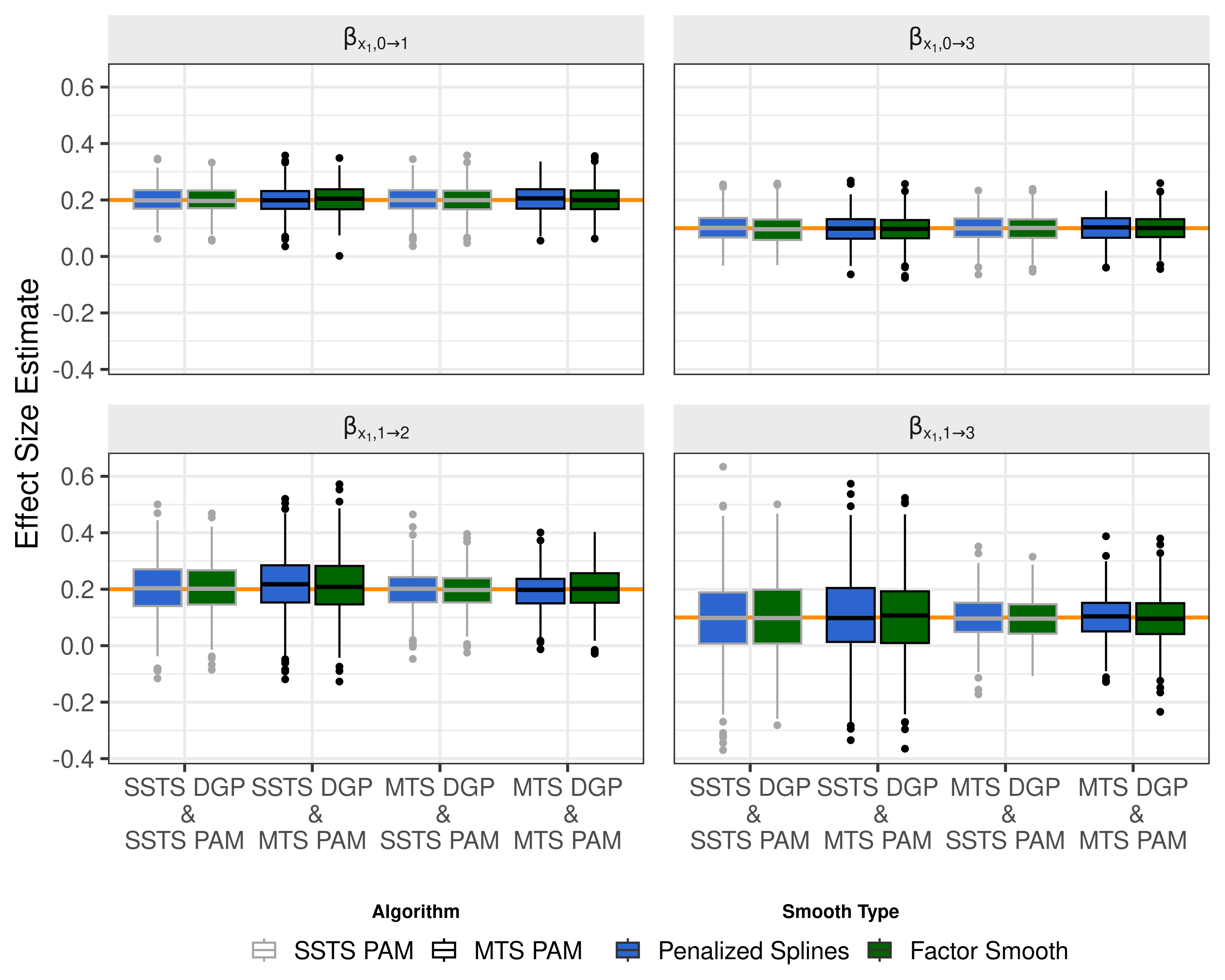}
\end{center}
\vspace{-0.5cm}
\caption{
Fixed effect estimates by transition from \emph{SSTS} and \emph{MTS} PAMs on data simulated from \emph{SSTS} and \emph{MTS} DGPs.
This figure illustrates first quartile, median and third quartile (boxes) of effect size estimates of a binary covariate $x_1$ on log-hazards across $500$ simulation runs for transitions $k \in \{0\rightarrow1, 0\rightarrow3, 1\rightarrow2, 1\rightarrow3\}$.
This is shown for \emph{SSTS} and \emph{MTS} PAMs, \emph{SSTS} and \emph{MTS} DGPs, and smooth types penalized splines versus factor smooth. 
Whiskers represent 1.5 times the interquartile range from the first and third quartile.
Orange lines denote true effect sizes $\beta_{x_1,k}$.
}
\label{fig:sim-ts-fe-boxplots}
\end{figure}

When including a fixed effect into the DGP, both PAMs -- regardless of DGP and smooth types -- achieve (close to) nominal fixed effect coverage for all transitions (Figure \ref{fig:sim-ts-fe-boxplots}, Table \ref{tab:sim-ts-fe-coverage}).

\subsection{Index event bias}
\label{ssec:sim-ieb}

\subsubsection{Setup}
\label{sssec:sim-ieb-setup}

For the simulations on the quantification of IEB when applying multi-state PAMs, we use the same simulation setup (state diagram, DGPs, models) as in the above simulations, except that we now extend the DGPs by adding two independent covariates: 
a risk factor of interest $x_1$ and another, potentially omitted risk factor $x_2$.
The risk factors are assumed to be either Bernoulli($0.5$)- or $N(0,1)$-distributed.
We consider small, medium and large effect size scenarios (Table \ref{tab:sim-ieb-effect-sizes}).

We first evaluate the correlation between the two risk factors in each non-absorbing state. 
As the two risk factors are independent in the healthy population (state 0) by design, we expect zero correlation in state 0 and a negative correlation in state 1, in line with IEB theory. 
Second, we evaluate the estimated effects of the risk factor of interest, $x_1$, on onset and progression log-hazards when only $x_1$ is included in the model (and $x_2$ is omitted).

We contrast this with the risk factor estimates after inclusion of $x_2$ into the model, which is expected to remove any bias.

\subsubsection{Results}
\label{sssec:sim-ieb-results}

\begin{figure}[!ht]
\centering
\begin{subfigure}{0.7\linewidth}
\includegraphics[width=\linewidth]{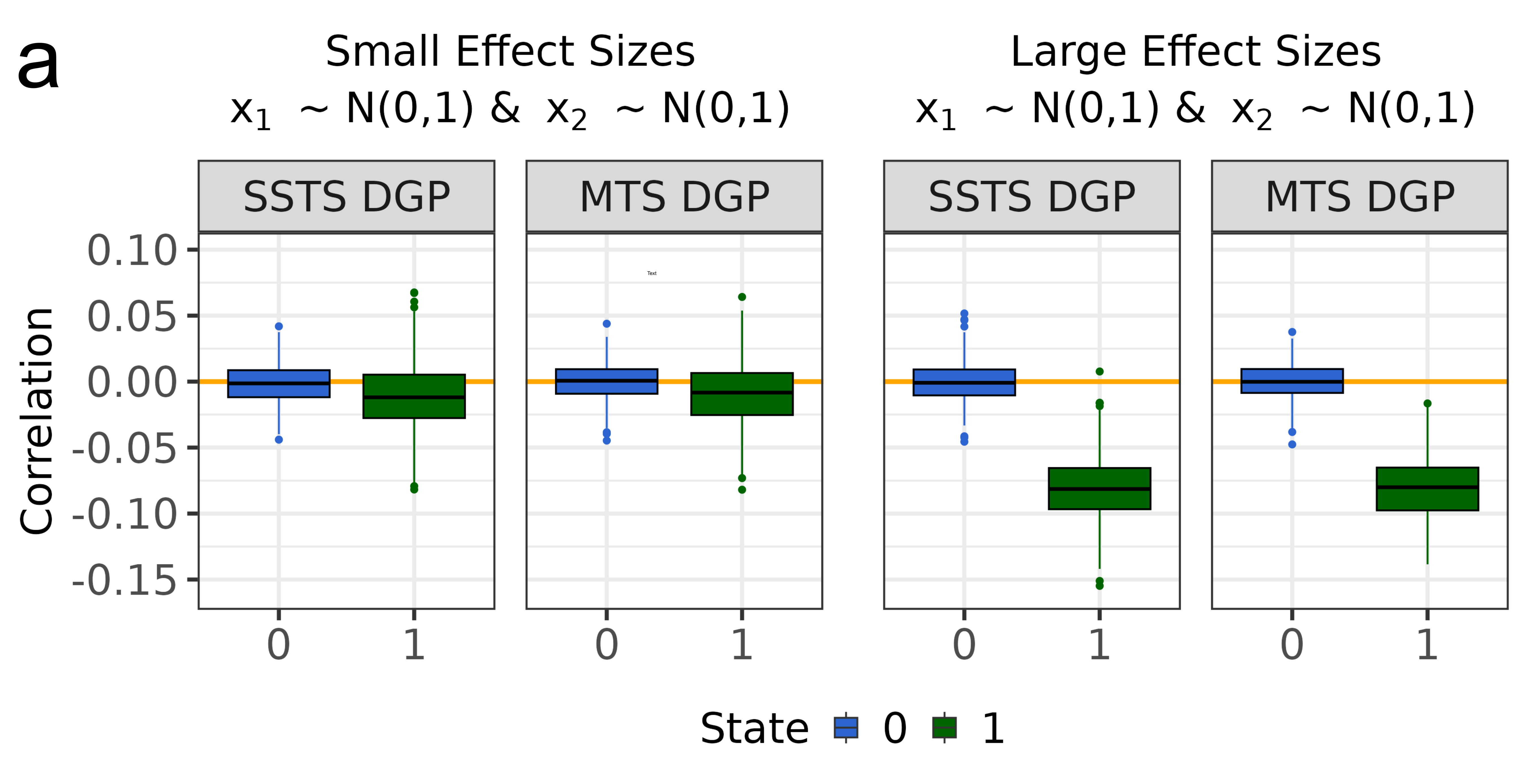}
\end{subfigure}
\begin{subfigure}{0.98\linewidth}
\includegraphics[width=\linewidth]{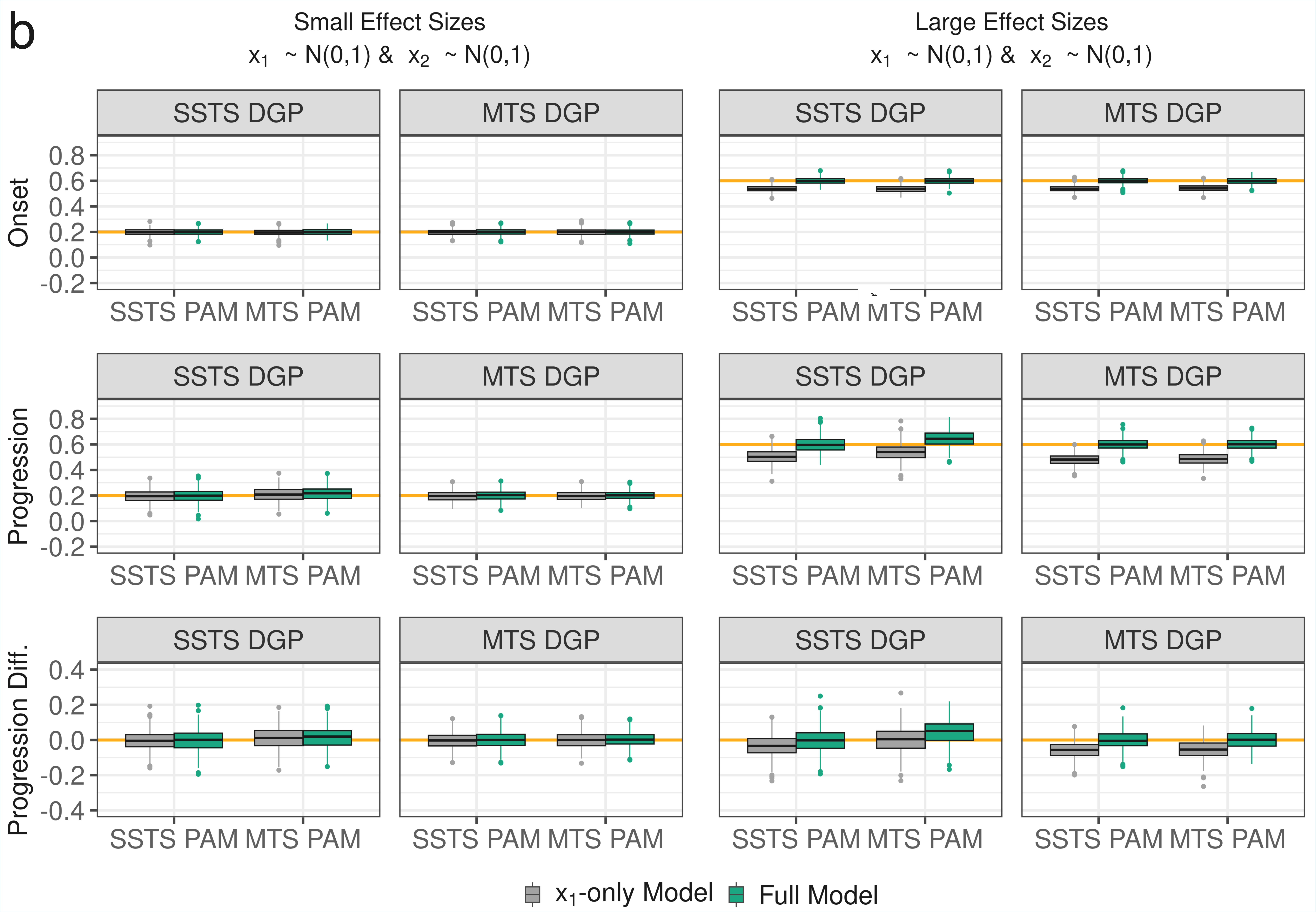}
\end{subfigure}
\caption{
Induced negative correlation between risk factors and bias in effect size estimates in simulated multi-state data.
Panel a shows Pearson correlation coefficients between two risk factors $x_1$ and $x_2$ in states $0$ and $1$, simulated as independent risk factors in the healthy population under \emph{SSTS} and \emph{MTS} DGPs.
Panel b shows effect size estimates of risk factor $x_1$ on the log-hazards of disease onset ($\hat{\beta}_{x_1,0\rightarrow1}$; first row), disease progression ($\hat{\beta}_{x_1,1\rightarrow2}$; second row), and their respective difference (third row). 
The underlying data are again simulated from \emph{SSTS} and \emph{MTS} DGPs and analyzed with \emph{SSTS} and \emph{MTS} PAMs. 
The first ($x_1$-only) model includes only the risk factor of interest, whereas the second (full) model includes both risk factors. 
Orange lines denote true effect sizes.
The full table of results on correlations, effect size estimates, and bias for all DGPs, models, risk factor distributions, and risk factor effect sizes is available at \url{https://github.com/survival-org/msm4diseaseHistories}.
}
\label{fig:sim-ieb-cor-coef}
\end{figure}

We observe correlation between $x_1$ and $x_2$ close to zero in state 0 across all risk factor distributions and effect size specifications, as expected (Figure \ref{fig:sim-ieb-cor-coef}a, Table \ref{tab:sim-ieb-cor}).
In state 1, we observe negative correlation between the risk factors that becomes more pronounced as risk factor effect sizes increase.

We find little to no bias in settings where risk factor effect sizes are small, even in state 1 with a small induced negative correlation.


When considering large risk factor effect sizes, we find that the model including only $x_1$ yields biased effect estimates for both onset and progression log-hazards:
For two $N(0,1)$-distributed risk factors with large effect sizes, we observe mean estimated onset and progression effects as low as $0.54$ and $0.48$, respectively. 
This is a relative bias of $10\%$ and $20\%$, given the true risk factor associations with onset and progression, $\beta_{1,onset} = \beta_{1,progression} = 0.6$ (Figure \ref{fig:sim-ieb-cor-coef}b).
While the bias on $\beta_{1,onset}$ is OVB, the bias on $\beta_{1,progression}$ is attributable to both OVB and IEB.
When including both risk factors $x_1$ and $x_2$ into the model, the bias is removed in all scenarios.

We find no difference in the bias between \emph{SSTS} and \emph{MTS} PAMs - regardless of the DGPs. 
This highlights, again, the relatively little importance of the choice of time scales when it comes to the estimation of fixed effect (i.e., risk factor) hazards.

\subsection{Interval-censoring}
\label{ssec:sim-ic}

In this section, we present a simulation study on the impact of IC on the estimation of fixed effect hazards as well as  baseline (log-)hazards, cumulative hazards and survival probabilities.
In doing so, we evaluate multiple DGPs, IC mechanisms, models, and estimation points.

\subsubsection{Setup}
\label{sssec:sim-ic-setup}

To limit complexity, we here restrict to two states with one transition ($0 \rightarrow 1)$, which reflects a standard survival setting.
However, owing to the composite nature of multi-state models, these results directly carry over to more complex models, as in Sections \ref{ssec:sim-time-scales}, \ref{ssec:sim-ieb}, and \ref{sec:application}.\\

\noindent
We consider the following three DGPs for the analysis of baseline hazard estimation under IC:
\begin{itemize}
    \item a piecewise exponential DGP with baseline log-hazard of $-3.5+6*f_0(t)$, where $f_0(t)$ is the density of a Gamma distribution with shape $\alpha=8$ and scale $\sigma=0.5$;
    \item a Weibull DGP with shape $\sigma=1.5$ and log-hazard intercept $\beta_0=-3.5$;
    \item and a Weibull-based DGP as implemented in the \texttt{simIC\_weib()} function from the \textsf{R} package \texttt{icenReg} \citep{anderson2017icenreg}, with shape $\sigma=1.5$ and scale $\lambda = 2$.
\end{itemize}

\noindent
For the piecewise exponential and Weibull DGPs, we consider beta-distributed, uniformly-jittered, and equidistant observation times with $1-10$ visit times per subject as IC mechanisms.
The \texttt{icenReg}-based DGP employs a single uniform renewal IC mechanism.\\

\noindent
Since here we have standard survival data and only a single transition type ($0 \rightarrow 1$), we can benchmark the performance of PAMs under IC by considering a total of four models:
\begin{itemize}
    \item PAM (estimating the baseline hazard using penalized splines with $k=20$ basis functions);
    \item Cox PH regression (not applicable for estimating baseline (log-)hazard rates; equivalent to the Nelson-Aalen estimator \citep{aalen1978nonparametric} and the Breslow estimator \citep{breslow1974covariance} for baseline cumulative hazards and the survival function, respectively, in the absence of covariates);
    \item and two AFT models (Weibull and Generalized Gamma).
\end{itemize}
For each of these models, we consider interval mid- and end-points as well as the exact time points (i.e., no IC) as estimation points.
For AFT models, we additionally consider IC adjustment (i.e., direct incorporation of IC into the likelihood by considering both interval start and end point).\\

\noindent
To evaluate the impact of IC on the estimation of fixed effect hazards, we add a Bernoulli($0.5$)-distributed covariate $x_1$ to the log-hazard of the DGPs and the models, analogously to Section \ref{ssec:sim-time-scales}.
In this case, we do not consider the \texttt{icenReg}-based DGP (incorporation of covariates with a pre-specified fixed effect on log-hazards not possible) nor the Generalized Gamma AFT model (not a PH model).


\subsubsection{Results}
\label{sssec:sim-ic-results}

\begin{figure}[!ht]
\begin{center}
\includegraphics[width=0.95\linewidth]{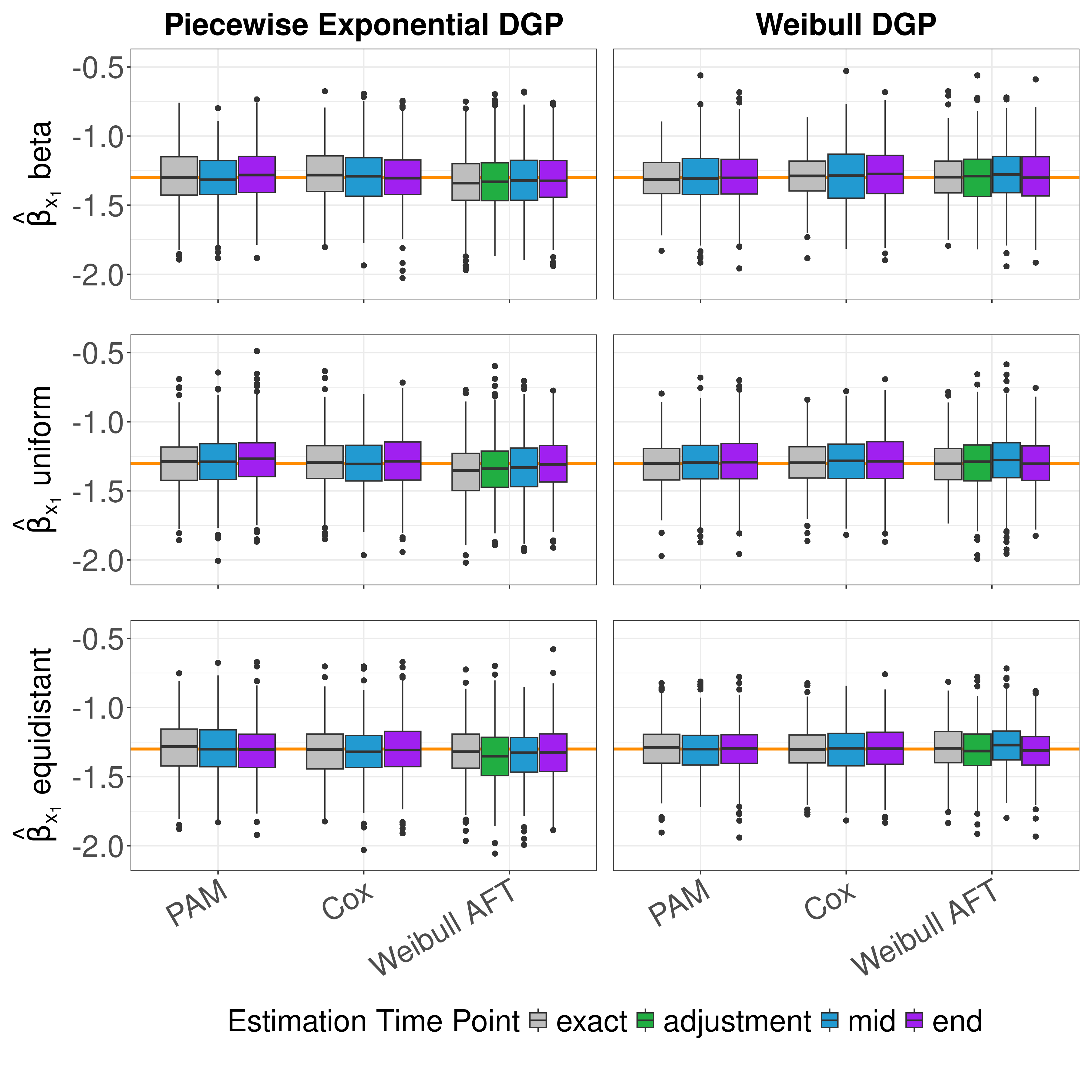}
\end{center}
\vspace{-0.5cm}
\caption{Boxplots of fixed effect estimates under interval-censoring. 
This figure shows boxplots of the estimated effect sizes of a \textit{Bernoulli}(0.5)-distributed covariate on the log-hazard.
Based on the simulation setup described in Section \ref{sssec:sim-ic-setup}, effect sizes are estimated using three models (x-axis) and three to four distinct estimation points (color coding), on data generated by two DGPs (columns) and three IC mechanisms (rows).
The orange line denotes the true effect on the log-hazard ($-1.3$).}
\label{fig:sim-ic-fe-boxplots}
\end{figure}

The fixed effect of covariate $x_1$ on the log-hazard under IC is estimated correctly across all DGPs, IC mechanisms, models, and estimation points (Figure \ref{fig:sim-ic-fe-boxplots}).
The estimation bias is negligible, with coverage at the nominal level (Tables \ref{tab:sim-ic-fe-bias} and \ref{tab:sim-ic-fe-coverage}).

The mean coverage of baseline (log-)hazards by PAMs with interval mid-point estimation is between 50\% and 75\%, depending on specifications.
PAMs with interval end-point perform consistently worse, due to a consistent time shift bias (Figure \ref{fig:sim-ic-bh-line-plots}).
The high coverage of PAMs estimated using exact time points (no IC) underlines the model's flexibility.
The Weibull and Generalized Gamma AFT models achieve nominal coverage of baseline (log-)hazards whenever the DGP is Weibull-based and IC adjustment is employed - as expected.
However, at the same time, their coverage is extremely low whenever the true DGP is not Weibull-based -- which highlights these models' limited flexibility in terms of estimating diverse baseline hazard shapes (Figure \ref{fig:sim-ic-bh-coverage-loghazards}, Table \ref{tab:sim-ic-bh-coverage-loghazard}).

For the piecewise exponential DGP, the coverage of baseline cumulative hazards and survival functions by PAMs is much higher, between 85\% and 97\%, depending on specifications.
For the piecewise exponential DGP, the Weibull AFT model is, by far, the worst, due to its limited flexibility; 
the performance of the more flexible generalized Gamma AFT model is on par with that of the Cox model (which here defaults to the Nelson-Aalen estimator).
For the Weibull DGP, the two AFT models with IC adjustment achieve nominal coverage (as expected), as does the PAM but not the Cox model (Figures \ref{fig:sim-ic-bh-coverage-cumu} and \ref{fig:sim-ic-bh-coverage-surv}, Tables \ref{tab:sim-ic-bh-coverage-cumu} and \ref{tab:sim-ic-bh-coverage-surv}).


\section{Application: CKD Onset and Progression in UK Biobank}
\label{sec:application}

In this section, we study hazards, transition probabilities, and risk factors of CKD onset and progression (cf. Figure \ref{fig:state-diagram-ckd}) in a large-scale UK Biobank (UKB) dataset \citep{gorski2025bias}.
Previous work on the same dataset has found the genetic variant rs77924615 (located in the \textit{UMOD} locus, well known for association with CKD \citep{kottgen2010new}) to be associated with kidney function decline over age and with CKD onset \citep{wiegrebe2024analyzing}. 

The multi-state approach presented here, which uses CKD stages instead of quantitative eGFR values, enables us to explicitly focus on CKD onset and progression of patients, as subjects who never get CKD -- the large majority \citep{wiegrebe2024analyzing} -- are considered to be censored (Table \ref{tab:ukb-descriptives}).


\subsection{UK Biobank data on history of CKD stages}
\label{ssec:ukb-data}

The UKB dataset curated by Gorski et al. \citep{gorski2025bias} contains $2,080,372$ eGFR measurements from $n=454,071$ subjects, collected from serum creatinine measurements from blood drawn at two study center assessments as well as from electronic health records (eHRs).
We include unrelated UKB participants of European ancestry (in line with Wiegrebe et al. \citep{wiegrebe2024analyzing}) with available eHR data.
We exclude eGFR measurements at age $<35$ years as well as between six months before and after acute kidney injury and nephrectomy, as in Wiegrebe et al. \citep{wiegrebe2024analyzing}.

We apply KDIGO definitions \citep{levin2013kidney} to define CKD stages: Healthy (eGFR $\geq$ 60 mL/min/1.73m$^2$), Mild CKD (eGFR $\in [30,60)$ mL/min/1.73m$^2$), Severe CKD (eGFR $\in [15,30)$ mL/min/1.73m$^2$), and ESKD (eGFR $<$ 15 mL/min/1.73m$^2$ or dialysis or kidney transplant).
The Death state is defined based on available death registries linked to the UK Biobank, which provide the exact date of death (along with primary and contributory causes) according to the ICD-10 system.
Healthy, Mild CKD, and Severe CKD are transient states. 
For the present analyses, the terminal endpoint of interest is the first occurrence of ESKD, whereas death before ESKD is the terminal competing event; this distinction is highlighted by the solid versus hatched gray shading in Figure \ref{fig:state-diagram-ckd}.
Both ESKD and Death are encoded as absorbing states, although ESKD is not biologically absorbing.
As in other first-event competing-risks settings, an event may be treated as terminal even when subsequent events remain possible \citep{putter2007competing}.
Post-ESKD pathways include dialysis, kidney transplantation, and comprehensive conservative care \citep{stevens2024kdigo}, with markedly different subsequent mortality across dialysis and transplantation populations \citep{wolfe1999comparison}; these pathways and subsequent mortality are outside the scope of the present model.

To account for natural fluctuations when measuring eGFR, we require two subsequent eGFR measurements to fall below the respective transition thresholds for transitions into Mild and Severe CKD. 
For computational reasons, we round subject age to years.
We impute missing interim CKD stages for "jump" transitions ($0 \rightarrow 2$, $0 \rightarrow 3$, and $1 \rightarrow 3$) -- which in the case of CKD are due to interval-censoring and not to some separate transition mechanism -- by linearly interpolating eGFR values. (This only applies to 213 subjects.)

The final multi-state dataset contains $142,667$ subjects and $19,293$ non-censoring transitions (Table \ref{tab:ukb-descriptives}).

\newcommand{\captionukbdescriptives}{Baseline characteristics and state transitions in the UKB dataset.}
\begin{table}[!ht]
\centering
\caption{\captionukbdescriptives}
\label{tab:ukb-descriptives}
\begin{tabular}{@{}lr@{}}
\toprule
\textbf{Characteristic} & \textbf{\begin{tabular}[b]{@{}l@{}}UKB Dataset \\ (n = 142,667)\end{tabular}} \\
\midrule
\multicolumn{2}{l}{\textit{Baseline Characteristics}} \\
Women, n (\%) & 76,705 (53.8) \\
Age at study entry & 56.0 (35.0--77.0) \\
Age at study exit & 65.0 (37.0--80.0) \\
\midrule
\multicolumn{2}{l}{\textit{Event Counts Overall}} \\
Transitions (incl. censoring) & 148,757 \\
Transitions (excl. censoring) & 19,293 \\
CKD-related transitions & 6,660 \\
Death transitions & 12,633 \\
\midrule
\multicolumn{2}{l}{\textit{Event Counts by Transition}} \\
Healthy \textrightarrow Mild CKD & 6,079 \\
Mild CKD \textrightarrow Severe CKD & 436 \\
Severe CKD \textrightarrow ESKD & 145 \\
Healthy \textrightarrow Death & 11,153 \\
Mild CKD \textrightarrow Death & 1,351 \\
Severe CKD \textrightarrow Death & 129 \\
\midrule
\multicolumn{2}{l}{\textit{Censoring}} \\
Censoring from Healthy & 124,868 \\
Censoring from Mild CKD & 4,457 \\
Censoring from Severe CKD & 139 \\
\midrule
\multicolumn{2}{l}{\textit{Interval Length by From-State*}} \\
Overall & 1.44 (0.00--20.50) \\
Healthy & 1.46 (0.00--20.50) \\
Mild CKD & 0.68 (0.00--18.90) \\
Severe CKD & 0.27 (0.00-5.64) \\
\bottomrule
\multicolumn{2}{l}{\footnotesize Ages and interval lengths are reported in years, as median (minimum--maximum).}\\
\multicolumn{2}{l}{\footnotesize CKD: Chronic Kidney Disease; ESKD: End-Stage Kidney Disease.}\\
\multicolumn{2}{l}{\footnotesize *Interval length: time between subsequent visits, i.e., evaluations of CKD stage.}
\end{tabular}
\end{table}

\subsection{Model specification}
\label{ssec:ukb-model-specification}

We draw on the conclusions from Section \ref{sec:simulations} to specify an \emph{SSTS} and an \emph{MTS} PAM with interval mid-point as estimation point.
Here, we choose chronological \textit{age} to be the primary time scale due to the non-informativeness of the study entry time point \citep{wiegrebe2024analyzing}. The baseline hazard model equation of the \emph{SSTS} PAM for transitions $k$ (cf. Figure \ref{fig:state-diagram-ckd}; Equation \eqref{eq:baseline-hazard-ssts}) is
\begin{equation}
    h_{0,k}^{ssts}(age) = \exp \Big( \beta_{0,k} + f_k^{ssts}(age) \Big).
\label{eq:pam-ssts-ukb}
\end{equation}
For the \emph{MTS} PAM, the baseline hazard model equation (cf. Equation \eqref{eq:baseline-hazard-mts}) is
\begin{equation}
\begin{aligned}
    h_{0,k}^{mts}(age, t_1, t_2) &= \exp\Big(
      \beta_{0,k} + 
      \mathbb{1}_{\{0\rightarrow1,1\rightarrow2,2\rightarrow3\}}(k) \cdot f_{03}^{mts}(age) +\\
      & \mathbb{1}_{\{0\rightarrow4,1\rightarrow4,2\rightarrow4\}}(k) \cdot \tilde{f}_{04}^{mts}(age) +\\
      & \mathbb{1}_{\{1\rightarrow2, 2\rightarrow3\}}(k) \cdot f_{13}^{mts}(t_1) 
      + \mathbb{1}_{\{1\rightarrow4, 2\rightarrow4\}}(k) \cdot \tilde{f}_{14}^{mts}(t_1) +\\
      & \mathbb{1}_{\{2\rightarrow3\}}(k) \cdot f_{23}^{mts}(t_2) 
      + \mathbb{1}_{\{2\rightarrow4\}}(k) \cdot \tilde{f}_{24}^{mts}(t_2)
    \Big),
\end{aligned}
\label{eq:pam-mts-ukb}
\end{equation}
%
where $t_1$ (time since onset of Mild CKD) and $t_2$ (time since progression to Severe CKD) are additional time scales (see Figure \ref{fig:sample-trajectories}).

To adjust for dependent left-truncation, we include the two relevant state-entry times: $age_1$ (age at onset of Mild CKD) and $age_2$ (age at progression to Severe CKD); cf. Equation \eqref{eq:baseline-hazard-left-truncation}.

The stratification helper variables required for model identifiability, as defined in Section \ref{sec:pam}, are $trans_{\text{after}\_1\_\text{exact}}$ and $trans_{\text{after}\_2\_\text{exact}}$ for the \emph{SSTS} PAM;
and $trans_{\text{after}\_0}$, $trans_{\text{after}\_1}$, and $trans_{\text{after}\_2}$ for the \emph{MTS} PAM.\\



\noindent
For IEB-related analyses (Section \ref{sssec:ukb-results-ieb}), we include the following risk factors in the model (stratified by transition; cf. Equation \eqref{eq:baseline-hazard-general}):
(i) the genetic variant rs77924615 (effect allele \textit{G});
(ii) sex; 
(iii) a polygenic risk score (PGS) to control for other genetic variants associated with kidney function \citep{stanzick2023kidneygps, wiegrebe2024analyzing}; 
(iv) binary diabetes and smoking status; 
and (v) body mass index (BMI) and urine albumin-to-creatinine ratio (uACR), both from UKB study center visits only.

All risk factors, except genetics and sex, are time-varying.
The effects of time scales ($age$, $t_1$, $t_2$), state-entry times ($age_1$, $age_2$), and quantitative risk factors (PGS, BMI, uACR) are estimated via penalized splines ($k=20$).

\subsection{Results}
\label{ssec:ukb-results}

\subsubsection{Estimation of baseline hazards and state-entry times}
\label{sssec:ukb-results-time-scales}

First, we find that goodness of fit is better for the baseline \emph{SSTS} PAM than for the baseline \emph{MTS} PAM according to the Akaike Information Criterion ($202,412$ versus $202,445$).
This is in line with results from our simulations (Section \ref{sssec:sim-time-scales-results}) regarding the robustness of the \emph{SSTS} PAM even under model misspecification.\\

\begin{figure}[!htbp]
\begin{center}
\includegraphics[width=0.97\linewidth]{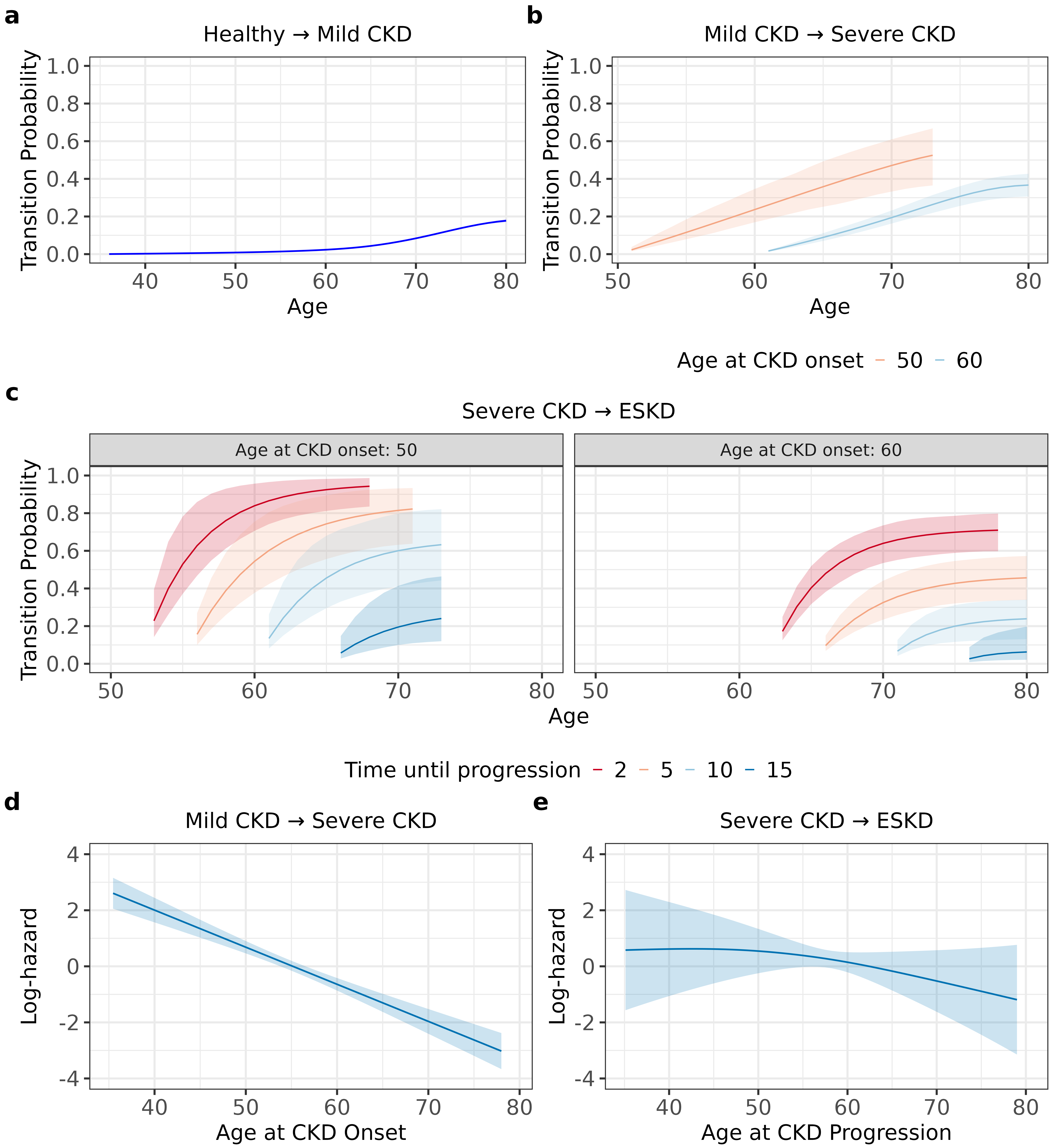}
\end{center}
\vspace{-0.5cm}
\caption{
Transition probabilities and state-entry time effects for CKD onset and progression in UK Biobank.
This figure shows transition probabilities (a-c) and the effects of state-entry times on transition log-hazards (d+e), estimated using an \emph{SSTS} PAM on the UK Biobank data introduced in Section \ref{ssec:ukb-data}.
Panel a shows "Healthy $\rightarrow$ Mild CKD" transition probabilities over age.
Panel b shows "Mild CKD $\rightarrow$ Severe CKD" transition probabilities over age for different values of age at onset of Mild CKD (lines).
Panel c shows "Severe CKD $\rightarrow$ ESKD" transition probabilities over age for different values of age at onset of Mild CKD (facets) and time until progression to Severe CKD (lines).
Panels d and e show the zero-centered smooth effects of state-entry times (age at onset of Mild CKD and age at progression to Severe CKD) on the log-hazards of the respective transitions "Mild CKD $\rightarrow$ Severe CKD" and "Severe CKD $\rightarrow$ ESKD".
Confidence intervals are based on the Bayesian posterior covariance matrix for the coefficients (cf. Section \ref{sec:pam}).}
\label{fig:ukb-baseline-overview}
\end{figure}

\noindent
Second, we consider estimated log-hazards and transition probabilities (Figures \ref{fig:ukb-baseline-overview}a-c and \ref{fig:ukb-baseline-01-04}-\ref{fig:ukb-baseline-24-slice}).
\emph{SSTS} and \emph{MTS} PAMs mostly produce similar estimates of baseline log-hazards and transition probabilities.

Log-hazards for the transition "Healthy $\rightarrow$ Mild CKD" start increasing at age 40 years, further accelerate from age 60 on, and plateau at around age 75. 
The corresponding transition probabilities are close to zero until age 50, then increase. 
For the transition "Healthy $\rightarrow$ Death", log-hazards rise steeply up to age 50, then almost plateau until almost age 65, followed by another increase thereafter. 
The corresponding direct transition probabilities increase exponentially, starting at age 50. 

For transitions out of Mild CKD, evaluation does not only depend on the single time dimension of chronological \textit{age},  but also on $age_1$ (age at onset of Mild CKD) and/or $t_1$ (time since onset of Mild CKD), with $age = age_1 + t_1$; see subjects B and C of Figure \ref{fig:sample-trajectories} with $age_1 = age_{onset}$ and $t_1 = time_{onset}$.
The "Mild CKD $\rightarrow$ Severe CKD" transition probabilities are highest for subjects with early CKD onset (i.e., small $age_1$) and, for a given $age_1$, increase over \textit{age}.
Direct "Mild CKD $\rightarrow$ Death" transition probabilities again increase sharply with \textit{age} (starting at around 50), independently of $age_1$ (Figures \ref{fig:ukb-baseline-12-14-contour} and \ref{fig:ukb-baseline-12-14-slice}).

For transitions out of Severe CKD, we have additional dependence upon $age_2$ (age at progression to Severe CKD) and/or $t_2$ (time since progression to Severe CKD), with $age = age_2 + t_2$; see subject B of Figure \ref{fig:sample-trajectories} with $age_2 = age_{progression}$ and $t_2 = time_{progression}$.
Transition probabilities for "Severe CKD $\rightarrow$ ESKD" rise with decreasing $age_1$ and $age_2$; that is, they are highest for subjects with early transitions into Mild and Severe CKD.
Given $age_1$ and $age_2$, the transition probabilities increase with \textit{age} (Figures \ref{fig:ukb-baseline-23-contour} and \ref{fig:ukb-baseline-23-slice}).
"Severe CKD $\rightarrow$ Death" transitions show a similar pattern as direct "Mild CKD $\rightarrow$ Death" transitions (Figures \ref{fig:ukb-baseline-24-contour} and \ref{fig:ukb-baseline-24-slice}).\\

\noindent
We can now answer the first question from the very beginning:
For a subject with onset of Mild CKD at age 50 and progression to Severe CKD at age 60, the probability of developing ESKD increases from 15\% at age 61 to $>60\%$ over the course of the next 10 years.
Over the same period of time, the probability of dying rises from 5\% to almost 30\%.\\

\noindent
Finally, we observe that the effects of state-entry times (here: age at CKD onset and progression) are mostly linear, though not always (Figures \ref{fig:ukb-baseline-overview}d+e and \ref{fig:ukb-left-truncation}).
This underlines the importance of the modeling flexibility provided by this PAM-based framework. 

\subsubsection{Index event bias and genetic effect estimation}
\label{sssec:ukb-results-ieb}

To investigate the effect of rs77924615 on CKD onset and progression, we use an \emph{SSTS} PAM because we found this model robust to model misspecification (Section \ref{sec:simulations}) and with better goodness of fit than the \emph{MTS} PAM on this dataset (Section \ref{sssec:ukb-results-time-scales}).
In the naive model for testing this variant (omitting other risk factors), we observe a strong effect of rs77924615 on CKD onset log-hazards ($+0.28$), while its effect on log-hazards for progression to Severe CKD is seemingly protective ($-0.12$; Table \ref{tab:ukb-models-risk-factors}).\\

\newcommand{\captionukbmodelsriskfactors}{Effect size estimates of the genetic variant rs77924615 on the log-hazards of CKD onset and progression in UK Biobank.
This table shows effect size estimates along with standard errors and P-values for the genetic variant rs77924615 (effect allele \textit{G}) on the log-hazards of the transitions 0$\rightarrow$1 ("Healthy $\rightarrow$ Mild CKD"), 1$\rightarrow$2 ("Mild CKD $\rightarrow$ Severe CKD"), and 2$\rightarrow$3 ("Severe CKD $\rightarrow$ ESKD"), in the UK Biobank dataset introduced in Section \ref{ssec:ukb-data}.
The models are estimated using an \emph{SSTS} PAM as specified in Section \ref{ssec:ukb-model-specification}, with additional risk factors as stated in the column Risk Factors.
Quantitative risk factors (PGS, BMI, uACR) are modeled via penalized splines ($k=20$).
}
\begin{table}[!ht]
\caption{\captionukbmodelsriskfactors}
\label{tab:ukb-models-risk-factors}
\centering
\begin{tabular}[t]{lcrrr}
\toprule
Risk Factors & Transition & Coefficient & SE & P-value\\
\midrule
\rowcolor{gray!10}  & 0→1 & 0.284 & 0.025 & $<$0.001 \\
\rowcolor{gray!10}  & 1→2 & -0.121 & 0.090 & 0.181 \\
\rowcolor{gray!10} \multirow{-3}{*}{\shortstack[l]{\cellcolor{gray!10}G only\\(AIC: 202,984)}} & 2→3 & -0.311 & 0.146 & 0.034 \\
 & 0→1 & 0.284 & 0.025 & $<$0.001 \\
 & 1→2 & -0.111 & 0.091 & 0.222 \\
\multirow{-3}{*}{\shortstack[l]{G + Sex\\(AIC: 202,275)}} & 2→3 & -0.330 & 0.146 & 0.024 \\
\rowcolor{gray!10}  & 0→1 & 0.282 & 0.025 & $<$0.001 \\
\rowcolor{gray!10}  & 1→2 & -0.113 & 0.091 & 0.214 \\
\rowcolor{gray!10} \multirow{-3}{*}{\cellcolor{gray!10}\shortstack[l]{G + Sex + PGS\\(AIC: 201,845)}} & 2→3 & -0.330 & 0.146 & 0.024 \\
 & 0→1 & 0.283 & 0.025 & $<$0.001 \\
 & 1→2 & -0.107 & 0.091 & 0.237 \\
\multirow{-3}{*}{\shortstack[l]{G + Sex + PGS + Diabetes\\(AIC: 201,644)}} & 2→3 & -0.331 & 0.146 & 0.024 \\
\rowcolor{gray!10}  & 0→1 & 0.280 & 0.026 & $<$0.001 \\
\rowcolor{gray!10}  & 1→2 & -0.086 & 0.095 & 0.364 \\
\rowcolor{gray!10} \multirow{-3}{*}{\cellcolor{gray!10}\shortstack[l]{G + Sex + PGS + Diabetes\\ + Smoking\\(AIC: 184,976)}} & 2→3 & -0.243 & 0.175 & 0.166 \\
 & 0→1 & 0.288 & 0.026 & $<$0.001 \\
 & 1→2 & -0.111 & 0.095 & 0.243 \\
\multirow{-3}{*}{\shortstack[l]{G + Sex + PGS + Diabetes\\ + Smoking + BMI\\AIC: (182,436)}} & 2→3 & -0.230 & 0.174 & 0.187 \\
\rowcolor{gray!10}  & 0→1 & 0.278 & 0.026 & $<$0.001 \\
\rowcolor{gray!10}  & 1→2 & 0.045 & 0.100 & 0.654 \\
\rowcolor{gray!10} \multirow{-3}{*}{\cellcolor{gray!10}\shortstack[l]{G + Sex + PGS + Diabetes\\ + Smoking + BMI + uACR\\(AIC: 175,181)}} & 2→3 & -0.159 & 0.178 & 0.371 \\
\bottomrule
\end{tabular}
\parbox{\linewidth}{\footnotesize
\begin{itemize}[leftmargin=*, noitemsep, label=\textcolor{white}{\textbullet}]
 \item G: Genetic variant rs77924615 (effect allele \textit{G}).
 \item 0: Healthy; 1: Mild CKD; 2: Severe CKD; 3: ESKD.
 \item AIC: Akaike Information Criterion.
\end{itemize}
}
\end{table}

\noindent
However, these naive estimates are potentially subject to IEB due to the omission of other risk factors (cf. Section \ref{ssec:sim-ieb}).
For example, we observe a higher prevalence of diabetes and larger values of uACR for rs77924615 genotype AA compared to the other genotypes in state 2 (Table \ref{tab:ukb-risk-factor-distributions}).



The most straightforward approach is to directly adjust the model for other risk factors, under the assumption that all relevant other risk factors are known and observable and, thus, no unmeasured confounding remains after adjusting for them. 
Here we include sex, PGS, diabetes, smoking, BMI, and uACR (see Section \ref{ssec:ukb-model-specification}).
Inclusion of these risk factors leaves the genetic effect on CKD onset largely unchanged, while substantially attenuating the genetic effect on progression to Severe CKD from $-0.12$ to $+0.05$ (Table \ref{tab:ukb-models-risk-factors}).
Additionally, the stepwise inclusion of the above risk factors also improves the model fit in terms of the Akaike Information Criterion.
So while there is sound evidence for a hazard-increasing effect of the \textit{UMOD} genetic variant on CKD onset, there seems to be neither a detrimental nor a protective effect on CKD progression.
This answers the second question posed at the very beginning of this paper.\\

\noindent
An alternative to the direct adjustment for other risk factors is to use propensity scores to compute subject- and state-specific stabilizing weights for the estimation of a weighted PAM (Appendix \ref{app-ssec:ukb-ps}).
This weighting approach produces similar results as the adjustment approach regarding the estimation of rs77924615 effects on CKD onset and progression (Table \ref{tab:ukb-adjustment-weighting}).
However, we note that the incorporation of time-varying covariates requires the estimation of time-varying weights and, while feasible, is much more complex \citep{hernan2000marginal, hernan2009observation, hernan2020causal}.

\subsubsection{Evaluating the impact of interval-censoring on the CKD multi-state model estimation}
\label{sssec:ukb-results-ic}

Due to the dense eHRs in our dataset (median time between eGFR measurements: 1.44 years), the overall degree of interval-censoring is limited.
Furthermore, death time points are not affected by interval-censoring as the corresponding information stems from exactly recorded death registries.
As expected, we observe informative observation \citep{sperrin2017informative, bible2024accounting} as interval length is related to CKD stage (median time between observations: 1.46, 0.68, and 0.27 years for Healthy, Mild CKD, and Severe CKD, respectively).
This means that less healthy individuals -- i.e., those who are more likely to have CKD-related events -- tend to have more detailed eHR trajectories and, hence, smaller interval lengths (cf. Section \ref{ssec:challenges-ic}).
As suggested in the literature \citep[cf.][Section 5.4]{cook2018multistate}, Poisson processes can be used to handle informative observation -- and our PAM-based framework is inherently linked to a Poisson process via likelihood equivalence.
In addition, the flexibility of our non-Markov framework allows for further mitigation of any potential biases from informative observation via direct incorporation of (history-dependent, time-varying) characteristics of the observation process into the model -- for example in the form of the number of visits or mean inter-visit time.

We perform sensitivity analyses on the results in Section \ref{sssec:ukb-results-ieb} by using two different estimation points -- interval mid- versus end-points -- as event times and evaluating the impact on risk factor effect size estimates.
As can be seen in Table \ref{tab:ukb-ic}, estimates are similar for mid- and end-point estimation -- as expected based on the results from Section \ref{sssec:sim-ic-results}.

\section{Discussion}
\label{sec:discussion}

In this work, we investigated the analysis of multi-stage disease histories via multi-state models, along with the resulting statistical challenges and their impact on model estimation. 
Via simulation studies and a real-world application in UKB, we demonstrated that most of these challenges can be addressed within a PAM-based modeling framework.
This framework facilitates:
adjustment for dependent left-truncation by including state-entry times; 
accommodation of multiple time scales while ensuring identifiability;
seamless incorporation of time-varying covariates (and effects) to account for OVB and IEB; 
and flexible non-linear effect estimation.\\

\noindent
Regarding the choice of time scales, a stratified single time scale proved more robust to model misspecification in terms of baseline hazard estimation than using multiple time scales.
The \emph{SSTS} PAM also showed a better fit on the empirical data from UKB.
Interestingly, this is also in line with the leukemia example by Iacobelli and Carstensen \citep{iacobelli2013multiple}, whose final model eventually does not contain multiple time scales either. 
Therefore, a relatively simple (single time scale) model appears to suffice even for estimation of hazards from complex DGPs with multiple time scales.
In terms of estimating the hazards of fixed effects (e.g., risk factors), both approaches were found to be robust.

As all non-initial states in a multi-state model are restricted to index events, IEB is of particular importance.
We demonstrated the induction of negative correlation within a "diseased subpopulation" between risk factors that are independent in the healthy population.
This caused attenuation bias when one of the risk factors was omitted from the model.
We then showed that this bias can be corrected for by including the omitted risk factor into the model or by using a propensity score-based weighting approach.
The PAM framework facilitates flexible covariate adjustment as well as weighting.

Furthermore, we showed that PAMs -- using interval mid-points as estimation points -- recovered fixed effect hazards very well even with interval-censored data, making this framework particularly suitable to disease histories derived from longitudinal data.\\

\noindent
On UKB data, \emph{SSTS} and \emph{MTS} PAMs mostly produced similar estimates of CKD onset and progression hazards and transition probabilities.
In particular, we found progression risk to Severe CKD to be highest for subjects with early onset of Mild CKD and to further rise with age.
From there, the risk of further progressing to ESKD was highest for subjects with early onset of Mild CKD and progression to Severe CKD, again increasing with age.
We also found the \textit{G} allele of the genetic variant rs77924615 (\emph{UMOD} locus) to increase the risk of CKD onset, in line with literature \citep{devuyst2018umod, wiegrebe2024analyzing}.
Importantly, we observed an initial protective effect of this \textit{G} allele on progression into Severe CKD and ESKD;
yet this effect disappeared after adjustment for other risk factors of CKD and CKD progression.
The initial protective effect thus seems to be attributable to IEB.\\

\noindent
However, we also identified limitations, which in turn provide opportunities for future research.

First, the individual simulation setups could be expanded even further.
The simulations regarding choice of time scales and IEB, for instance, could be enhanced by also considering other DGPs beyond the piecewise exponential distribution -- though we do note that the latter one is particularly flexible and that the simulations on interval-censoring underlined the ability of PAMs to adequately estimate multiple distinct baseline hazards.
In addition, it would be interesting to expand the IEB simulations to scenarios with dependence structures between risk factors in the healthy population.
In particular, for specific application settings it could be informative to directly use empirical joint distributions of risk factors for simulations.

Second, the recovery of baseline hazards by PAMs under interval-censoring is still suboptimal.
This is because direct incorporation of interval-censoring into the PEM likelihood, while maintaining its proportionality to a Poisson likelihood, remains a challenge.
In this context, it would also be of interest to conduct further simulations regarding the impact of interval-censoring within complex multi-state model settings, because interval-censored transition times out of the initial state also induce censoring in the observed state-entry times of subsequent states.

Third, our proposed handling of IEB relies upon other risk factors being observable and on the assumption of no unmeasured confounding remaining after risk factor adjustment.
Both of these are often not (or only partially) the case: In our UKB dataset, for instance, the risk factors BMI and uACR are available only from study center assessments, but not from eHRs;
and in addition, it is very likely that further (unknown, unaccounted for) risk factors exist, which is why some degree of unmeasured confounding cannot be ruled out here.
And while off-the-shelf bias correction methods for linear and binary outcomes exist \citep{dudbridge2019adjustment, mahmoud2022robust, donovan2025application}, their applicability to multi-state models is still to be explored.

Finally, as UKB participants are rather healthy \citep{fry2017comparison}, CKD progressions in this UKB dataset are relatively rare.
This, in turn, makes the estimation of (especially small-scale, genetic) risk factors more difficult.
It would thus be interesting to conduct similar CKD onset and progression analyses in less healthy cohorts.

\newpage

\section*{Data availability}
The UK Biobank analyses in Section \ref{sec:application} were conducted under the application numbers 20272 and 23940.
UK Biobank is a publicly accessible database. Individual participant data from UKB is available via the UK Biobank resource.

\section*{Code availability}
Multi-state PAMs (including all necessary preprocessing) are implemented in the \textsf{R} package \texttt{pammtools} \citep{piller2026pammtools}, which uses \texttt{mgcv::gam()/bam()} \citep{wood2015package} for efficient and flexible estimation.
All code required for running the simulations (Section \ref{sec:simulations}), for the UK Biobank analyses (Section \ref{sec:application}), as well as for creating the corresponding figures and tables, is available from the GitHub repository 
\url{https://github.com/survival-org/msm4diseaseHistories}.




\clearpage


\appendix
\setcounter{table}{0}
\setcounter{figure}{0}
\renewcommand{\thetable}{A.\arabic{table}}
\renewcommand{\thefigure}{A.\arabic{figure}}
\section{Appendix}
\label{app}

\subsection{Additional information on the flexible modeling of disease histories using piecewise exponential additive models}
\label{app-sec:pam}

\subsubsection{Transformation to piecewise exponential data}
\label{app-ssec:ped}
The preprocessing step to transform standard survival data into the PED format starts by partitioning the follow-up time into $J$ intervals $(0=a_0,a_1]$, $(a_1,a_2]$, $\ldots$, $(a_{j-1},a_j]$, $\ldots$, $(a_{J-1},a_J]$ using cut points $a_0 < a_1 < \cdots < a_{J-1} < a_J$. 
$I_j:=(a_{j-1},a_j]$ denotes the $j$-th time interval and $J_i \in \{1, \dots, J\}$ indexes the time interval for which $y_i$ was observed ($y_i \in I_{J_i} = (a_{J_i-1},a_{J_i}]$).
Using this partitioning, the original dataset is transformed into the long-form PED, which contains one row for each unique combination of subject $i$ and interval-at-risk $j$ ($\forall j \in \{1, \dots, J_i\}$), as well as the following newly introduced variables:
\begin{itemize}
    \item an event indicator $d_{ij} = \begin{cases} 1 & \text{if } y_i \in I_j \wedge \ d_i = 1\\ 0 & \mbox{else} \end{cases}$,
    \item the time at risk $y_{ij} = \begin{cases} a_j - a_{j-1} & \text{if } a_j < y_i\\ y_i - a_{j-1} & \text{if } a_{j-1} < y_i \leq a_j\end{cases}$,
    \item and an "offset" $o_{ij} = \log (y_{ij})$.
\end{itemize}
Note that accounting for left-truncation, time-varying covariates, and time-varying effects all require similar data transformations and can therefore seamlessly be incorporated into this data preprocessing step \citep{piller2025reduction}. 
To capture changes in time-varying covariates accurately, additional time cut points can be added at times where the time-varying covariates change \citep{bender2018pammtools}.

PEMs then use $d_{ij}\sim Po(\mu_{ij})$ as outcome of a Poisson regression model with expected conditional event rate $\mu_{ij}=h_{ij}y_{ij}$ and logarithmized time-at-risk entering as offset $o_{ij}$, to estimate the hazard rate $h_{ij}$ as the product of a piecewise constant baseline hazard rate (for each interval) and the exponential of a linear predictor.
PEMs can be shown to be equivalent to the Cox Proportional Hazards model under certain assumptions \citep{whitehead1980fitting}, but since they do not rely on the non-parametric estimation of the baseline hazard (due to assuming piecewise-constant hazards), they are fully parametric with respect to parameter estimation.
Still, we here refer to PEMs as semi-parametric approach, analogously to the Cox model, because no parametric event time distribution is assumed - in contrast to parametric approaches such as Accelerated Failure Time models \citep{kalbfleisch2002statistical}.

\subsubsection{Augmentation of piecewise exponential data for multi-state models}
\label{app-ssec:ped-msm}

To simplify the multi-state process, we begin by analyzing each state separately, thereby reducing the problem to a set of nested competing risks models. Specifically, for each state $\ell$, we assume there are $q_{\ell}$ possible transitions - each corresponding to a competing risk. For each of these $q_{\ell}$ competing risks (transitions), we construct a separate dataset. In the dataset corresponding to transition $k$, we replace the event indicator $d_{ij}$ by a cause-specific event indicator:
\[
d_{ijk} = 
\begin{cases}
1 & \text{if } d_{ij} = 1 \text{ and } d_{ik} = 1, \\
0 & \text{otherwise}
\end{cases}
\]
where $d_{ik}$ is a binary indicator of whether the event was of type $k$. 
Effectively, in the $k$-th dataset all other transitions $\tilde{k} \neq k$ are thus treated as censored. 
This data transformation step can also be interpreted as an augmentation procedure involving counterfactual transitions, where, for each possible transition $k$, we consider what would happen under the assumption that only $k$ were observable.
As subjects enter transition-specific risk sets at different points in time, the multi-state transformation accounts for this left-truncation by including the state entry time in addition to the transition time $y_{ij}$. 
In case of multiple visits to a state (due to recurrent events), episode $e$ can be added to the transformed data as additional column.
Finally, all state-specific datasets are stacked.

Following the example in \citep{piller2026pammtools}, which extends the partition of the follow-up from single-endpoint to multi-state analysis, Table \ref{tab:ped-transformation} displays the PED transformation for patients A and B from Figure \ref{fig:sample-trajectories}. The follow-up (measured on the age scale) is partitioned into intervals, with cut points $a_0=0, a_1=35, a_2=50, a_3=55, a_4=60, a_5=75$ set as the event and left-truncation times. For each interval $I_{j}$, $d_{ij}$ indicates whether a transition to state $\text{state}_{ij}$ occurs at the end of the interval, $y_{ij}$ is the time at risk within the interval, and $\text{age}_{ij}$ is the age at the end of the interval. In addition to age, the time-varying covariates $t_1$ and $t_2$ encode the time elapsed since onset of mild CKD and since progression to severe CKD, respectively, and are set to $0$ prior to the corresponding state entry.

Three aspects of this example are essential to multi-state PED transformations. First, both patients are left-truncated, entering the study at ages 55 and 35, respectively. Consequently, all intervals prior to each patient's age at study entry are removed from the PED. Second, patient A is censored at age 72, which does not coincide with an interval boundary. As a result, the time at risk $y_{ij}$ in interval $j=3$ is shorter than the full interval length. Third, for every occupied state, all competing transitions that could have occurred but did not are included as additional rows, ensuring that the set of patients at risk for each transition is correctly represented. The cause-specific event indicator is set to $0$ in these rows to reflect that the corresponding transition did not take place.





\subsection{Additional information on CKD onset and progression analyses in UK Biobank}
\label{app-sec:application}

\subsubsection{Propensity score-based correction for index event bias}
\label{app-ssec:ukb-ps}

The idea of the weighting approach is to create subject- and state-specific (i.e., transition-specific) weights that reflect how likely a subject is to have a certain rs77924615 genotype, given the values of their other risk factors.
These weights can then be fed as an argument into the PAM estimation algorithm.

To calculate the weights, we first compute the marginal probability of a subject having their true rs77924615 genotype (versus not having it), separately for each from-state.
This is identical to the state-specific relative frequency of that genotype.
Second, we compute the propensity score -- the probability of a subject having their true rs77924615 genotype (versus not having it), given their set of risk factors and the from-state -- via multinomial logistic regression.
The weights are calculated by dividing the subject- and state-specific relative frequency of the genotype by the respective propensity score.
A large weight means that a subject, given their risk factors and the from-state, is unlikely to have a certain genotype;
such an observation is accordingly up-weighted during model estimation.

\newpage

\subsection{Supplementary figures}
\label{app-sec:figures}

\begin{figure}[!ht]
\begin{center}
\includegraphics[width=0.9\linewidth]{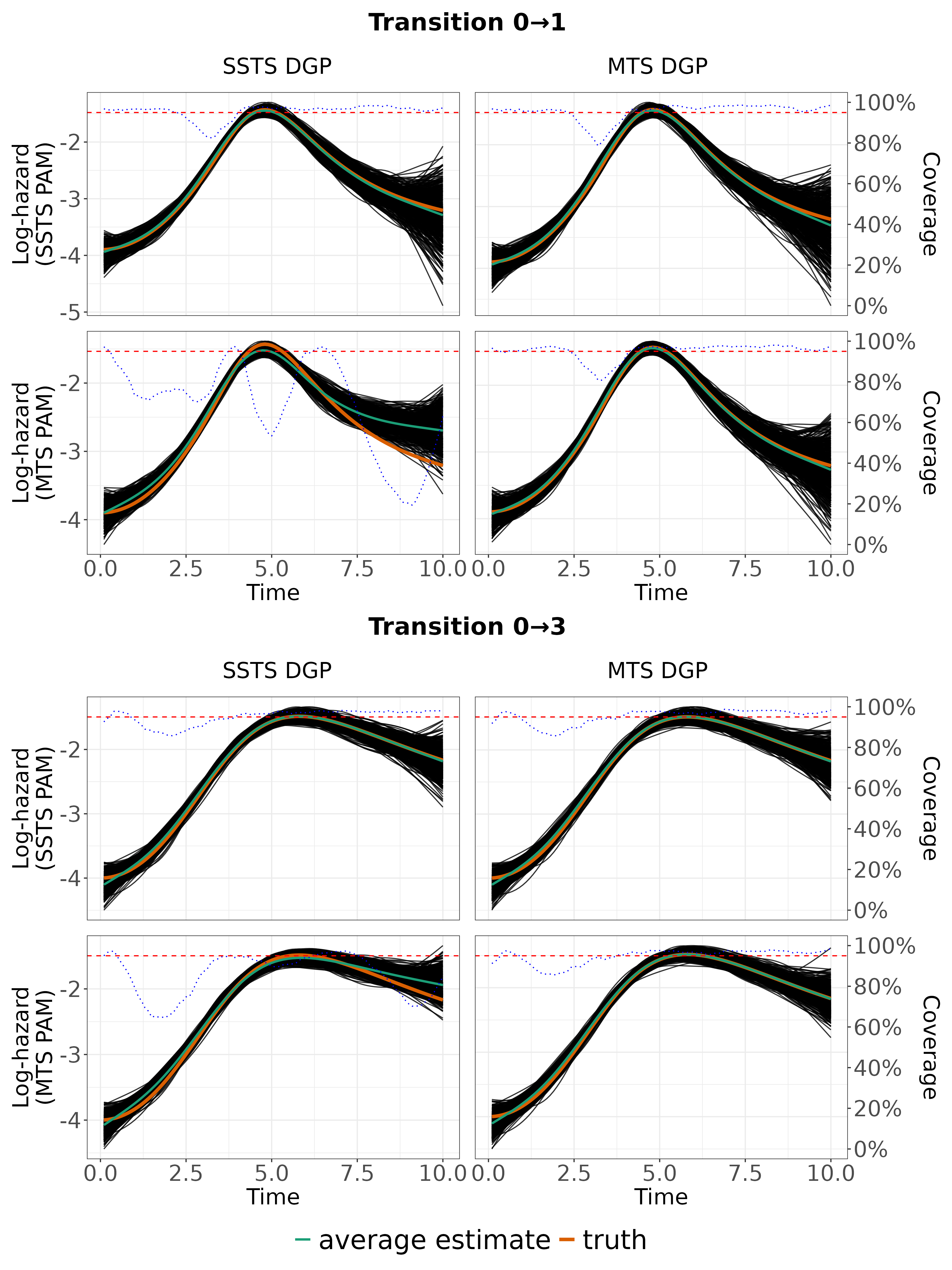}
\end{center}
\vspace{-0.5cm}
\caption{Line plots and coverage of \emph{SSTS} and \emph{MTS} PAMs on data simulated from \emph{SSTS} and \emph{MTS} DGPs. 
This figure shows line plots (solid black lines) and point-wise coverage (dotted blue line) of baseline log-hazards for the transitions $0 \rightarrow 1$ (top) and $0 \rightarrow 3$ (bottom), estimated on data simulated from an \emph{SSTS} DGP (left) and an \emph{MTS} DGP (right), using either an \emph{SSTS} PAM or an \emph{MTS} PAM (only penalized splines smooth for illustration).
The green and orange lines represent average estimates and the true log-hazard, respectively.
The dashed red line corresponds to nominal (95\%) coverage.}
\label{fig:sim-ts-bh-line-plots}
\end{figure}

\begin{figure}[!ht]
\begin{center}
\includegraphics[width=1.0\linewidth]{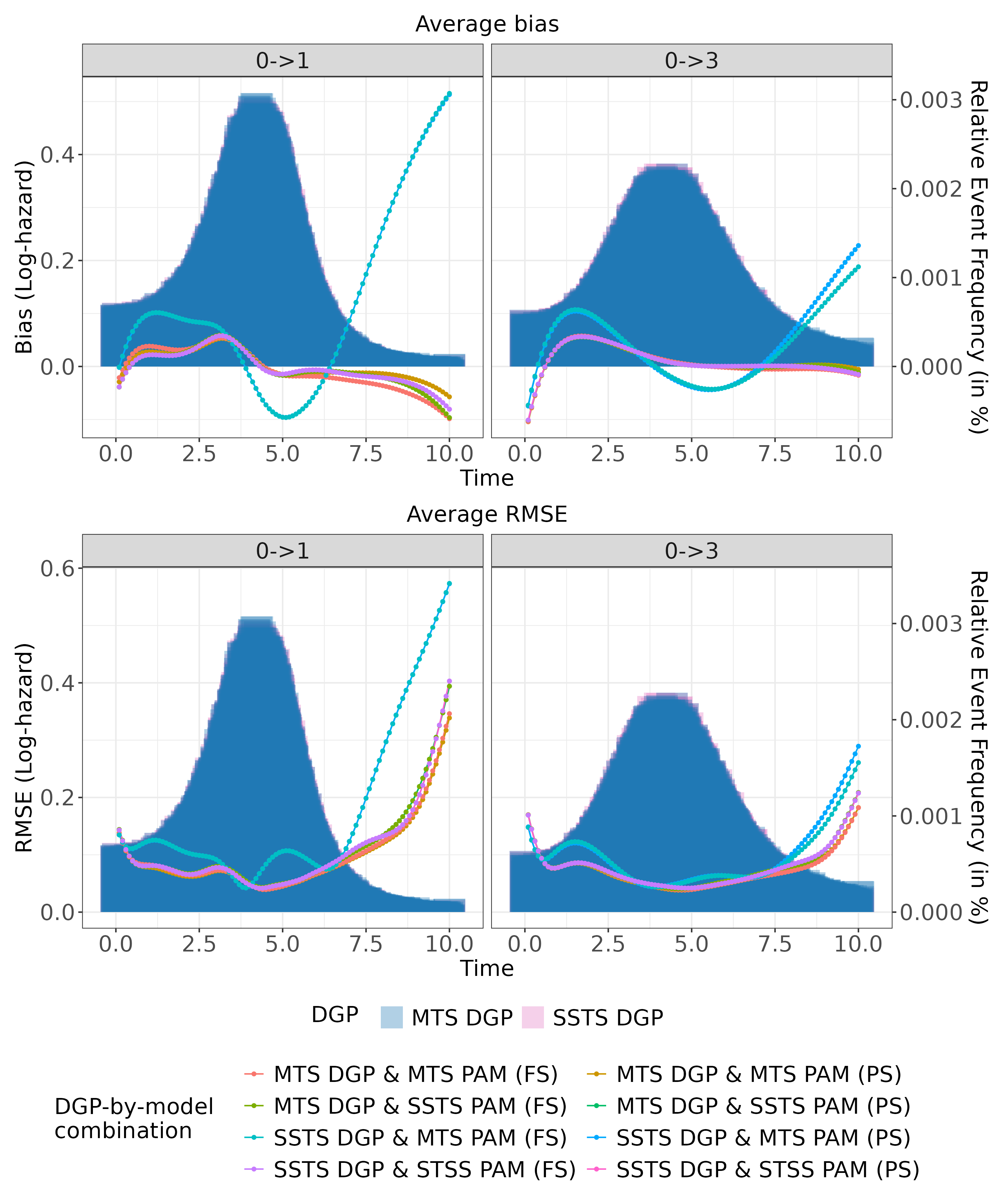}
\end{center}
\vspace{-0.5cm}
\caption{Bias and RMSE of \emph{SSTS} and \emph{MTS} PAM-based log-hazards over time on data simulated from \emph{SSTS} and \emph{MTS} DGPs. This figure shows log-hazard bias and RMSE (left y-axis) over time (x-axis) and averaged across simulation runs for the transitions $0 \rightarrow 1$ (top) and $0 \rightarrow 3$ (bottom), estimated on data simulated from an \emph{SSTS} DGP and an \emph{MTS} DGP, using either an \emph{SSTS} PAM or an \emph{MTS} PAM and either a penalized splines smooth or a factor smooth.
In addition, this figure shows relative event frequencies over time (averaged across simulation runs; right y-axis) for these two transitions and each of the two DGPs.}
\label{fig:sim-ts-bh-bias_rmse}
\end{figure}

\begin{figure}[!ht]
\begin{center}
\includegraphics[width=1.0\linewidth]{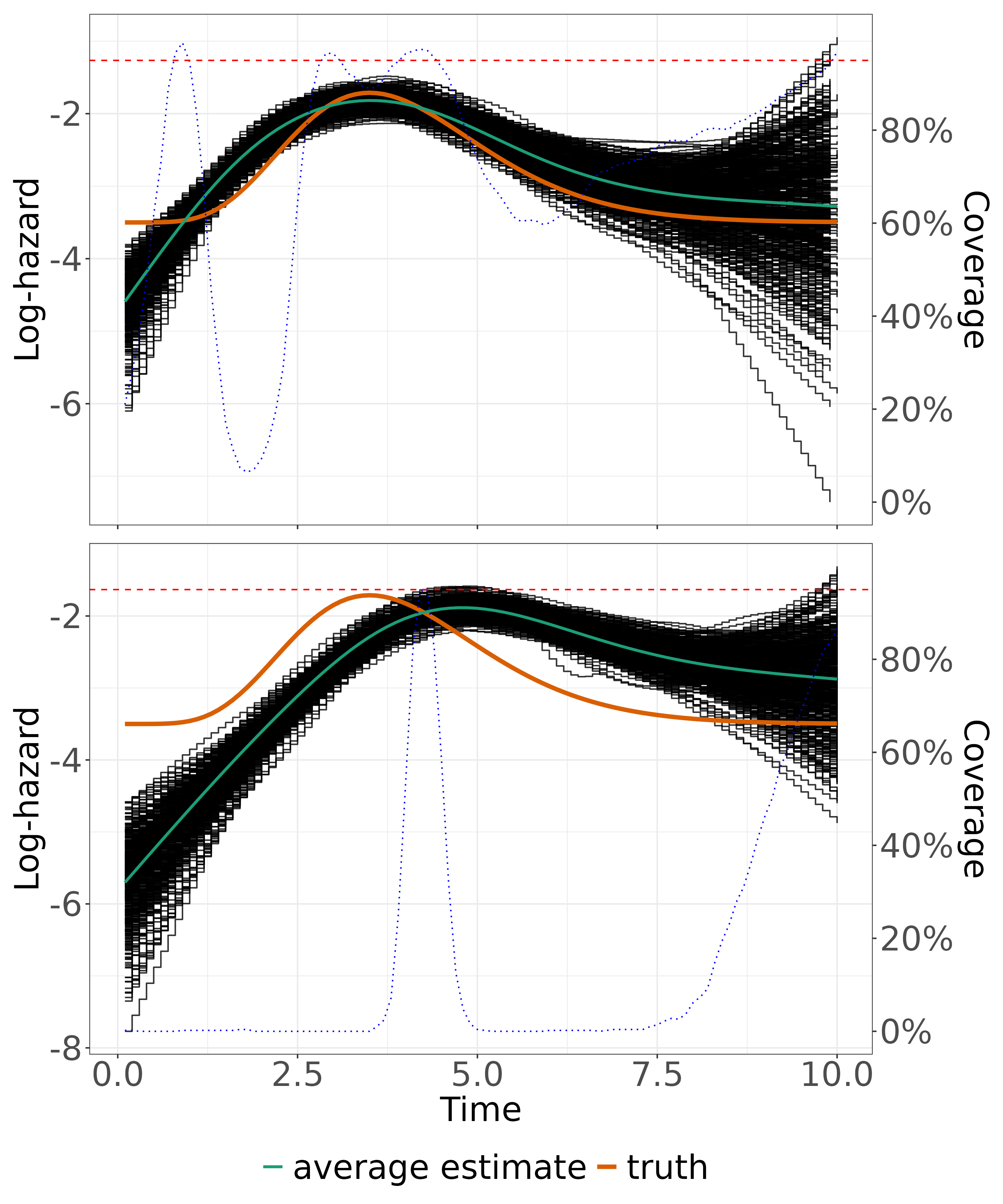}
\end{center}
\vspace{-0.5cm}
\caption{Line plots and point-wise coverage of PAM-based log-hazards over time under interval-censoring using interval mid- and end-points. 
This figure shows line plots (in black) of log-hazards on data simulated from a piecewise exponential DGP with beta IC mechanism, estimated using PAMs with interval mid-points (top) and end-points (bottom) as estimation points.
The green and orange lines represent average estimates and the true log-hazard, respectively.
The dotted blue line denotes pointwise coverage.
The dashed red line corresponds to nominal (95\%) coverage.}
\label{fig:sim-ic-bh-line-plots}
\end{figure}

\begin{figure}[!ht]
\begin{center}
\includegraphics[width=1.0\linewidth]{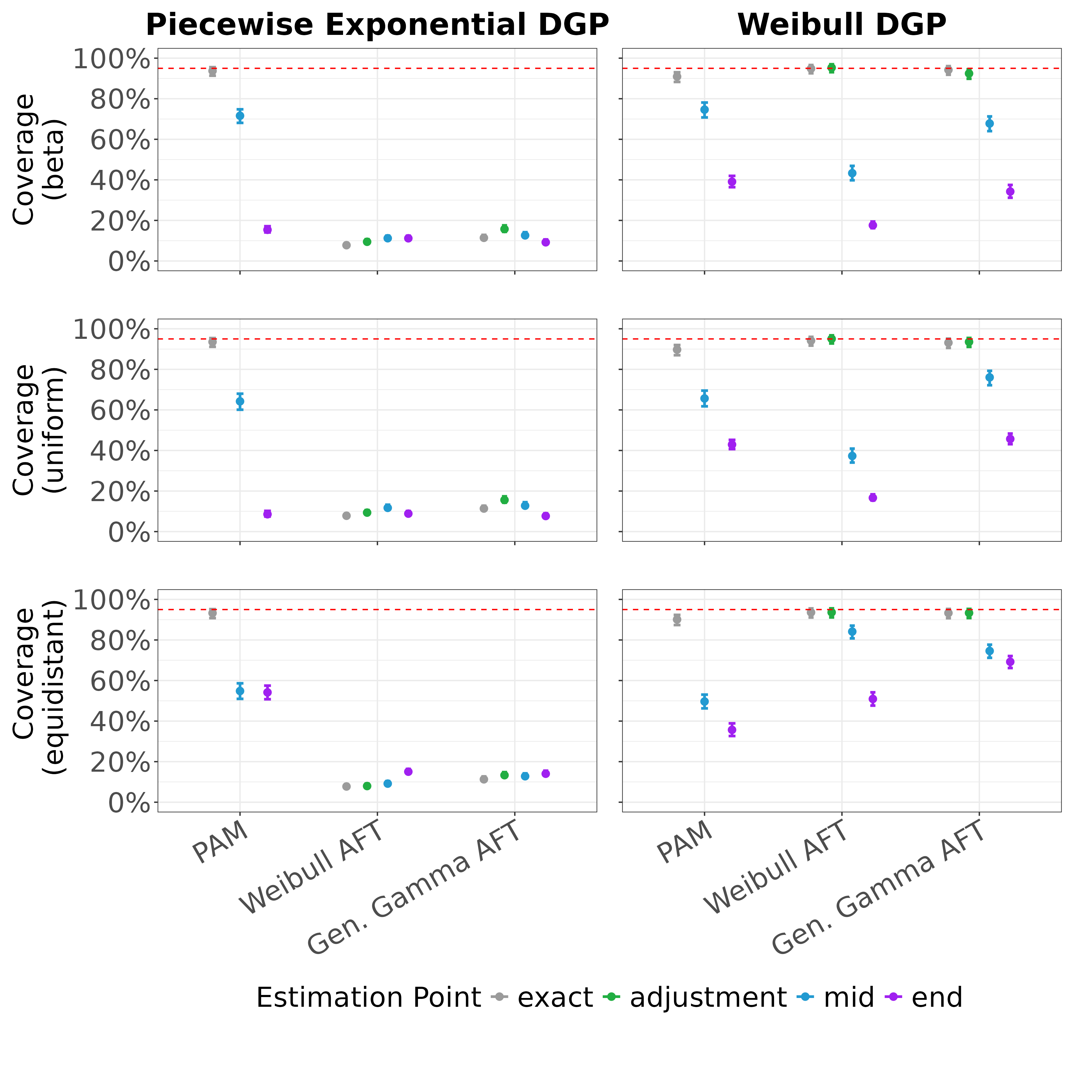}
\end{center}
\vspace{-0.5cm}
\caption{Mean coverage of baseline log-hazards under interval-censoring. 
This figure shows mean baseline log-hazard coverage of three models (x-axis) using three to four distinct estimation points (color coding) on data generated by two DGPs (columns) and three IC mechanisms (rows), based on the simulation setup described in Section \ref{sssec:sim-ic-setup}.
For each model and combination of DGP and IC mechanism, the dot and error bars represent mean coverage and 95\% confidence intervals (from an exact binomial test), computed pointwise across all simulation runs and then averaged over time points.
The dashed red line corresponds to nominal (95\%) coverage.
The results from the \texttt{icenReg}-based DGP are similar to those from the Weibull-based DGP with uniform IC mechanism (see Table \ref{tab:sim-ic-bh-coverage-loghazard}).
}
\label{fig:sim-ic-bh-coverage-loghazards}
\end{figure}

\begin{figure}[!ht]
\begin{center}
\includegraphics[width=1.0\linewidth]{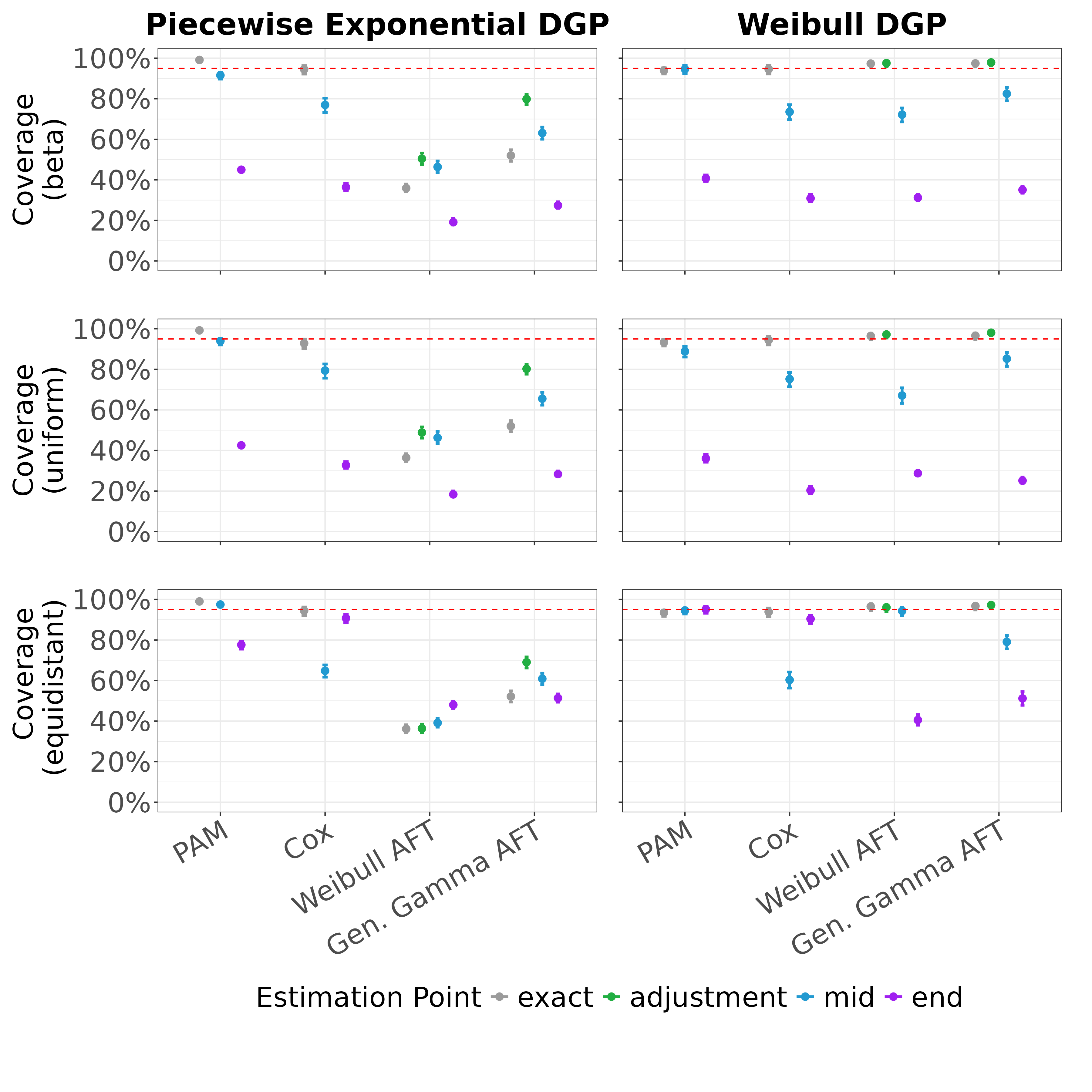}
\end{center}
\vspace{-0.5cm}
\caption{Mean coverage of baseline cumulative hazards under interval-censoring. 
This figure shows mean baseline cumulative hazard coverage of four models (x-axis) using three to four distinct estimation points (color coding) on data generated by two DGPs (columns) and three IC mechanisms (rows), based on the simulation setup described in Section \ref{sssec:sim-ic-setup}.
For each model and combination of DGP and IC mechanism, the dot and error bars represent mean coverage and 95\% confidence intervals (from an exact binomial test), computed pointwise across all simulation runs and then averaged over time points.
The dashed red line corresponds to nominal (95\%) coverage.
The results from the \texttt{icenReg}-based DGP are similar to those from the Weibull-based DGP with uniform IC mechanism (see Table \ref{tab:sim-ic-bh-coverage-loghazard}).
}
\label{fig:sim-ic-bh-coverage-cumu}
\end{figure}

\begin{figure}[!ht]
\begin{center}
\includegraphics[width=1.0\linewidth]{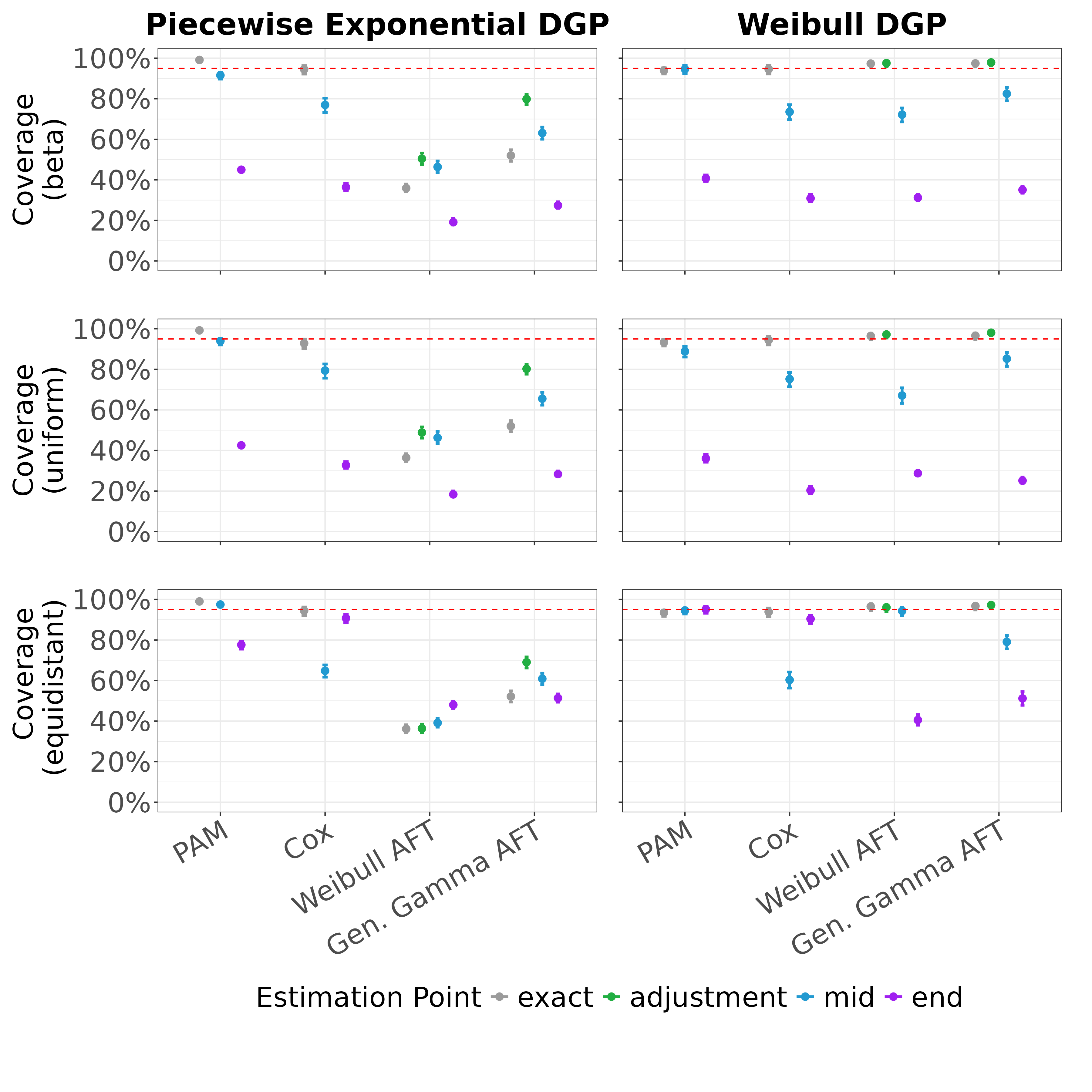}
\end{center}
\vspace{-0.5cm}
\caption{Mean coverage of baseline survival functions under interval-censoring. 
This figure shows mean baseline survival function coverage of four models (x-axis) using three to four distinct estimation points (color coding) on data generated by two DGPs (columns) and three IC mechanisms (rows), based on the simulation setup described in Section \ref{sssec:sim-ic-setup}.
For each model and combination of DGP and IC mechanism, the dot and error bars represent mean coverage and 95\% confidence intervals (from an exact binomial test), computed pointwise across all simulation runs and then averaged over time points.
The dashed red line corresponds to nominal (95\%) coverage.
The results from the \texttt{icenReg}-based DGP are similar to those from the Weibull-based DGP with uniform IC mechanism (see Table \ref{tab:sim-ic-bh-coverage-loghazard}).
}
\label{fig:sim-ic-bh-coverage-surv}
\end{figure}

\begin{figure}[!ht]
\begin{center}
\includegraphics[width=0.7\linewidth]{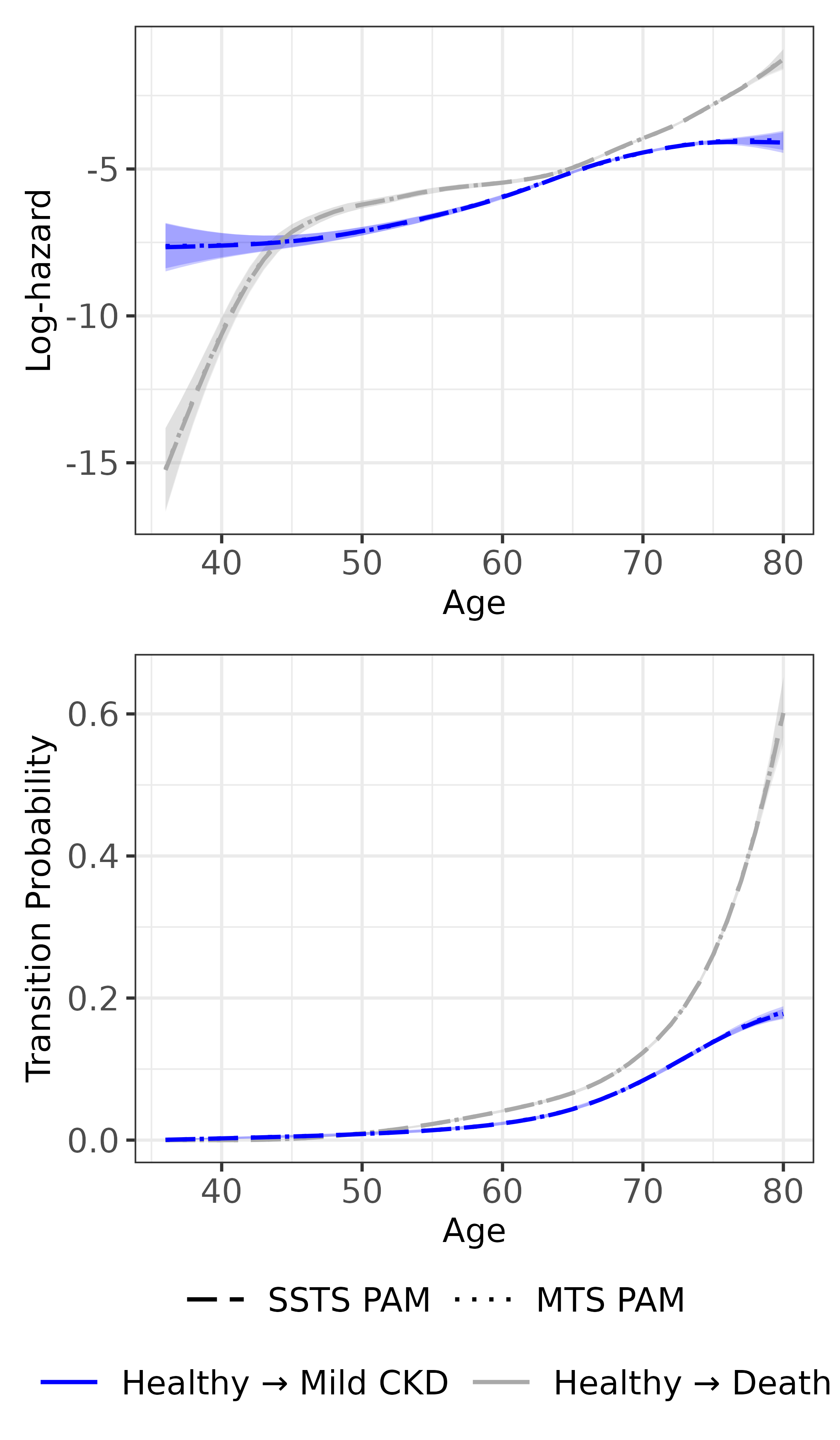}
\end{center}
\vspace{-0.5cm}
\caption{Predicted log-hazards and transition probabilities over age for transitions "Healthy $\rightarrow$ Mild CKD" and "Healthy $\rightarrow$ Death" in UK Biobank.
This figure shows log-hazards (top) and transition probabilities (bottom) for the transitions from state Healthy into states Mild CKD (grey lines) and Death (blue lines), estimated using an \emph{SSTS} PAM (dashed lines) and an \emph{MTS} PAM (dotted lines) on the UK Biobank data introduced in Section \ref{ssec:ukb-data}.
Shades depict 95\% confidence bands.
}
\label{fig:ukb-baseline-01-04}
\end{figure}

\begin{figure}[!ht]
\begin{center}
\includegraphics[width=0.6\linewidth]{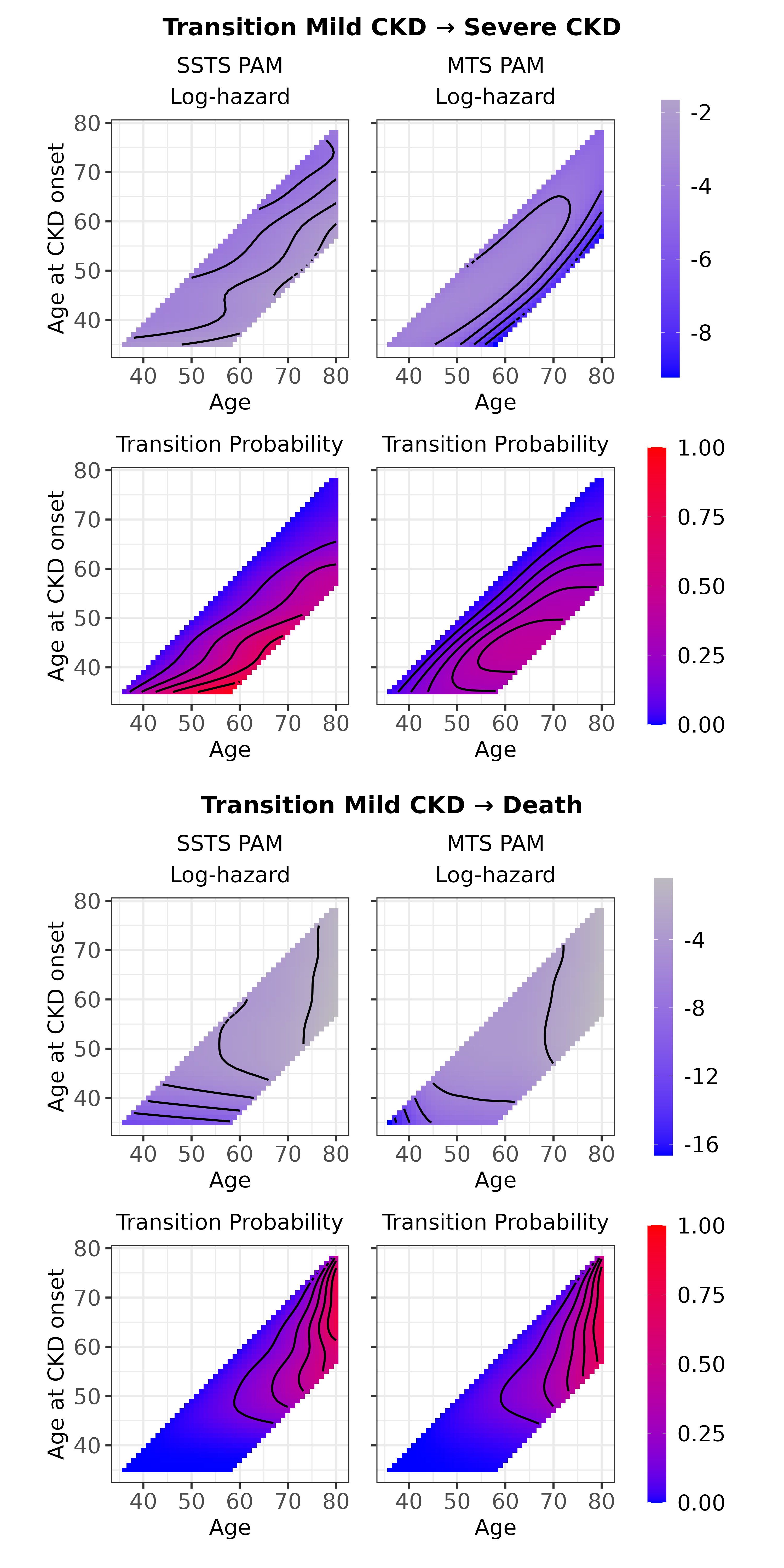}
\end{center}
\vspace{-0.5cm}
\caption{Contour plots of predicted log-hazards and transition probabilities over age and age at CKD onset for transitions "Mild CKD $\rightarrow$ Severe CKD" and CKD $\rightarrow$ Death in UK Biobank.
For the direct transitions from Mild CKD to Severe CKD (top) and Death (bottom), this figure shows contour plots of log-hazards (top rows) and transition probabilities (bottom rows) over age and age at CKD onset, estimated using an \emph{SSTS} PAM (left column) and an \emph{MTS} PAM (right column) on the UK Biobank data introduced in Section \ref{ssec:ukb-data}.
By definition, age must exceed age at CKD onset.
Prediction horizons within Mild CKD are restricted to the empirically observed range (time since progression to CKD $\in [0,23]$ years).
Grey lines represent contour lines.
}
\label{fig:ukb-baseline-12-14-contour}
\end{figure}

\begin{figure}[!ht]
\begin{center}
\includegraphics[width=1.0\linewidth]{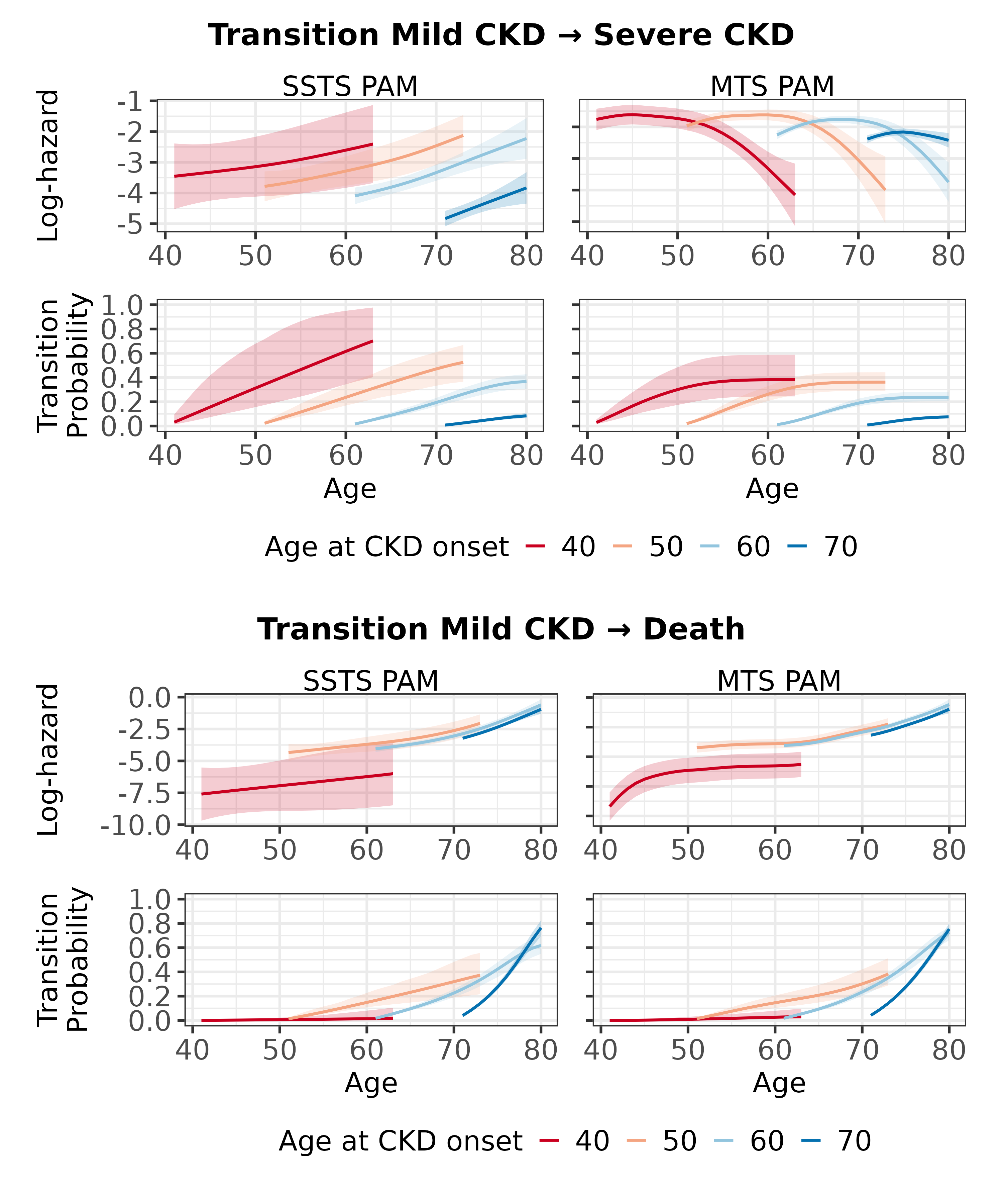}
\end{center}
\vspace{-0.5cm}
\caption{Slice plots of predicted log-hazards and transition probabilities for transitions "Mild CKD $\rightarrow$ Severe CKD" and CKD $\rightarrow$ Death over age for different values of age at CKD onset in UK Biobank.
For the direct transitions from Mild CKD to Severe CKD (top) and Death (bottom), this figure shows slice plots of log-hazards (top rows) and transition probabilities (bottom rows) over age for four distinct values of age at CKD onset, estimated using an \emph{SSTS} PAM (left column) and an \emph{MTS} PAM (right column) on the UK Biobank data introduced in Section \ref{ssec:ukb-data}.
Shades depict 95\% confidence bands.
By definition, age must exceed age at CKD onset.
Prediction horizons within Mild CKD are restricted to the empirically observed range (time since CKD onset $\in [0,23]$ years).
}
\label{fig:ukb-baseline-12-14-slice}
\end{figure}

\begin{figure}[!ht]
\begin{center}
\includegraphics[width=0.95\linewidth]{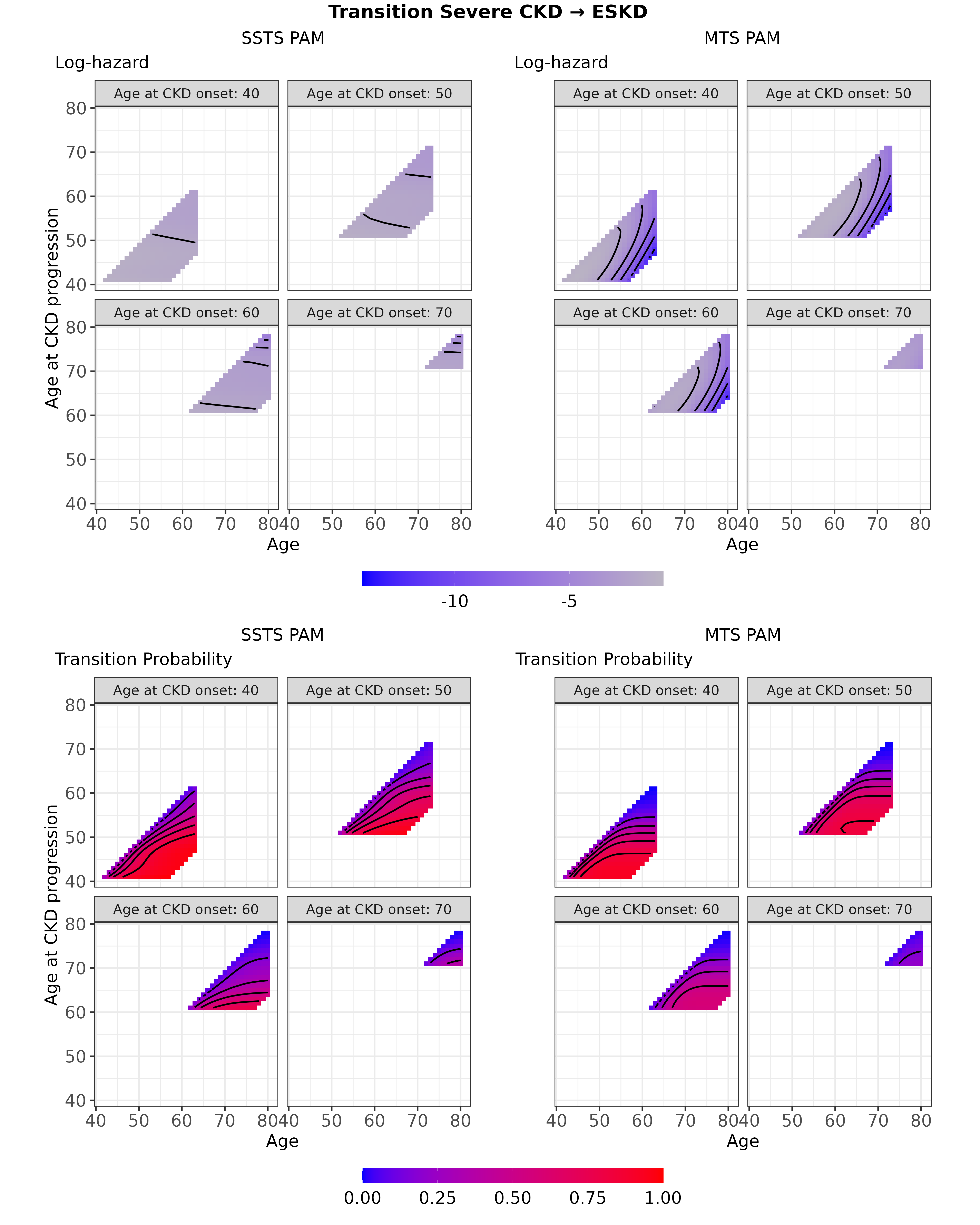}
\end{center}
\vspace{-0.5cm}
\caption{Contour plots of predicted log-hazards and transition probabilities for transition "Severe CKD $\rightarrow$ ESKD" over age and age at CKD progression for different values of age at CKD onset in UK Biobank.
For the transition "Severe CKD $\rightarrow$ ESKD", this figure shows contour plots of log-hazards (top) and transition probabilities (bottom) over age and age at progression to Severe CKD for four distinct values of age at CKD onset, estimated using an \emph{SSTS} PAM (left column) and an \emph{MTS} PAM (right column) on the UK Biobank data introduced in Section \ref{ssec:ukb-data}.
By definition, age must exceed age at CKD onset and progression and age at CKD progression must exceed age at CKD onset.
Prediction horizons within Severe CKD are restricted to the empirically observed range (time since CKD progression $\in [0,16]$ years).
Grey lines represent contour lines.
}
\label{fig:ukb-baseline-23-contour}
\end{figure}

\begin{figure}[!ht]
\begin{center}
\includegraphics[width=1.0\linewidth]{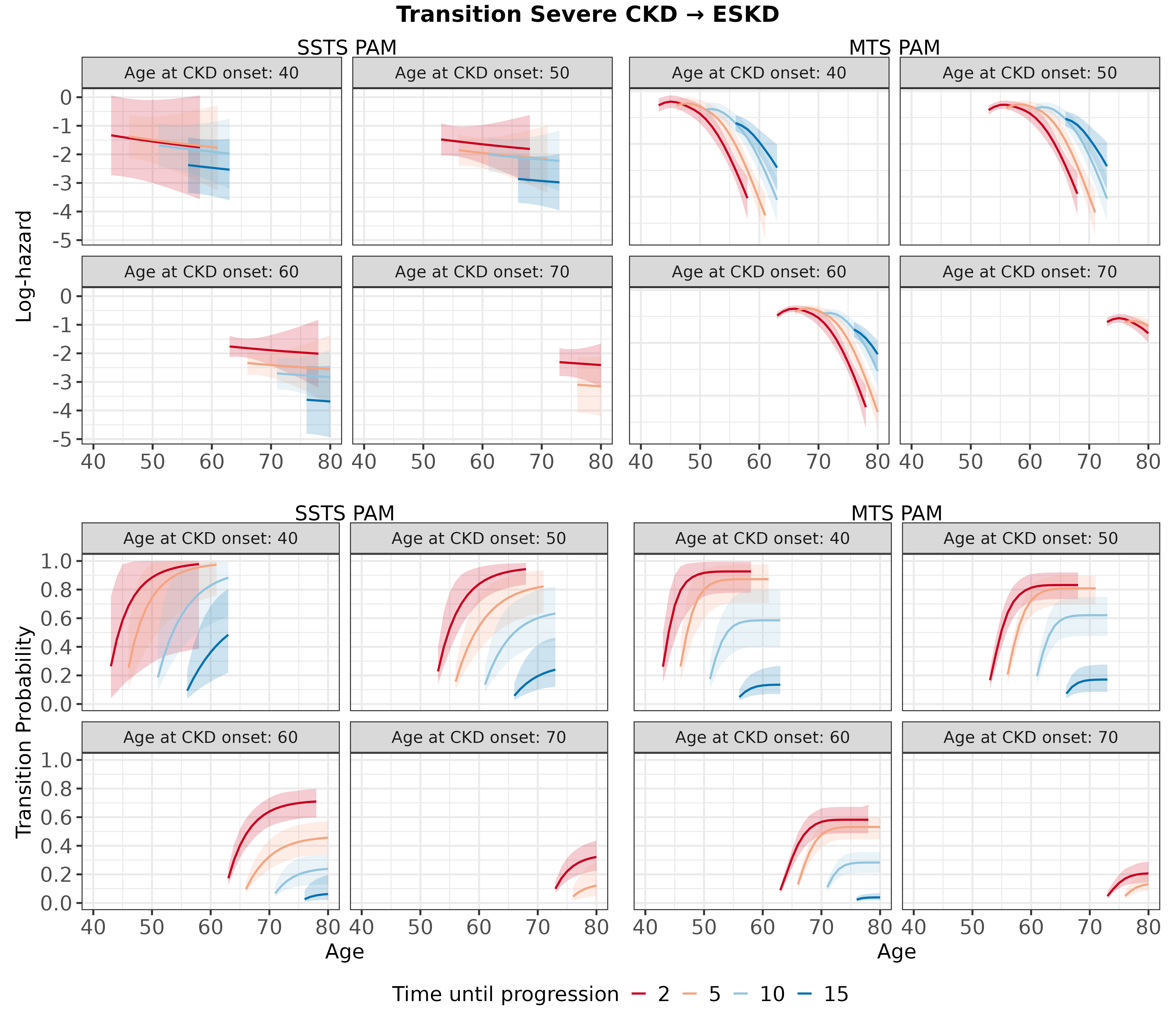}
\end{center}
\vspace{-0.5cm}
\caption{Slice plots of predicted log-hazards and transition probabilities for transition "Severe CKD $\rightarrow$ ESKD" over age for different values of age at CKD onset and time until CKD progression in UK Biobank.
For the transition "Severe CKD $\rightarrow$ ESKD", this figure shows slice plots of log-hazards (top) and transition probabilities (bottom) over age, for four distinct values of age at CKD onset (facets) and time until CKD progression (colors), estimated using an \emph{SSTS} PAM (left column) and an \emph{MTS} PAM (right column) on the UK Biobank data introduced in Section \ref{ssec:ukb-data}.
Shades depict 95\% confidence bands.
By definition, age must exceed age at CKD onset and progression and age at CKD progression must exceed age at CKD onset.
Prediction horizons within Severe CKD are restricted to the empirically observed range (time since CKD progression $\in [0,16]$ years).
}
\label{fig:ukb-baseline-23-slice}
\end{figure}

\begin{figure}[!ht]
\begin{center}
\includegraphics[width=0.95\linewidth]{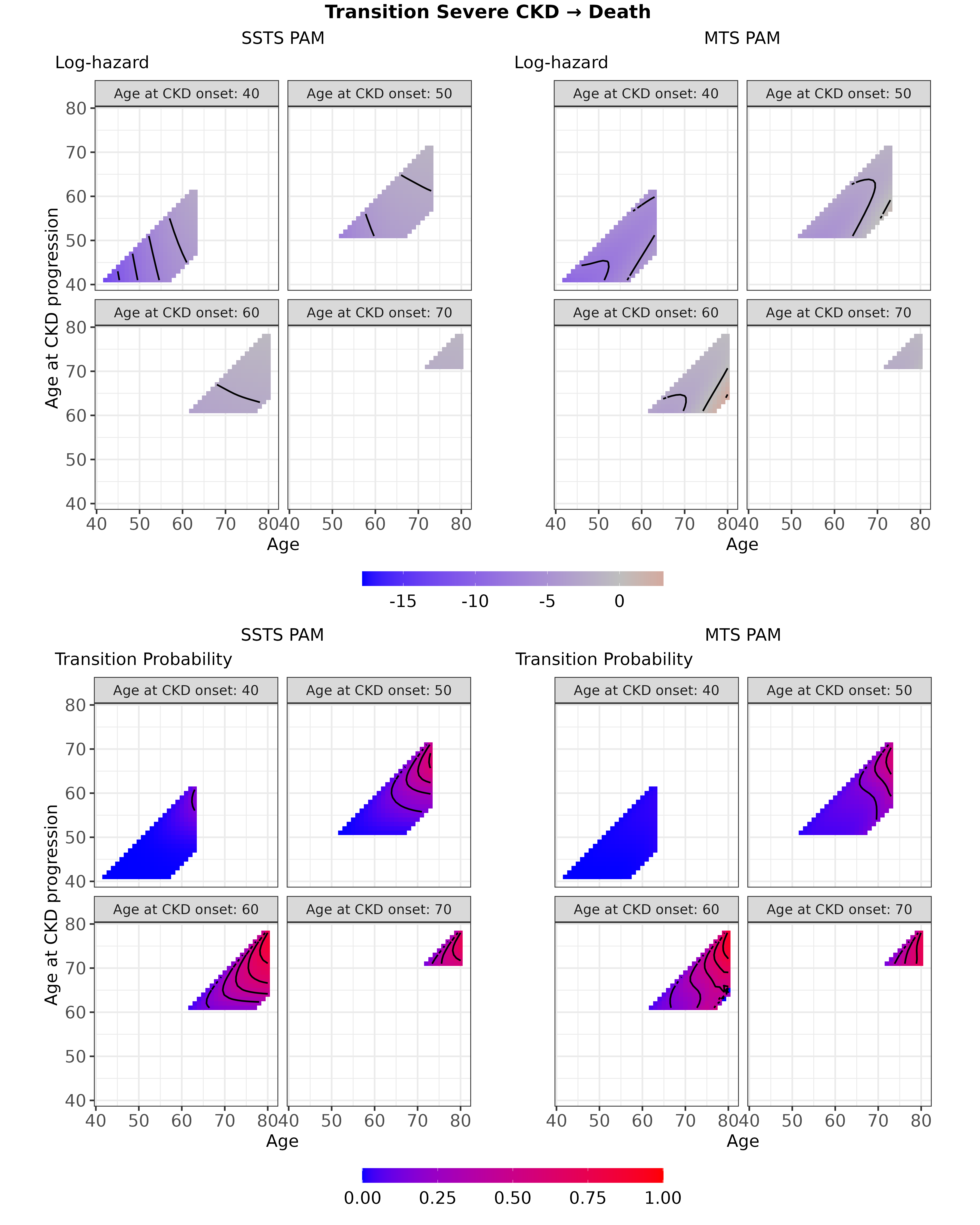}
\end{center}
\vspace{-0.5cm}
\caption{Contour plots of predicted log-hazards and transition probabilities for transition "Severe CKD $\rightarrow$ Death" over age and age at CKD progression for different values of age at CKD onset in UK Biobank.
For the transition "Severe CKD $\rightarrow$ Death", this figure shows contour plots of log-hazards (top) and transition probabilities (bottom) over age and age at progression to Severe CKD for four distinct values of age at CKD onset, estimated using an \emph{SSTS} PAM (left column) and an \emph{MTS} PAM (right column) on the UK Biobank data introduced in Section \ref{ssec:ukb-data}.
By definition, age must exceed age at CKD onset and progression and age at CKD progression must exceed age at CKD onset.
Prediction horizons within Severe CKD are restricted to the empirically observed range (time since CKD progression $\in [0,16]$ years).
Grey lines represent contour lines.
}
\label{fig:ukb-baseline-24-contour}
\end{figure}

\begin{figure}[!ht]
\begin{center}
\includegraphics[width=1.0\linewidth]{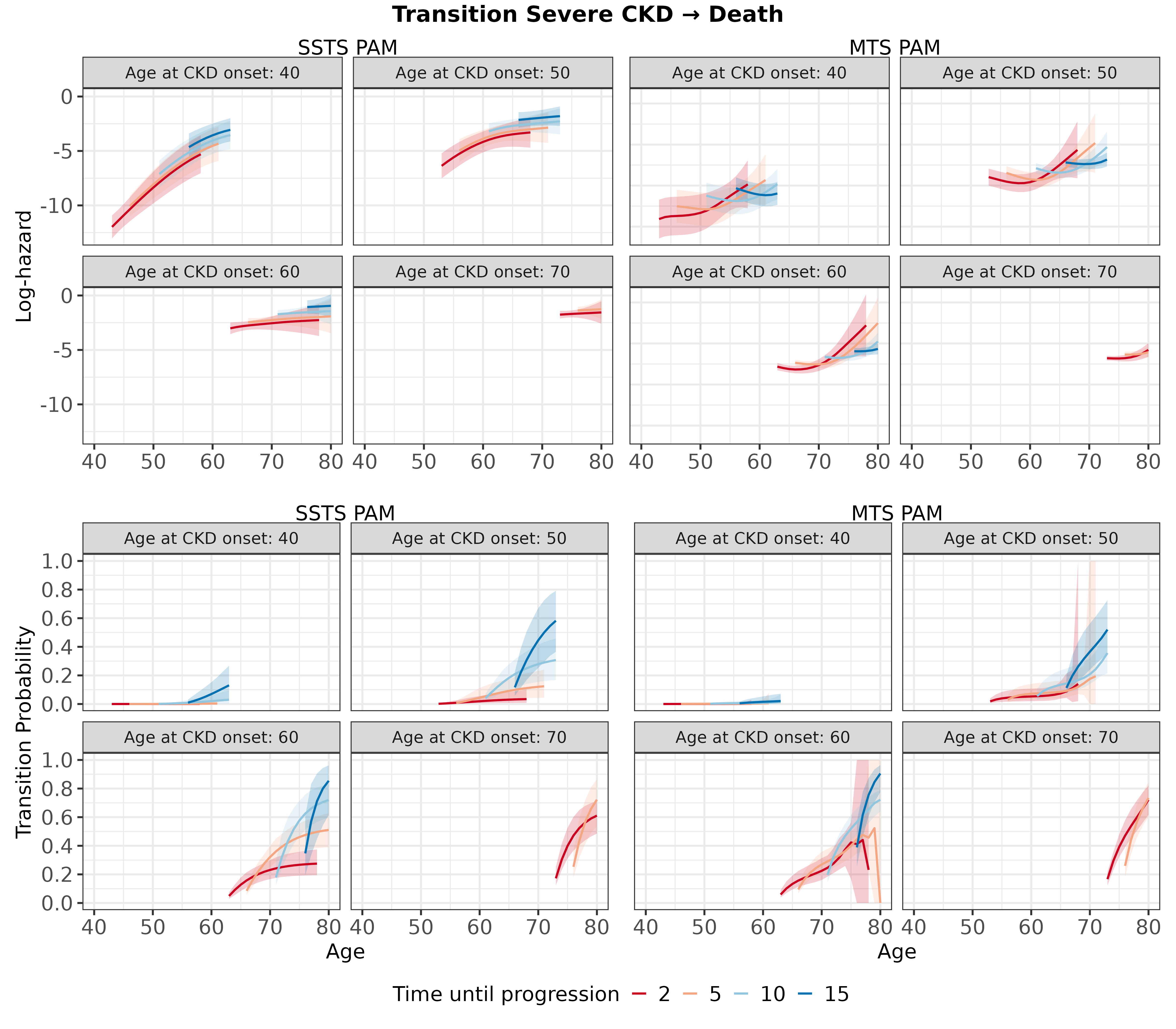}
\end{center}
\vspace{-0.5cm}
\caption{Slice plots of predicted log-hazards and transition probabilities for transition "Severe CKD $\rightarrow$ Death" over age for different values of age at CKD onset and time until CKD progression in UK Biobank.
For the transition "Severe CKD $\rightarrow$ Death", this figure shows contour plots of log-hazards (top) and transition probabilities (bottom) over age, for four distinct values of age at CKD onset (facets) and time until CKD progression (colors), estimated using an \emph{SSTS} PAM (left column) and an \emph{MTS} PAM (right column) on the UK Biobank data introduced in Section \ref{ssec:ukb-data}.
Shades depict 95\% confidence bands.
By definition, age must exceed age at CKD onset and progression and age at CKD progression must exceed age at CKD onset.
Prediction horizons within Severe CKD are restricted to the empirically observed range (time since CKD progression $\in [0,16]$ years).
}
\label{fig:ukb-baseline-24-slice}
\end{figure}

\begin{figure}[!ht]
\centering
\begin{subfigure}[b]{0.48\linewidth}
    \includegraphics[width=\linewidth]{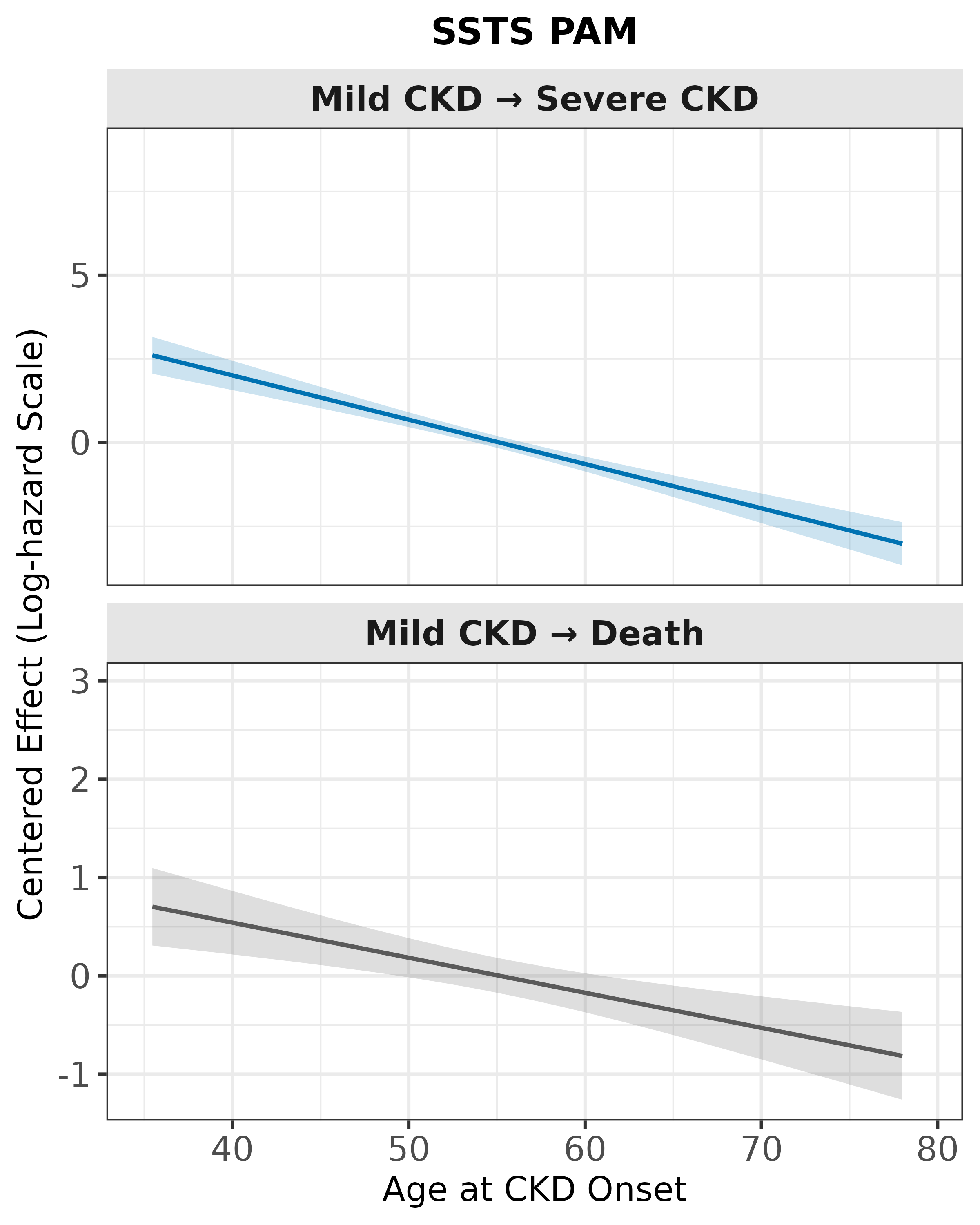}
\end{subfigure}
\hfill
\begin{subfigure}[b]{0.48\linewidth}
    \includegraphics[width=\linewidth]{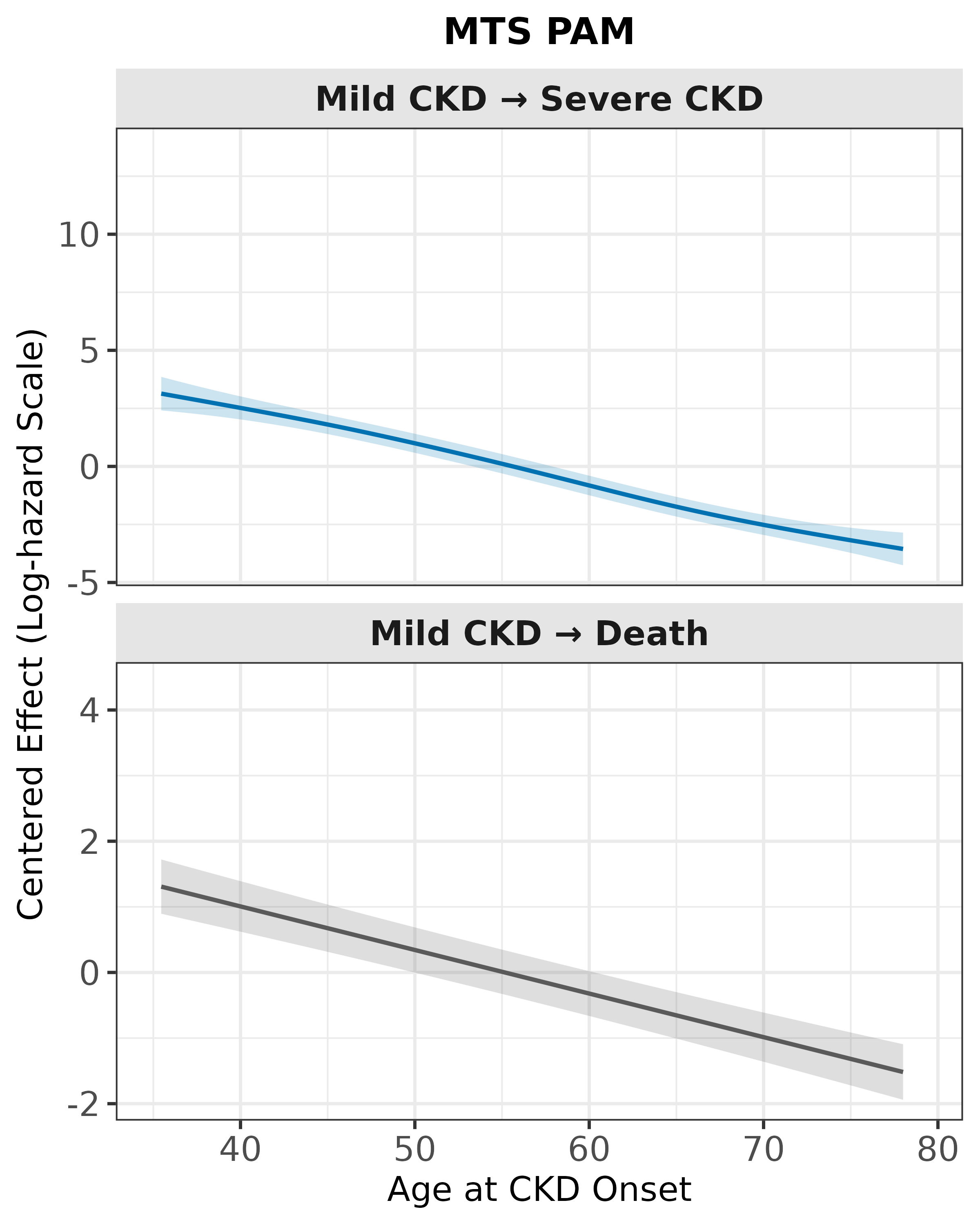}
\end{subfigure}
\begin{subfigure}[b]{0.48\linewidth}
    \includegraphics[width=\linewidth]{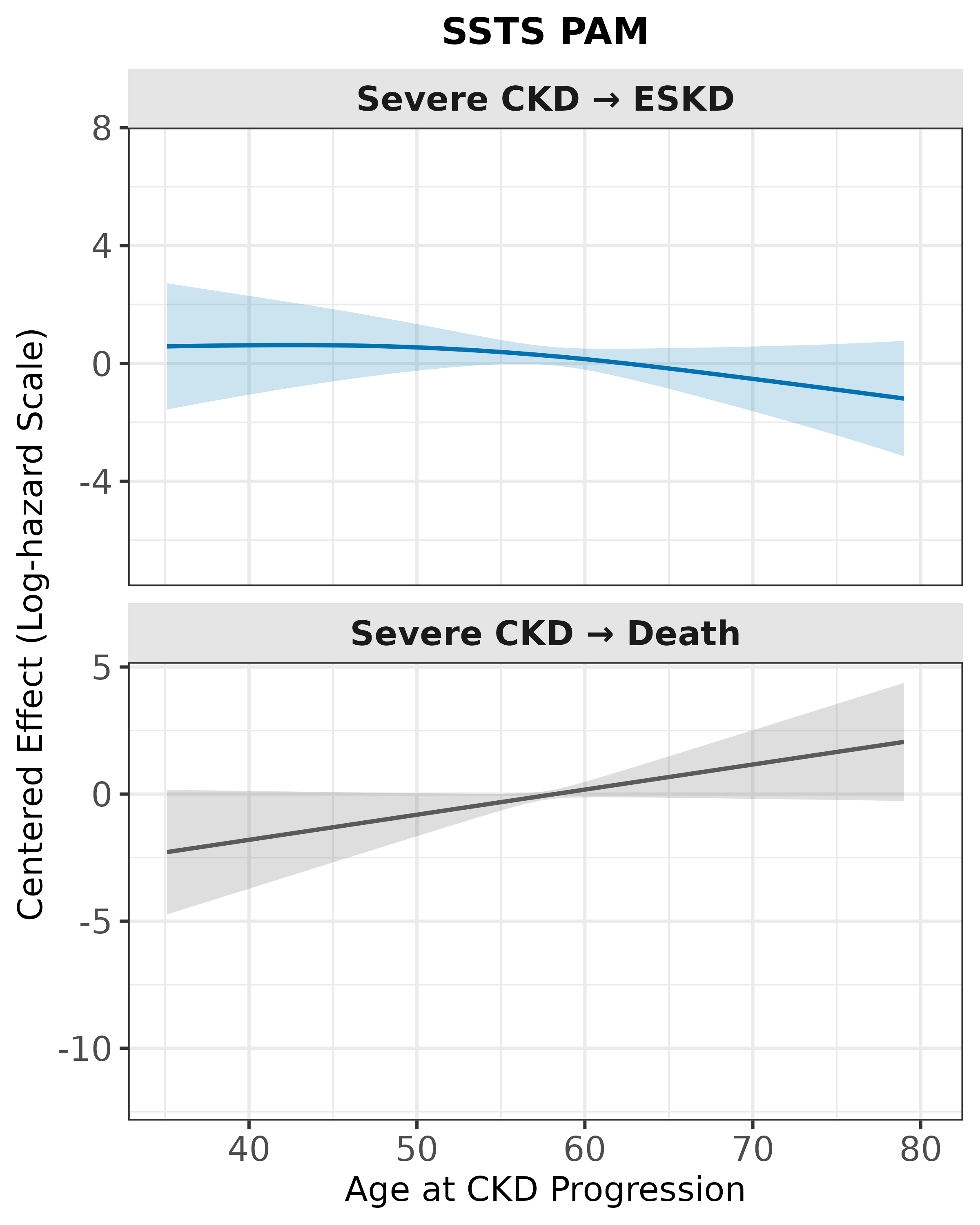}
\end{subfigure}
\hfill
\begin{subfigure}[b]{0.48\linewidth}
    \includegraphics[width=\linewidth]{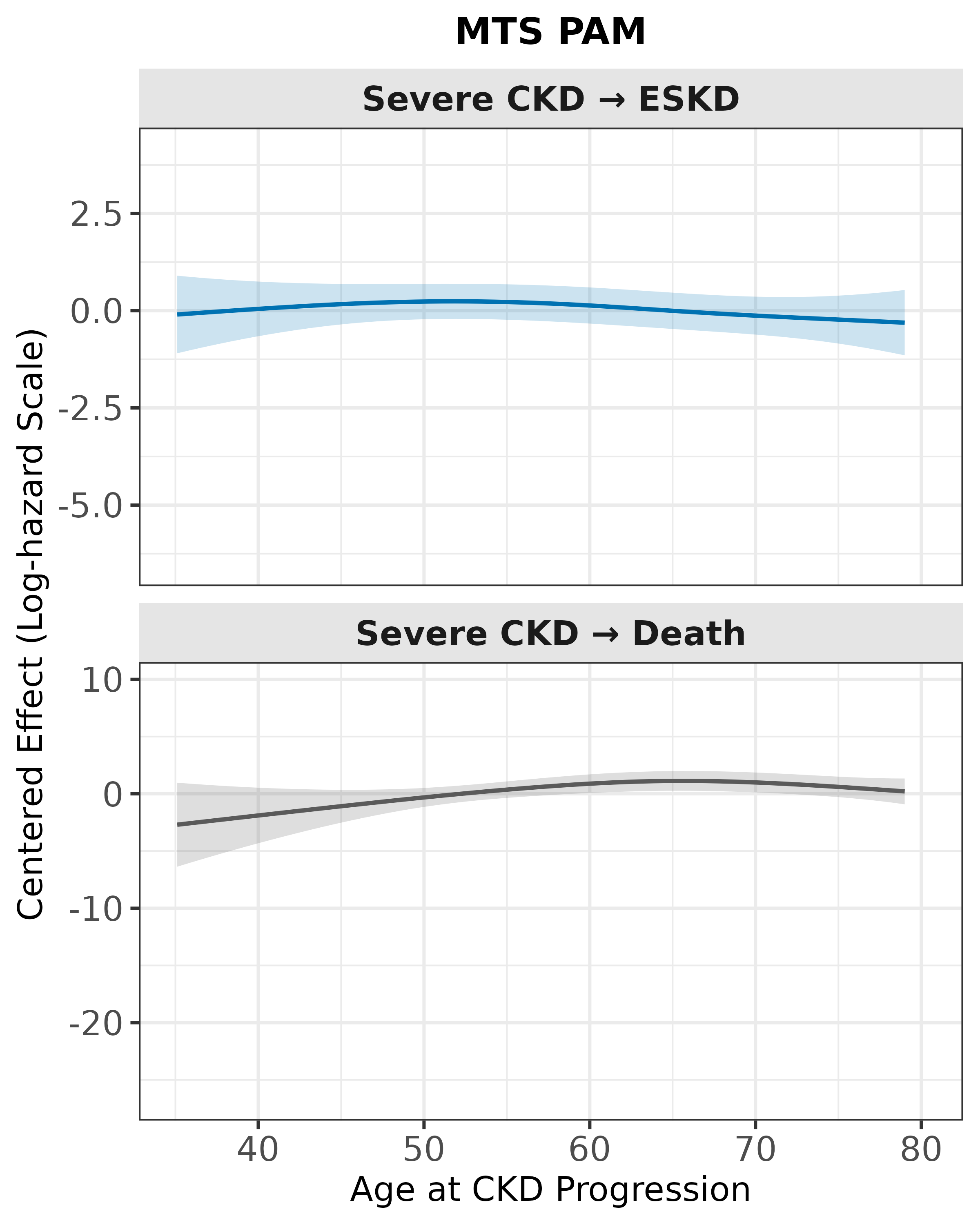}
\end{subfigure}
\caption{
Centered effects of state-entry times on transitions out of states Mild and Severe CKD in UK Biobank.
This figure shows centered effect plots for the effect of state-entry times on log-hazards, estimated with an \emph{SSTS} PAM (left) or an \emph{MTS} PAM, on UK Biobank data.
The two top rows correspond to effects of left-truncation variable age at CKD onset on the log-hazards of transitions out of state Mild CKD (to Severe CKD or Death).
The two bottom rows correspond to effects of left-truncation variable age at CKD progression on the log-hazards of transitions out of state Severe CKD (to ESKD or Death).
}
\label{fig:ukb-left-truncation}
\end{figure}

\clearpage

\subsection{Supplementary tables}
\label{app-sec:tables}

\newcommand{\captionpeddata}{Illustration of the PED transformation for subjects A and B of Figure \ref{fig:sample-trajectories}. Starting from a standard multi-state survival dataset containing ID $i$, study entry age $\text{age}_\text{entry}$, observed time $\text{age}_i$, transition indicator $\text{state}_i$ (with state numbers corresponding to mild CKD (1), severe CKD (2), ESKD (3), and death (4)), and event indicator $\text{d}_i$, the follow-up is partitioned into time intervals (here, with event and left-truncation times $a_0=0, a_1=35, a_2=50, a_3=55, a_4=60, a_5=75$). Age  $age_{ij}$ at end of time interval $j$ and alternative time scales $\text{t}_1$ (time since mild CKD onset) and $\text{t}_2$ (time since severe CKD onset) are added.}
\begin{table}[!ht]
    \caption{\captionpeddata}
    \label{tab:ped-transformation}
    \centering
    \resizebox{\textwidth}{!}{%
    \begin{tabular}{
          @{}p{5.75cm}@{}
          @{}p{1.5cm}@{}
          @{}p{9cm}@{}
      }
        \begin{minipage}{\linewidth}
            \centering
            \begin{tabular}{|c|c|c|c|c|}
                \hline
                \textbf{i} & $\textbf{age}_\text{entry}$ & $\textbf{age}_i$ & $\textbf{state}_i$ & $\textbf{d}_i$ \\ \hline
                A & 55 & 72 & cens & 0 \\ \hline
                B & 35 & 50 & 1 & 1 \\ \hline
                B & 35 & 60 & 2 & 1 \\ \hline
                B & 35 & 75 & 3 & 1 \\ \hline
            \end{tabular}
        \end{minipage}
        &
        \begin{minipage}{\linewidth}
            \centering
            \scriptsize
            PED\\[1.2em]
            \scalebox{4}{$\rightarrow$} \\
        \end{minipage}
        &
        \begin{minipage}{\linewidth}
            \centering
            \begin{tabular}{|c|c|c|c|c|c|c|c|c|}
                \hline
                \textbf{i} & \textbf{j} & $\textbf{I}_j$ & \textbf{state}$_{ij}$ & $\textbf{d}_{ij}$ & \textbf{y}$_{ij}$ & \textbf{age}$_{ij}$ & \textbf{t}$_1$ & \textbf{t}$_2$ \\ \hline
                A & 2 & (55,60] & 1 & 0 & 5 & 60 & 0  & 0 \\ \hline
                A & 2 & (55,60] & 4 & 0 & 5 & 60 & 0  & 0 \\ \hline
                A & 3 & (60,75] & 1 & 0 & 12 & 72 & 0  & 0 \\ \hline
                A & 3 & (60,75] & 4 & 0 & 12 & 72 & 0  & 0 \\ \hline
                B & 1 & (35,50] & 1 & 1 & 15 & 50 & 0  & 0 \\ \hline
                B & 1 & (35,50] & 4 & 0 & 15 & 50 & 0  & 0 \\ \hline
                B & 2 & (50,55] & 2 & 0 & 5 & 55 & 5 & 0 \\ \hline
                B & 2 & (50,55] & 4 & 0 & 5 & 55 & 5 & 0 \\ \hline
                B & 3 & (55,60] & 2 & 1 & 5 & 60 & 10 & 0 \\ \hline
                B & 3 & (55,60] & 4 & 0 & 5 & 60 & 10 & 0 \\ \hline
                B & 4 & (60,75] & 3 & 1 & 15 & 75 & 25 & 15\\ \hline
                B & 4 & (60,75] & 4 & 0 & 15 & 75 & 25 & 15\\ \hline
            \end{tabular}
        \end{minipage}
    \end{tabular}%
    }
\end{table}

\newcommand{\captiontsdgp}{Details on data generating processes. This table provides detailed information on the baseline hazard functions and fixed effect parameters of the data generating processes employed in Section \ref{ssec:sim-time-scales} (see Equations \eqref{eq:baseline-hazard-ssts-dgp}, \eqref{eq:baseline-hazard-mts-dgp} and \eqref{eq:hazard-covariate-dgp}).
The \emph{SSTS} DGP has one time-varying baseline log-hazard $f_k^{ssts}(t)$ per transition $k \in \{0\rightarrow1, 0\rightarrow3, 1\rightarrow2, 1\rightarrow3\}$, i.e., a total of four.
For the \emph{MTS} DGP, we also have four time scale functions:
Functions $f_0^{mts}(t)$ and $f_1^{mts}(t_1)$ represent the time-varying baseline log-hazard of time since study entry on transition $0 \rightarrow 1$ (\textit{onset}; first time scale) and of time since onset on transition $1 \rightarrow 2$ (\textit{progression}; second time scale).
Functions $\tilde{f}_0^{mts}(t)$ and $\tilde{f}_1^{mts}(t_1)$ represent the time-varying baseline log-hazard of time since study entry on transition $0 \rightarrow 3$ (first time scale) and of time since onset on $1 \rightarrow 3$ (second time scale).
For transitions out of state $1$, the components $f_0^{mts}(t)$ and $f_1^{mts}(t_1)/\tilde{f}_1^{mts}(t_1)$ are then added up.
$f_{\text{until}\_1}(t_{\text{until}\_1})$ and $\tilde{f}_{\text{until}\_1}(t_{\text{until}\_1})$ are shared across the two DGPs and represent the effect of the time until onset $t_{\text{until}\_1}$ on transitions $1 \rightarrow 2$ and $1 \rightarrow 3$, respectively. 
The fixed effect parameters $\beta_{0,k}$ and $\beta_{x_1,k}$ are transition-specific intercepts and covariate fixed effects on the log-hazard level, shared across the two DGPs.}
\begin{table}[H]
\centering
\caption{\captiontsdgp}
\label{tab:ts-dgp-details}
\begin{tabular}{r l}
\toprule
\textbf{Parameter} & \textbf{Definition} \\
\midrule
\multicolumn{2}{l}{\textit{Time-Dependent Baseline Hazard Functions}} \\
$f_{0\rightarrow1}^{ssts}(t)$ & $= \frac{0.10 t^2}{0.7 + 0.04 (\max\{0, t - 3\})^3}$ \\
$f_{0\rightarrow3}^{ssts}(t)$ & $= \frac{0.15 t^2}{0.9 + 0.01 (\max\{0, t - 1\})^3}$ \\
$f_{1\rightarrow2}^{ssts}(t)$ & $= 0.48 e^{-0.10t}$ \\
$f_{1\rightarrow3}^{ssts}(t)$ & $= 0.16 e^{-0.30t}$ \\
\addlinespace
$f_{02}^{mts}(t)$ & $= f_{0\rightarrow1}^{ssts}(t)$ \\
$f_{12}^{mts}(t_1)$ & $= 0.32 e^{-0.15t_1}$ \\
$\tilde{f}_{03}^{mts}(t)$ & $= \tilde{f}_{0\rightarrow3}^{ssts}(t)$ \\
$\tilde{f}_{13}^{mts}(t_1)$ & $= 0.14 e^{-0.25t_1}$ \\
\addlinespace
$f_{\text{entry}\_1}(t_{\text{entry}\_1})$ & $= 2.50 e^{-0.60t_{\text{entry}\_1}}$ \\
$\tilde{f}_{\text{entry}\_1}(t_{\text{entry}\_1})$ & $= 0.14 e^{-0.25t_{\text{entry}\_1}}$ \\
\midrule
\multicolumn{2}{l}{\textit{Fixed Effect Parameters}} \\
$\beta_{0,0\rightarrow1}$ & $= -3.9$ \\
$\beta_{0,0\rightarrow3}$ & $= -4.0$ \\
$\beta_{0,1\rightarrow2}$ & $= -3.4$ \\
$\beta_{0,1\rightarrow3}$ & $= -3.4$ \\
\addlinespace
$\beta_{x_1,0\rightarrow1}$ & $= 0.2$ \\
$\beta_{x_1,0\rightarrow3}$ & $= 0.1$ \\
$\beta_{x_1,1\rightarrow2}$ & $= 0.2$ \\
$\beta_{x_1,1\rightarrow3}$ & $= 0.1$ \\
\bottomrule
\end{tabular}
\end{table}

\newcommand{\captiontsbhcoverage}{Mean coverage of baseline log-hazards, cumulative hazards, and transition probabilities estimated using \emph{SSTS} and \emph{MTS} PAMs on data simulated from \emph{SSTS} and \emph{MTS} DGPs.}
    \begin{table*}[htbp]
    \centering
    \caption{\captiontsbhcoverage}
    \label{tab:sim-ts-bh-coverage}
    \begin{sideways}
    \footnotesize
    \setlength{\tabcolsep}{2.5pt}
    \begin{tabular}{lccccccccc}
    \toprule
    & & \multicolumn{4}{c}{SSTS DGP} & \multicolumn{4}{c}{MTS DGP} \\
    \cmidrule(lr){3-6} \cmidrule(lr){7-10}
    & & \multicolumn{2}{c}{SSTS PAM} & \multicolumn{2}{c}{MTS PAM} & \multicolumn{2}{c}{SSTS PAM} & \multicolumn{2}{c}{MTS PAM} \\
    \cmidrule(lr){3-4} \cmidrule(lr){5-6} \cmidrule(lr){7-8} \cmidrule(lr){9-10}
    Transition & \shortstack[l]{Quantity of \\ Interest} & \shortstack[c]{Penalized\\Splines} & \shortstack[c]{Factor\\Smooth} & \shortstack[c]{Penalized\\Splines} & \shortstack[c]{Factor\\Smooth} & \shortstack[c]{Penalized\\Splines} & \shortstack[c]{Factor\\Smooth} & \shortstack[c]{Penalized\\Splines} & \shortstack[c]{Factor\\Smooth} \\
    \midrule
    0 $\rightarrow$ 1 & \makecell[l]{Log-hazard} & \makecell[t]{0.95\\(0.93; 0.97)} & \makecell[t]{0.95\\(0.93; 0.97)} & \makecell[t]{0.69\\(0.65; 0.72)} & \makecell[t]{0.68\\(0.65; 0.72)} & \makecell[t]{0.95\\(0.93; 0.97)} & \makecell[t]{0.95\\(0.93; 0.97)} & \makecell[t]{0.95\\(0.93; 0.97)} & \makecell[t]{0.95\\(0.93; 0.96)} \\
    0 $\rightarrow$ 1 & \makecell[l]{Cumulative\\Hazard} & \makecell[t]{1.00\\(0.99; 1.00)} & \makecell[t]{1.00\\(0.99; 1.00)} & \makecell[t]{0.97\\(0.95; 0.98)} & \makecell[t]{0.97\\(0.95; 0.98)} & \makecell[t]{0.99\\(0.98; 1.00)} & \makecell[t]{0.99\\(0.98; 1.00)} & \makecell[t]{0.99\\(0.98; 1.00)} & \makecell[t]{0.99\\(0.98; 1.00)} \\
    0 $\rightarrow$ 1 & \makecell[l]{Transition\\Probability} & \makecell[t]{0.91\\(0.88; 0.93)} & \makecell[t]{0.90\\(0.88; 0.93)} & \makecell[t]{0.81\\(0.77; 0.84)} & \makecell[t]{0.81\\(0.77; 0.84)} & \makecell[t]{0.90\\(0.87; 0.93)} & \makecell[t]{0.90\\(0.88; 0.93)} & \makecell[t]{0.90\\(0.87; 0.92)} & \makecell[t]{0.90\\(0.87; 0.92)} \\
    0 $\rightarrow$ 3 & \makecell[l]{Log-hazard} & \makecell[t]{0.95\\(0.93; 0.97)} & \makecell[t]{0.95\\(0.93; 0.97)} & \makecell[t]{0.86\\(0.83; 0.88)} & \makecell[t]{0.87\\(0.84; 0.89)} & \makecell[t]{0.95\\(0.93; 0.97)} & \makecell[t]{0.95\\(0.93; 0.97)} & \makecell[t]{0.95\\(0.93; 0.96)} & \makecell[t]{0.95\\(0.93; 0.96)} \\
    0 $\rightarrow$ 3 & \makecell[l]{Cumulative\\Hazard} & \makecell[t]{0.99\\(0.98; 1.00)} & \makecell[t]{0.99\\(0.98; 1.00)} & \makecell[t]{0.97\\(0.96; 0.98)} & \makecell[t]{0.97\\(0.96; 0.98)} & \makecell[t]{0.99\\(0.98; 0.99)} & \makecell[t]{0.99\\(0.98; 0.99)} & \makecell[t]{0.99\\(0.98; 0.99)} & \makecell[t]{0.99\\(0.98; 0.99)} \\
    0 $\rightarrow$ 3 & \makecell[l]{Transition\\Probability} & \makecell[t]{0.91\\(0.89; 0.94)} & \makecell[t]{0.92\\(0.89; 0.94)} & \makecell[t]{0.87\\(0.84; 0.89)} & \makecell[t]{0.86\\(0.83; 0.89)} & \makecell[t]{0.92\\(0.89; 0.94)} & \makecell[t]{0.92\\(0.89; 0.94)} & \makecell[t]{0.92\\(0.89; 0.94)} & \makecell[t]{0.92\\(0.89; 0.94)} \\
    1 $\rightarrow$ 2 & \makecell[l]{Log-hazard} & \makecell[t]{0.95\\(0.93; 0.97)} & \makecell[t]{0.96\\(0.94; 0.97)} & \makecell[t]{0.54\\(0.51; 0.57)} & \makecell[t]{0.54\\(0.51; 0.57)} & \makecell[t]{0.96\\(0.94; 0.97)} & \makecell[t]{0.93\\(0.91; 0.95)} & \makecell[t]{0.96\\(0.93; 0.97)} & \makecell[t]{0.94\\(0.91; 0.96)} \\
    1 $\rightarrow$ 2 & \makecell[l]{Cumulative\\Hazard} & \makecell[t]{0.96\\(0.94; 0.98)} & \makecell[t]{0.97\\(0.95; 0.98)} & \makecell[t]{0.70\\(0.67; 0.72)} & \makecell[t]{0.70\\(0.68; 0.72)} & \makecell[t]{0.98\\(0.96; 0.99)} & \makecell[t]{0.96\\(0.95; 0.97)} & \makecell[t]{0.97\\(0.95; 0.98)} & \makecell[t]{0.95\\(0.93; 0.97)} \\
    1 $\rightarrow$ 2 & \makecell[l]{Transition\\Probability} & \makecell[t]{0.93\\(0.91; 0.95)} & \makecell[t]{0.95\\(0.92; 0.96)} & \makecell[t]{0.70\\(0.67; 0.73)} & \makecell[t]{0.70\\(0.67; 0.73)} & \makecell[t]{0.94\\(0.92; 0.96)} & \makecell[t]{0.94\\(0.91; 0.95)} & \makecell[t]{0.93\\(0.91; 0.95)} & \makecell[t]{0.94\\(0.91; 0.95)} \\
    1 $\rightarrow$ 3 & \makecell[l]{Log-hazard} & \makecell[t]{0.93\\(0.91; 0.95)} & \makecell[t]{0.95\\(0.92; 0.96)} & \makecell[t]{0.63\\(0.59; 0.66)} & \makecell[t]{0.63\\(0.60; 0.66)} & \makecell[t]{0.95\\(0.93; 0.97)} & \makecell[t]{0.97\\(0.95; 0.98)} & \makecell[t]{0.96\\(0.94; 0.97)} & \makecell[t]{0.97\\(0.95; 0.98)} \\
    1 $\rightarrow$ 3 & \makecell[l]{Cumulative\\Hazard} & \makecell[t]{0.95\\(0.92; 0.96)} & \makecell[t]{0.96\\(0.94; 0.98)} & \makecell[t]{0.80\\(0.77; 0.82)} & \makecell[t]{0.79\\(0.77; 0.82)} & \makecell[t]{0.97\\(0.96; 0.98)} & \makecell[t]{0.98\\(0.97; 0.99)} & \makecell[t]{0.97\\(0.96; 0.99)} & \makecell[t]{0.98\\(0.96; 0.99)} \\
    1 $\rightarrow$ 3 & \makecell[l]{Transition\\Probability} & \makecell[t]{0.91\\(0.88; 0.94)} & \makecell[t]{0.94\\(0.91; 0.96)} & \makecell[t]{0.79\\(0.75; 0.82)} & \makecell[t]{0.80\\(0.77; 0.82)} & \makecell[t]{0.94\\(0.92; 0.96)} & \makecell[t]{0.95\\(0.93; 0.97)} & \makecell[t]{0.94\\(0.91; 0.96)} & \makecell[t]{0.95\\(0.93; 0.97)} \\
    \bottomrule
    \end{tabular}
    \normalsize
    \end{sideways}
    \end{table*}

\newcommand{\captionfecoverage}{Mean coverage of fixed effect log-hazards estimated using \emph{SSTS} and \emph{MTS} PAMs on data simulated from \emph{SSTS} and \emph{MTS} DGPs.}
\begin{table*}[htbp]
\centering
\caption{\captionfecoverage}
\label{tab:sim-ts-fe-coverage}
\begin{sideways}
\footnotesize
\setlength{\tabcolsep}{3pt}
\begin{tabular}{lcccccccc}
\toprule
& \multicolumn{4}{c}{SSTS DGP} & \multicolumn{4}{c}{MTS DGP} \\
\cmidrule(lr){2-5} \cmidrule(lr){6-9}
& \multicolumn{2}{c}{SSTS PAM} & \multicolumn{2}{c}{MTS PAM} & \multicolumn{2}{c}{SSTS PAM} & \multicolumn{2}{c}{MTS PAM} \\
\cmidrule(lr){2-3} \cmidrule(lr){4-5} \cmidrule(lr){6-7} \cmidrule(lr){8-9}
Transition & \shortstack[c]{Penalized\\Splines} & \shortstack[c]{Factor\\Smooth} & \shortstack[c]{Penalized\\Splines} & \shortstack[c]{Factor\\Smooth} & \shortstack[c]{Penalized\\Splines} & \shortstack[c]{Factor\\Smooth} & \shortstack[c]{Penalized\\Splines} & \shortstack[c]{Factor\\Smooth} \\
\midrule
$\beta_{x_1,0\rightarrow 1}$ & \makecell[t]{0.97\\(0.95; 0.98)} & \makecell[t]{0.95\\(0.93; 0.97)} & \makecell[t]{0.95\\(0.93; 0.97)} & \makecell[t]{0.95\\(0.93; 0.97)} & \makecell[t]{0.94\\(0.92; 0.96)} & \makecell[t]{0.95\\(0.93; 0.97)} & \makecell[t]{0.97\\(0.95; 0.98)} & \makecell[t]{0.95\\(0.93; 0.97)} \\
$\beta_{x_1,0\rightarrow 3}$ & \makecell[t]{0.99\\(0.98; 1.00)} & \makecell[t]{0.99\\(0.98; 1.00)} & \makecell[t]{0.99\\(0.98; 1.00)} & \makecell[t]{0.99\\(0.98; 1.00)} & \makecell[t]{1.00\\(0.99; 1.00)} & \makecell[t]{0.99\\(0.98; 1.00)} & \makecell[t]{1.00\\(0.99; 1.00)} & \makecell[t]{1.00\\(0.99; 1.00)} \\
$\beta_{x_1,1\rightarrow 2}$ & \makecell[t]{0.97\\(0.95; 0.98)} & \makecell[t]{0.98\\(0.96; 0.99)} & \makecell[t]{0.95\\(0.92; 0.97)} & \makecell[t]{0.95\\(0.93; 0.97)} & \makecell[t]{0.97\\(0.96; 0.99)} & \makecell[t]{0.99\\(0.97; 0.99)} & \makecell[t]{0.98\\(0.97; 0.99)} & \makecell[t]{0.98\\(0.96; 0.99)} \\
$\beta_{x_1,1\rightarrow 3}$ & \makecell[t]{0.96\\(0.94; 0.98)} & \makecell[t]{0.98\\(0.96; 0.99)} & \makecell[t]{0.95\\(0.93; 0.97)} & \makecell[t]{0.96\\(0.94; 0.98)} & \makecell[t]{0.98\\(0.97; 0.99)} & \makecell[t]{0.99\\(0.98; 1.00)} & \makecell[t]{0.98\\(0.96; 0.99)} & \makecell[t]{0.97\\(0.95; 0.98)} \\
\bottomrule
\end{tabular}
\normalsize
\end{sideways}
\end{table*}

\newcommand{\captioniebeffectsizes}{Details on effect sizes of risk factors for index event bias simulations.
The setup for index event bias simulations is based on the simulations in Section \ref{ssec:sim-time-scales}, in particular Equations \eqref{eq:baseline-hazard-ssts-dgp} and \eqref{eq:baseline-hazard-mts-dgp}. 
The only difference now is that two risk factors $x_1$ and $x_2$ are incorporated into the log-hazard of the respective DGP (identical for \emph{SSTS} and \emph{MTS} DGP).
This table provides details on the corresponding fixed effect sizes.
}
\begin{table}[htbp]
\centering
\caption{\captioniebeffectsizes}
\label{tab:sim-ieb-effect-sizes}
\begin{tabular}{p{0.38\textwidth} >{\centering\arraybackslash}p{0.14\textwidth} >{\centering\arraybackslash}p{0.14\textwidth} >{\centering\arraybackslash}p{0.14\textwidth}}
\toprule
\multicolumn{1}{c}{} & \multicolumn{3}{c}{Effect Sizes} \\
\cmidrule(lr){2-4}
\multicolumn{1}{l}{Coefficient} & \multicolumn{1}{c}{Small} & \multicolumn{1}{c}{Medium} & \multicolumn{1}{c}{Large} \\
\midrule
\multicolumn{4}{l}{\textbf{Transition-specific intercepts}} \\
$\beta_{x_0,0\rightarrow 1}$ & -3.9 & -3.9 & -3.9 \\
$\beta_{x_0,0\rightarrow 3}$ & -4.0 & -4.0 & -4.0 \\
$\beta_{x_0,1\rightarrow 2}$ & -3.4 & -3.4 & -4.4 \\
$\beta_{x_0,1\rightarrow 3}$ & -3.4 & -3.4 & -3.4 \\
\midrule
\multicolumn{4}{l}{\textbf{Transition-specific effect sizes of risk factor }$x_1$} \\
$\beta_{x_1,0\rightarrow 1}$ & 0.2 & 0.4 & 0.6 \\
$\beta_{x_1,0\rightarrow 3}$ & 0.0 & 0.0 & 0.0 \\
$\beta_{x_1,1\rightarrow 2}$ & 0.2 & 0.4 & 0.6 \\
$\beta_{x_1,1\rightarrow 3}$ & 0.0 & 0.0 & 0.0 \\
\midrule
\multicolumn{4}{l}{\textbf{Transition-specific effect sizes of risk factor }$x_2$} \\
$\beta_{x_2,0\rightarrow 1}$ & 0.2 & 0.4 & 0.6 \\
$\beta_{x_2,0\rightarrow 3}$ & 0.0 & 0.0 & 0.0 \\
$\beta_{x_2,1\rightarrow 2}$ & 0.2 & 0.4 & 0.6 \\
$\beta_{x_2,1\rightarrow 3}$ & 0.0 & 0.0 & 0.0 \\
\bottomrule
\end{tabular}
\end{table}

\newcommand{\captioniebcor}{Correlation between two independent risk factors in simulated general population and diseased subpopulation data.
This table shows mean correlation and empirical 95\% confidence intervals (computed across simulation runs) between two risk factors $x_1$ and $x_2$ in states $0$ and $1$ from simulated multi-state data using either an \emph{SSTS} DGP or an \emph{MTS} DGP, described in Section \ref{sssec:sim-ieb-setup}.
The two risk factors are, by construction, independent in the general population (i.e., in state $0$), which is why any (negative) correlation in state $1$ is induced by the restriction of the population to the index event (here: transition $0 \rightarrow 1$).
}
\begin{table}[ht!]
 \centering
 \setlength{\tabcolsep}{2pt}
 \caption{\captioniebcor}
 \label{tab:sim-ieb-cor}
 \footnotesize
 \begin{tabular}{llclcccc}
 \toprule
  & & \textbf{Effect} & \multicolumn{2}{c}{\textbf{STSS DGP}} & \multicolumn{2}{c}{\textbf{MTS DGP}} \\
 \cmidrule(lr){4-5} \cmidrule(lr){6-7}
 \textbf{Dist. \(x_1\)} & \textbf{Dist. \(x_2\)} & \textbf{Sizes} & \textbf{State 0} & \textbf{State 1} & \textbf{State 0} & \textbf{State 1} \\
 \midrule
 \(\text{Ber}(0.5)\) & \(\text{Ber}(0.5)\) & Small & \makecell{0.00 \\ (-0.03; 0.03)} & \makecell{-0.00 \\ (-0.05; 0.04)} & \makecell{0.00 \\ (-0.03; 0.03)} & \makecell{-0.00 \\ (-0.05; 0.04)} \\
\(\text{Ber}(0.5)\) & \(\text{Ber}(0.5)\) & Medium & \makecell{0.00 \\ (-0.03; 0.03)} & \makecell{-0.01 \\ (-0.05; 0.03)} & \makecell{0.00 \\ (-0.03; 0.03)} & \makecell{-0.01 \\ (-0.05; 0.03)} \\
\(\text{Ber}(0.5)\) & \(\text{Ber}(0.5)\) & Large & \makecell{-0.00 \\ (-0.03; 0.03)} & \makecell{-0.03 \\ (-0.07; 0.02)} & \makecell{-0.00 \\ (-0.03; 0.02)} & \makecell{-0.02 \\ (-0.07; 0.02)} \\
\(\text{Ber}(0.5)\) & \(N(0,1)\) & Small & \makecell{-0.00 \\ (-0.03; 0.03)} & \makecell{-0.01 \\ (-0.05; 0.05)} & \makecell{0.00 \\ (-0.03; 0.03)} & \makecell{-0.00 \\ (-0.05; 0.04)} \\
\(\text{Ber}(0.5)\) & \(N(0,1)\) & Medium & \makecell{-0.00 \\ (-0.03; 0.03)} & \makecell{-0.02 \\ (-0.06; 0.03)} & \makecell{0.00 \\ (-0.03; 0.03)} & \makecell{-0.02 \\ (-0.06; 0.03)} \\
\(\text{Ber}(0.5)\) & \(N(0,1)\) & Large & \makecell{-0.00 \\ (-0.03; 0.03)} & \makecell{-0.05 \\ (-0.09; -0.01)} & \makecell{0.00 \\ (-0.03; 0.03)} & \makecell{-0.04 \\ (-0.09; 0.00)} \\
\(N(0,1)\) & \(\text{Ber}(0.5)\) & Small & \makecell{-0.00 \\ (-0.03; 0.03)} & \makecell{-0.01 \\ (-0.05; 0.04)} & \makecell{-0.00 \\ (-0.03; 0.03)} & \makecell{-0.00 \\ (-0.06; 0.04)} \\
\(N(0,1)\) & \(\text{Ber}(0.5)\) & Medium & \makecell{0.00 \\ (-0.03; 0.03)} & \makecell{-0.02 \\ (-0.07; 0.03)} & \makecell{0.00 \\ (-0.03; 0.03)} & \makecell{-0.02 \\ (-0.07; 0.03)} \\
\(N(0,1)\) & \(\text{Ber}(0.5)\) & Large & \makecell{0.00 \\ (-0.03; 0.03)} & \makecell{-0.04 \\ (-0.09; -0.00)} & \makecell{0.00 \\ (-0.03; 0.03)} & \makecell{-0.04 \\ (-0.09; 0.00)} \\
\(N(0,1)\) & \(N(0,1)\) & Small & \makecell{-0.00 \\ (-0.03; 0.03)} & \makecell{-0.01 \\ (-0.06; 0.04)} & \makecell{0.00 \\ (-0.03; 0.03)} & \makecell{-0.01 \\ (-0.06; 0.04)} \\
\(N(0,1)\) & \(N(0,1)\) & Medium & \makecell{-0.00 \\ (-0.03; 0.03)} & \makecell{-0.04 \\ (-0.09; 0.01)} & \makecell{0.00 \\ (-0.03; 0.03)} & \makecell{-0.04 \\ (-0.09; 0.01)} \\
\(N(0,1)\) & \(N(0,1)\) & Large & \makecell{-0.00 \\ (-0.03; 0.03)} & \makecell{-0.08 \\ (-0.13; -0.03)} & \makecell{0.00 \\ (-0.03; 0.03)} & \makecell{-0.08 \\ (-0.12; -0.03)} \\
 \bottomrule
 \normalsize
 \end{tabular}
 \end{table}

\newcommand{\captionicbhcoverageloghazard}{Mean coverage of baseline log-hazards under interval-censoring.
This table shows mean baseline log-hazard coverage for the IC simulation settings described in Section \ref{sssec:sim-ic-setup}.
Mean coverage and 95\% CIs (from an exact binomial test) are computed pointwise across all simulation runs, then averaged over time points.
}
\begin{table*}[htbp]
\centering
\caption{\captionicbhcoverageloghazard}
\label{tab:sim-ic-bh-coverage-loghazard}
\begin{sideways}
\footnotesize
\setlength{\tabcolsep}{3pt}
\begin{tabular}{llccccccc}
\toprule
& & \multicolumn{3}{c}{Piecewise Exponential DGP} & \multicolumn{3}{c}{Weibull DGP} & \multicolumn{1}{c}{icenReg DGP} \\
\cmidrule(lr){3-5} \cmidrule(lr){6-8} \cmidrule(lr){9-9}
Model & \shortstack[l]{Estimation \\ Time Point} & Beta & Uniform & Equidistant & Beta & Uniform & Equidistant & Uniform \\
\midrule
PAM & Exact & \makecell{0.94\\(0.91; 0.96)} & \makecell{0.94\\(0.91; 0.95)} & \makecell{0.93\\(0.91; 0.95)} & \makecell{0.91\\(0.88; 0.93)} & \makecell{0.90\\(0.87; 0.92)} & \makecell{0.90\\(0.87; 0.92)} & \makecell{0.95\\(0.86; 0.96)} \\
PAM & Mid & \makecell{0.72\\(0.68; 0.75)} & \makecell{0.64\\(0.60; 0.68)} & \makecell{0.55\\(0.51; 0.59)} & \makecell{0.75\\(0.71; 0.78)} & \makecell{0.66\\(0.62; 0.70)} & \makecell{0.50\\(0.46; 0.53)} & \makecell{0.71\\(0.65; 0.73)} \\
PAM & End & \makecell{0.15\\(0.14; 0.17)} & \makecell{0.09\\(0.07; 0.10)} & \makecell{0.54\\(0.51; 0.57)} & \makecell{0.39\\(0.36; 0.42)} & \makecell{0.43\\(0.41; 0.45)} & \makecell{0.36\\(0.33; 0.39)} & \makecell{0.43\\(0.41; 0.45)} \\
\addlinespace
Gen. Gamma AFT & Exact & \makecell{0.11\\(0.11; 0.13)} & \makecell{0.11\\(0.11; 0.13)} & \makecell{0.11\\(0.10; 0.13)} & \makecell{0.94\\(0.92; 0.96)} & \makecell{0.93\\(0.91; 0.95)} & \makecell{0.93\\(0.91; 0.95)} & \makecell{0.96\\(0.89; 0.98)} \\
\addlinespace
Weibull AFT & Exact & \makecell{0.08\\(0.07; 0.09)} & \makecell{0.08\\(0.07; 0.09)} & \makecell{0.08\\(0.07; 0.09)} & \makecell{0.95\\(0.93; 0.97)} & \makecell{0.94\\(0.92; 0.96)} & \makecell{0.94\\(0.91; 0.96)} & \makecell{0.92\\(0.84; 0.96)} \\
\addlinespace
Gen. Gamma AFT & Adjustment & \makecell{0.16\\(0.14; 0.18)} & \makecell{0.16\\(0.14; 0.17)} & \makecell{0.13\\(0.12; 0.15)} & \makecell{0.92\\(0.90; 0.95)} & \makecell{0.94\\(0.91; 0.95)} & \makecell{0.93\\(0.91; 0.95)} & \makecell{0.97\\(0.94; 0.98)} \\
\addlinespace
Weibull AFT & Adjustment & \makecell{0.09\\(0.09; 0.11)} & \makecell{0.09\\(0.09; 0.11)} & \makecell{0.08\\(0.07; 0.09)} & \makecell{0.95\\(0.93; 0.97)} & \makecell{0.95\\(0.93; 0.97)} & \makecell{0.94\\(0.91; 0.96)} & \makecell{0.95\\(0.92; 0.97)} \\
\addlinespace
Gen. Gamma AFT & Mid & \makecell{0.13\\(0.12; 0.14)} & \makecell{0.13\\(0.12; 0.15)} & \makecell{0.13\\(0.12; 0.14)} & \makecell{0.68\\(0.64; 0.71)} & \makecell{0.76\\(0.72; 0.79)} & \makecell{0.75\\(0.71; 0.78)} & \makecell{0.37\\(0.34; 0.46)} \\
\addlinespace
Weibull AFT & Mid & \makecell{0.11\\(0.11; 0.13)} & \makecell{0.12\\(0.11; 0.13)} & \makecell{0.09\\(0.08; 0.10)} & \makecell{0.43\\(0.40; 0.47)} & \makecell{0.37\\(0.34; 0.41)} & \makecell{0.84\\(0.81; 0.87)} & \makecell{0.33\\(0.25; 0.41)} \\
\addlinespace
Gen. Gamma AFT & End & \makecell{0.09\\(0.08; 0.11)} & \makecell{0.08\\(0.07; 0.09)} & \makecell{0.14\\(0.13; 0.16)} & \makecell{0.34\\(0.31; 0.37)} & \makecell{0.46\\(0.43; 0.48)} & \makecell{0.69\\(0.66; 0.72)} & \makecell{0.42\\(0.40; 0.44)} \\
\addlinespace
Weibull AFT & End & \makecell{0.11\\(0.10; 0.12)} & \makecell{0.09\\(0.08; 0.10)} & \makecell{0.15\\(0.14; 0.16)} & \makecell{0.18\\(0.16; 0.19)} & \makecell{0.17\\(0.15; 0.18)} & \makecell{0.51\\(0.48; 0.54)} & \makecell{0.21\\(0.18; 0.23)} \\
\bottomrule
\end{tabular}
\normalsize
\end{sideways}
\end{table*}

\newcommand{\captionicbhcoveragecumu}{Mean coverage of baseline cumulative hazards under interval-censoring.
This table shows mean baseline cumulative hazard coverage for the IC simulation settings described in Section \ref{sssec:sim-ic-setup}.
Mean coverage and 95\% CIs (from an exact binomial test) are computed pointwise across all simulation runs and then averaged over time points.
}
\begin{table*}[htbp]
\centering
\caption{\captionicbhcoveragecumu}
\label{tab:sim-ic-bh-coverage-cumu}
\begin{sideways}
\footnotesize
\setlength{\tabcolsep}{3pt}
\begin{tabular}{llccccccc}
\toprule
& & \multicolumn{3}{c}{Piecewise Exponential DGP} & \multicolumn{3}{c}{Weibull DGP} & \multicolumn{1}{c}{icenReg DGP} \\
\cmidrule(lr){3-5} \cmidrule(lr){6-8} \cmidrule(lr){9-9}
Model & \shortstack[l]{Estimation \\ Time Point} & Beta & Uniform & Equidistant & Beta & Uniform & Equidistant & Uniform \\
\midrule
PAM & Exact & \makecell{0.99\\(0.98; 0.99)} & \makecell{0.99\\(0.98; 1.00)} & \makecell{0.99\\(0.98; 0.99)} & \makecell{0.94\\(0.92; 0.95)} & \makecell{0.93\\(0.91; 0.95)} & \makecell{0.93\\(0.92; 0.95)} & \makecell{0.98\\(0.90; 0.99)} \\
PAM & Mid & \makecell{0.92\\(0.90; 0.93)} & \makecell{0.94\\(0.92; 0.95)} & \makecell{0.97\\(0.96; 0.98)} & \makecell{0.95\\(0.92; 0.96)} & \makecell{0.89\\(0.86; 0.91)} & \makecell{0.95\\(0.93; 0.96)} & \makecell{0.83\\(0.78; 0.84)} \\
PAM & End & \makecell{0.45\\(0.44; 0.46)} & \makecell{0.43\\(0.42; 0.44)} & \makecell{0.78\\(0.75; 0.79)} & \makecell{0.41\\(0.39; 0.42)} & \makecell{0.36\\(0.34; 0.38)} & \makecell{0.95\\(0.93; 0.97)} & \makecell{0.16\\(0.15; 0.18)} \\
\addlinespace
Cox & Exact & \makecell{0.95\\(0.92; 0.96)} & \makecell{0.93\\(0.90; 0.95)} & \makecell{0.94\\(0.92; 0.96)} & \makecell{0.95\\(0.92; 0.96)} & \makecell{0.94\\(0.92; 0.96)} & \makecell{0.94\\(0.91; 0.96)} & \makecell{0.76\\(0.72; 0.83)} \\
Cox & Mid & \makecell{0.77\\(0.73; 0.80)} & \makecell{0.79\\(0.76; 0.83)} & \makecell{0.65\\(0.62; 0.68)} & \makecell{0.74\\(0.70; 0.77)} & \makecell{0.75\\(0.71; 0.78)} & \makecell{0.60\\(0.56; 0.64)} & \makecell{0.46\\(0.42; 0.54)} \\
Cox & End & \makecell{0.36\\(0.35; 0.38)} & \makecell{0.33\\(0.31; 0.35)} & \makecell{0.91\\(0.88; 0.93)} & \makecell{0.31\\(0.29; 0.33)} & \makecell{0.20\\(0.19; 0.22)} & \makecell{0.90\\(0.88; 0.92)} & \makecell{0.01\\(0.01; 0.02)} \\
\addlinespace
Gen. Gamma AFT & Exact & \makecell{0.52\\(0.49; 0.55)} & \makecell{0.52\\(0.49; 0.55)} & \makecell{0.52\\(0.49; 0.55)} & \makecell{0.97\\(0.96; 0.99)} & \makecell{0.97\\(0.95; 0.98)} & \makecell{0.97\\(0.95; 0.98)} & \makecell{0.99\\(0.93; 1.00)} \\
\addlinespace
Weibull AFT & Exact & \makecell{0.36\\(0.34; 0.38)} & \makecell{0.36\\(0.35; 0.38)} & \makecell{0.36\\(0.34; 0.38)} & \makecell{0.97\\(0.96; 0.98)} & \makecell{0.96\\(0.95; 0.98)} & \makecell{0.97\\(0.95; 0.98)} & \makecell{0.95\\(0.87; 0.98)} \\
\addlinespace
Gen. Gamma AFT & Adjustment & \makecell{0.80\\(0.77; 0.82)} & \makecell{0.80\\(0.78; 0.82)} & \makecell{0.69\\(0.66; 0.72)} & \makecell{0.98\\(0.96; 0.99)} & \makecell{0.98\\(0.96; 0.99)} & \makecell{0.97\\(0.95; 0.98)} & \makecell{0.99\\(0.98; 1.00)} \\
\addlinespace
Weibull AFT & Adjustment & \makecell{0.50\\(0.48; 0.53)} & \makecell{0.49\\(0.46; 0.52)} & \makecell{0.36\\(0.34; 0.38)} & \makecell{0.98\\(0.96; 0.99)} & \makecell{0.97\\(0.95; 0.98)} & \makecell{0.96\\(0.94; 0.98)} & \makecell{0.98\\(0.96; 0.99)} \\
\addlinespace
Gen. Gamma AFT & Mid & \makecell{0.63\\(0.60; 0.66)} & \makecell{0.66\\(0.62; 0.69)} & \makecell{0.61\\(0.58; 0.64)} & \makecell{0.82\\(0.79; 0.86)} & \makecell{0.85\\(0.82; 0.88)} & \makecell{0.79\\(0.76; 0.82)} & \makecell{0.48\\(0.44; 0.57)} \\
\addlinespace
Weibull AFT & Mid & \makecell{0.46\\(0.44; 0.49)} & \makecell{0.46\\(0.43; 0.49)} & \makecell{0.39\\(0.37; 0.41)} & \makecell{0.72\\(0.69; 0.75)} & \makecell{0.67\\(0.63; 0.71)} & \makecell{0.94\\(0.92; 0.96)} & \makecell{0.53\\(0.45; 0.59)} \\
\addlinespace
Gen. Gamma AFT & End & \makecell{0.27\\(0.26; 0.29)} & \makecell{0.28\\(0.27; 0.30)} & \makecell{0.51\\(0.49; 0.53)} & \makecell{0.35\\(0.33; 0.37)} & \makecell{0.25\\(0.24; 0.27)} & \makecell{0.51\\(0.48; 0.55)} & \makecell{0.05\\(0.04; 0.07)} \\
\addlinespace
Weibull AFT & End & \makecell{0.19\\(0.18; 0.21)} & \makecell{0.18\\(0.17; 0.20)} & \makecell{0.48\\(0.46; 0.50)} & \makecell{0.31\\(0.30; 0.33)} & \makecell{0.29\\(0.28; 0.30)} & \makecell{0.41\\(0.38; 0.43)} & \makecell{0.14\\(0.12; 0.16)} \\
\bottomrule
\end{tabular}
\normalsize
\end{sideways}
\end{table*}

\newcommand{\captionicbhcoveragesurv}{Mean coverage of baseline survival functions under interval-censoring.
This table shows mean baseline survival function coverage for the IC simulation settings described in Section \ref{sssec:sim-ic-setup}.
Mean coverage and 95\% CIs (from an exact binomial test) are computed pointwise across all simulation runs and then averaged over time points.
}
\begin{table*}[htbp]
\centering
\caption{\captionicbhcoveragesurv}
\label{tab:sim-ic-bh-coverage-surv}
\begin{sideways}
\footnotesize
\setlength{\tabcolsep}{3pt}
\begin{tabular}{llccccccc}
\toprule
& & \multicolumn{3}{c}{Piecewise Exponential DGP} & \multicolumn{3}{c}{Weibull DGP} & \multicolumn{1}{c}{icenReg DGP} \\
\cmidrule(lr){3-5} \cmidrule(lr){6-8} \cmidrule(lr){9-9}
Model & \shortstack[l]{Estimation \\ Time Point} & Beta & Uniform & Equidistant & Beta & Uniform & Equidistant & Uniform \\
\midrule
PAM & Exact & \makecell{0.99\\(0.98; 0.99)} & \makecell{0.99\\(0.98; 1.00)} & \makecell{0.99\\(0.98; 0.99)} & \makecell{0.94\\(0.92; 0.95)} & \makecell{0.93\\(0.91; 0.95)} & \makecell{0.93\\(0.92; 0.95)} & \makecell{0.98\\(0.90; 0.99)} \\
PAM & Mid & \makecell{0.92\\(0.90; 0.93)} & \makecell{0.94\\(0.92; 0.95)} & \makecell{0.97\\(0.96; 0.98)} & \makecell{0.95\\(0.92; 0.96)} & \makecell{0.89\\(0.86; 0.91)} & \makecell{0.95\\(0.93; 0.96)} & \makecell{0.83\\(0.78; 0.84)} \\
PAM & End & \makecell{0.45\\(0.44; 0.46)} & \makecell{0.43\\(0.42; 0.44)} & \makecell{0.78\\(0.75; 0.79)} & \makecell{0.41\\(0.39; 0.42)} & \makecell{0.36\\(0.34; 0.38)} & \makecell{0.95\\(0.93; 0.97)} & \makecell{0.16\\(0.15; 0.18)} \\
\addlinespace
Cox & Exact & \makecell{0.95\\(0.92; 0.96)} & \makecell{0.93\\(0.90; 0.95)} & \makecell{0.94\\(0.92; 0.96)} & \makecell{0.95\\(0.92; 0.96)} & \makecell{0.94\\(0.92; 0.96)} & \makecell{0.94\\(0.91; 0.96)} & \makecell{0.76\\(0.72; 0.83)} \\
Cox & Mid & \makecell{0.77\\(0.73; 0.80)} & \makecell{0.79\\(0.76; 0.83)} & \makecell{0.65\\(0.62; 0.68)} & \makecell{0.74\\(0.70; 0.77)} & \makecell{0.75\\(0.71; 0.78)} & \makecell{0.60\\(0.56; 0.64)} & \makecell{0.46\\(0.42; 0.54)} \\
Cox & End & \makecell{0.36\\(0.35; 0.38)} & \makecell{0.33\\(0.31; 0.35)} & \makecell{0.91\\(0.88; 0.93)} & \makecell{0.31\\(0.29; 0.33)} & \makecell{0.20\\(0.19; 0.22)} & \makecell{0.90\\(0.88; 0.92)} & \makecell{0.01\\(0.01; 0.02)} \\
\addlinespace
Gen. Gamma AFT & Exact & \makecell{0.52\\(0.49; 0.55)} & \makecell{0.52\\(0.49; 0.55)} & \makecell{0.52\\(0.49; 0.55)} & \makecell{0.97\\(0.96; 0.99)} & \makecell{0.97\\(0.95; 0.98)} & \makecell{0.97\\(0.95; 0.98)} & \makecell{0.99\\(0.93; 1.00)} \\
\addlinespace
Weibull AFT & Exact & \makecell{0.36\\(0.34; 0.38)} & \makecell{0.36\\(0.35; 0.38)} & \makecell{0.36\\(0.34; 0.38)} & \makecell{0.97\\(0.96; 0.98)} & \makecell{0.96\\(0.95; 0.98)} & \makecell{0.97\\(0.95; 0.98)} & \makecell{0.95\\(0.87; 0.98)} \\
\addlinespace
Gen. Gamma AFT & Adjustment & \makecell{0.80\\(0.77; 0.82)} & \makecell{0.80\\(0.78; 0.82)} & \makecell{0.69\\(0.66; 0.72)} & \makecell{0.98\\(0.96; 0.99)} & \makecell{0.98\\(0.96; 0.99)} & \makecell{0.97\\(0.95; 0.98)} & \makecell{0.99\\(0.98; 1.00)} \\
\addlinespace
Weibull AFT & Adjustment & \makecell{0.50\\(0.48; 0.53)} & \makecell{0.49\\(0.46; 0.52)} & \makecell{0.36\\(0.34; 0.38)} & \makecell{0.98\\(0.96; 0.99)} & \makecell{0.97\\(0.95; 0.98)} & \makecell{0.96\\(0.94; 0.98)} & \makecell{0.98\\(0.96; 0.99)} \\
\addlinespace
Gen. Gamma AFT & Mid & \makecell{0.63\\(0.60; 0.66)} & \makecell{0.66\\(0.62; 0.69)} & \makecell{0.61\\(0.58; 0.64)} & \makecell{0.82\\(0.79; 0.86)} & \makecell{0.85\\(0.82; 0.88)} & \makecell{0.79\\(0.76; 0.82)} & \makecell{0.48\\(0.44; 0.57)} \\
\addlinespace
Weibull AFT & Mid & \makecell{0.46\\(0.44; 0.49)} & \makecell{0.46\\(0.43; 0.49)} & \makecell{0.39\\(0.37; 0.41)} & \makecell{0.72\\(0.69; 0.75)} & \makecell{0.67\\(0.63; 0.71)} & \makecell{0.94\\(0.92; 0.96)} & \makecell{0.53\\(0.45; 0.59)} \\
\addlinespace
Gen. Gamma AFT & End & \makecell{0.27\\(0.26; 0.29)} & \makecell{0.28\\(0.27; 0.30)} & \makecell{0.51\\(0.49; 0.53)} & \makecell{0.35\\(0.33; 0.37)} & \makecell{0.25\\(0.24; 0.27)} & \makecell{0.51\\(0.48; 0.55)} & \makecell{0.05\\(0.04; 0.07)} \\
\addlinespace
Weibull AFT & End & \makecell{0.19\\(0.18; 0.21)} & \makecell{0.18\\(0.17; 0.20)} & \makecell{0.48\\(0.46; 0.50)} & \makecell{0.31\\(0.30; 0.33)} & \makecell{0.29\\(0.28; 0.30)} & \makecell{0.41\\(0.38; 0.43)} & \makecell{0.14\\(0.12; 0.16)} \\
\bottomrule
\end{tabular}
\normalsize
\end{sideways}
\end{table*}

\newcommand{\captionicfebias}{Mean bias of fixed effect parameter estimates under interval-censoring. This table shows the mean bias (across simulation runs) in estimating the fixed effect coefficient $\beta_{x_1}$ of a \textit{Bernoulli}(0.5)-distributed covariate $x_1$ for three models and three to four estimation points on data generated from two DGPs with three IC mechanisms each, described in Section \ref{sssec:sim-ic-setup}.
}
\begin{table*}[htbp]
\centering
\caption{\captionicfebias}
\label{tab:sim-ic-fe-bias}
\begin{sideways}
\begin{tabular}{llcccccc}
\toprule
& & \multicolumn{3}{c}{Piecewise Exponential DGP} & \multicolumn{3}{c}{Weibull DGP} \\
\cmidrule(lr){3-5} \cmidrule(lr){6-8}
Model & \shortstack[l]{Estimation \\ Time Point} & Beta & Uniform & Equidistant & Beta & Uniform & Equidistant \\
\midrule
PAM & Exact & 0.00 & 0.00 & 0.01 & -0.01 & -0.00 & 0.00 \\
PAM & Mid & -0.01 & 0.01 & 0.01 & 0.00 & 0.01 & -0.00 \\
PAM & End & 0.02 & 0.03 & -0.01 & 0.00 & 0.01 & -0.00 \\
\addlinespace
Cox & Exact & 0.02 & 0.01 & -0.01 & 0.01 & 0.00 & -0.00 \\
Cox & Mid & 0.01 & 0.00 & -0.01 & 0.01 & 0.01 & -0.00 \\
Cox & End & -0.00 & 0.01 & -0.00 & 0.02 & 0.02 & 0.00 \\
\addlinespace
Weibull AFT & Exact & -0.04 & -0.06 & -0.02 & -0.00 & -0.00 & 0.00 \\
Weibull AFT & Adj. & -0.03 & -0.04 & -0.05 & 0.00 & 0.00 & -0.01 \\
Weibull AFT & Mid & -0.02 & -0.03 & -0.04 & 0.02 & 0.02 & 0.03 \\
Weibull AFT & End & -0.01 & -0.01 & -0.03 & 0.00 & 0.00 & -0.01 \\
\bottomrule
\end{tabular}
\end{sideways}
\end{table*}

\newcommand{\captionicfecoverage}{Mean coverage of fixed effect parameter estimates under interval-censoring. This table shows mean coverage (across simulation runs) along with 95\% confidence intervals (using an exact Binomial test) of the true fixed effect parameter $\beta_{x_1}$ of a \textit{Bernoulli}(0.5)-distributed covariate $x_1$ for three models and three to four estimation points on data generated from two DGPs with three IC mechanisms each, described in Section \ref{sssec:sim-ic-setup}.
}
\begin{table*}[htbp]
\centering
\caption{\captionicfecoverage}
\label{tab:sim-ic-fe-coverage}
\begin{sideways}
\footnotesize
\setlength{\tabcolsep}{4pt}
\begin{tabular}{llcccccc}
\toprule
& & \multicolumn{3}{c}{Piecewise Exponential DGP} & \multicolumn{3}{c}{Weibull DGP} \\
\cmidrule(lr){3-5} \cmidrule(lr){6-8}
Model & \shortstack[l]{Estimation \\ Time Point} & Beta & Uniform & Equidistant & Beta & Uniform & Equidistant \\
\midrule
PAM & Exact & \makecell{0.93\\(0.91; 0.95)} & \makecell{0.96\\(0.94; 0.97)} & \makecell{0.95\\(0.93; 0.97)} & \makecell{0.98\\(0.96; 0.99)} & \makecell{0.95\\(0.93; 0.97)} & \makecell{0.96\\(0.94; 0.98)} \\
PAM & Mid & \makecell{0.96\\(0.94; 0.98)} & \makecell{0.95\\(0.93; 0.97)} & \makecell{0.94\\(0.91; 0.96)} & \makecell{0.93\\(0.91; 0.95)} & \makecell{0.96\\(0.94; 0.98)} & \makecell{0.95\\(0.93; 0.97)} \\
PAM & End & \makecell{0.96\\(0.93; 0.97)} & \makecell{0.95\\(0.93; 0.97)} & \makecell{0.95\\(0.93; 0.97)} & \makecell{0.96\\(0.94; 0.97)} & \makecell{0.94\\(0.91; 0.96)} & \makecell{0.95\\(0.93; 0.97)} \\
\addlinespace
Cox & Exact & \makecell{0.97\\(0.95; 0.98)} & \makecell{0.95\\(0.93; 0.97)} & \makecell{0.94\\(0.92; 0.96)} & \makecell{0.94\\(0.92; 0.96)} & \makecell{0.96\\(0.94; 0.98)} & \makecell{0.96\\(0.93; 0.97)} \\
Cox & Mid & \makecell{0.95\\(0.93; 0.97)} & \makecell{0.96\\(0.93; 0.97)} & \makecell{0.96\\(0.94; 0.97)} & \makecell{0.96\\(0.94; 0.97)} & \makecell{0.96\\(0.94; 0.97)} & \makecell{0.94\\(0.92; 0.96)} \\
Cox & End & \makecell{0.95\\(0.92; 0.96)} & \makecell{0.95\\(0.93; 0.97)} & \makecell{0.95\\(0.92; 0.96)} & \makecell{0.96\\(0.93; 0.97)} & \makecell{0.97\\(0.95; 0.98)} & \makecell{0.95\\(0.92; 0.96)} \\
\addlinespace
Weibull AFT & Exact & \makecell{0.92\\(0.89; 0.94)} & \makecell{0.94\\(0.91; 0.96)} & \makecell{0.94\\(0.92; 0.96)} & \makecell{0.95\\(0.93; 0.97)} & \makecell{0.96\\(0.94; 0.97)} & \makecell{0.97\\(0.95; 0.98)} \\
Weibull AFT & Adj. & \makecell{0.95\\(0.93; 0.97)} & \makecell{0.95\\(0.92; 0.96)} & \makecell{0.93\\(0.90; 0.95)} & \makecell{0.94\\(0.92; 0.96)} & \makecell{0.95\\(0.92; 0.96)} & \makecell{0.93\\(0.91; 0.95)} \\
Weibull AFT & Mid & \makecell{0.92\\(0.89; 0.94)} & \makecell{0.93\\(0.90; 0.95)} & \makecell{0.94\\(0.92; 0.96)} & \makecell{0.95\\(0.92; 0.96)} & \makecell{0.96\\(0.93; 0.97)} & \makecell{0.96\\(0.94; 0.97)} \\
Weibull AFT & End & \makecell{0.95\\(0.93; 0.97)} & \makecell{0.95\\(0.93; 0.97)} & \makecell{0.94\\(0.92; 0.96)} & \makecell{0.95\\(0.93; 0.97)} & \makecell{0.96\\(0.94; 0.97)} & \makecell{0.94\\(0.92; 0.96)} \\
\bottomrule
\end{tabular}
\normalsize
\end{sideways}
\end{table*}

\newcommand{\captionukbriskfactordistributions}{
Summary statistics of risk factors by state and rs77924615 genotype in UK Biobank.
This table provides summary statistics for the six risk factors of CKD onset and progression introduced in Section \ref{sssec:ukb-results-ieb} (median (min-max) for quantitative, n (\%) for binary).
For each risk factor, rows correspond to the from-state (0: Healthy; 1: Mild CKD; 2: Severe CKD; see Figure \ref{fig:state-diagram-ckd}).
The three columns correspond to the different genotypes of the genetic variant rs77924615 residing in the \textit{UMOD} locus with effect allele G.
}
\begin{table}[!ht]
\footnotesize
\caption{\captionukbriskfactordistributions}
\label{tab:ukb-risk-factor-distributions}
\centering
\begin{tabular}[t]{llcccc}
\toprule
& & \multicolumn{3}{c}{Genotype of rs77924615} & \\
\cmidrule(lr){3-5}
Risk Factor & State & AA & AG/GA & GG & Total \\ & & (n=4,938) & (n=40,852) & (n=84,931) & (n=130,721) \\
\midrule
\cellcolor{gray!10} & \cellcolor{gray!10}0 & \cellcolor{gray!10}2,554 (53.3\%) & \cellcolor{gray!10}21,003 (53.5\%) & \cellcolor{gray!10}43,076 (53.4\%) & \cellcolor{gray!10}66,633 (53.4\%) \\
\cellcolor{gray!10} & \cellcolor{gray!10}1 & \cellcolor{gray!10}56 (43.4\%) & \cellcolor{gray!10}791 (52.8\%) & \cellcolor{gray!10}2,110 (52.0\%) & \cellcolor{gray!10}2,957 (52.0\%) \\
\multirow{-3}{*}{\cellcolor{gray!10}Sex (Women)} & \cellcolor{gray!10}2 & \cellcolor{gray!10}7 (50.0\%) & \cellcolor{gray!10}46 (47.4\%) & \cellcolor{gray!10}98 (41.7\%) & \cellcolor{gray!10}151 (43.6\%) \\
 & 0 & \makecell{580 \\ (423-860)} & \makecell{580 \\ (426-883)} & \makecell{582 \\ (417-884)} & \makecell{581 \\ (417-884)} \\
 & 1 & \makecell{574 \\ (465-776)} & \makecell{586 \\ (443-889)} & \makecell{587 \\ (444-850)} & \makecell{586 \\ (443-889)} \\
\multirow{-3}{*}{PGS} & 2 & \makecell{572 \\ (494-776)} & \makecell{580 \\ (455-791)} & \makecell{582 \\ (461-848)} & \makecell{580 \\ (455-848)} \\
\cellcolor{gray!10} & \cellcolor{gray!10}0 & \cellcolor{gray!10}386 (8.1\%) & \cellcolor{gray!10}3,042 (7.7\%) & \cellcolor{gray!10}6,135 (7.6\%) & \cellcolor{gray!10}9,563 (7.7\%) \\
\cellcolor{gray!10} & \cellcolor{gray!10}1 & \cellcolor{gray!10}25 (19.4\%) & \cellcolor{gray!10}291 (19.4\%) & \cellcolor{gray!10}773 (19.0\%) & \cellcolor{gray!10}1,089 (19.1\%) \\
\multirow{-3}{*}{\cellcolor{gray!10}Diabetes} & \cellcolor{gray!10}2 & \cellcolor{gray!10}6 (42.9\%) & \cellcolor{gray!10}26 (26.8\%) & \cellcolor{gray!10}67 (28.5\%) & \cellcolor{gray!10}99 (28.6\%) \\
 & 0 & 522 (10.9\%) & 3,944 (10.0\%) & 8,193 (10.2\%) & 12,659 (10.2\%) \\
 & 1 & 16 (12.4\%) & 151 (10.1\%) & 404 (9.9\%) & 571 (10.0\%) \\
\multirow{-3}{*}{Smoking} & 2 & 0 (0.0\%) & 15 (15.5\%) & 27 (11.5\%) & 42 (12.1\%) \\
\cellcolor{gray!10} & \cellcolor{gray!10}0 & \cellcolor{gray!10}\makecell{27.2 \\ (16.5-55.0)} & \cellcolor{gray!10}\makecell{27.0 \\ (14.3-67.3)} & \cellcolor{gray!10}\makecell{26.9 \\ (12.6-74.7)} & \cellcolor{gray!10}\makecell{27.0 \\ (12.6-74.7)} \\
\cellcolor{gray!10} & \cellcolor{gray!10}1 & \cellcolor{gray!10}\makecell{29.0 \\ (20.4-45.2)} & \cellcolor{gray!10}\makecell{28.9 \\ (17.8-66.2)} & \cellcolor{gray!10}\makecell{28.8 \\ (17.6-58.4)} & \cellcolor{gray!10}\makecell{28.8 \\ (17.6-66.2)} \\
\multirow{-3}{*}{\cellcolor{gray!10}BMI} & \cellcolor{gray!10}2 & \cellcolor{gray!10}\makecell{30.2 \\ (20.9-39.9)} & \cellcolor{gray!10}\makecell{28.2 \\ (18.4-48.8)} & \cellcolor{gray!10}\makecell{29.9 \\ (18.3-48.3)} & \cellcolor{gray!10}\makecell{29.6 \\ (18.3-48.8)} \\
 & 0 & \makecell{10.0 \\ (2.3-1859.2)} & \makecell{9.8 \\ (1.7-2918.6)} & \makecell{9.8 \\ (1.7-8531.1)} & \makecell{9.8 \\ (1.7-8531.1)} \\
 & 1 & \makecell{11.4 \\ (3.0-1611.7)} & \makecell{11.5 \\ (2.4-2918.6)} & \makecell{10.9 \\ (2.0-8531.1)} & \makecell{11.1 \\ (2.0-8531.1)} \\
\multirow{-3}{*}{uACR} & 2 & \makecell{78.1 \\ (5.2-1611.7)} & \makecell{25.4 \\ (3.3-2897.2)} & \makecell{26.6 \\ (3.4-8531.1)} & \makecell{26.6 \\ (3.3-8531.1)} \\
\midrule\midrule \\
 & 0 & 4,795 (97.1\%) & 39,256 (96.1\%) & 80,635 (94.9\%) & 124,686 (95.4\%) \\
 & 1 & 129 (2.6\%) & 1,499 (3.7\%) & 4,061 (4.8\%) & 5,689 (4.4\%) \\
\multirow{-3}{*}{n}  & 2 & 14 (0.3\%) & 97 (0.2\%) & 235 (0.3\%) & 346 (0.3\%) \\
\bottomrule
\end{tabular}
\normalsize
\end{table}

\newcommand{\captionukbadjustmentweighting}{Comparison of genetic effect size estimates under adjustment versus weighting approaches in UK Biobank.
This table compares effect size estimates of the genetic variant rs77924615 on CKD onset and progression in the UK Biobank dataset introduced in Section \ref{ssec:ukb-data}, using \emph{SSTS} PAMs as introduced in Section \ref{ssec:ukb-model-specification}, for four different scenarios of how to account for confounding risk factors: via direct adjustment in the model formula (Adjustment yes/no) and/or via a weighting approach using propensity scores from a multinomial logistic regression (see Appendix \ref{app-ssec:ukb-ps}).
Both approaches use all risk factors introduced in Section \ref{sssec:ukb-results-ieb}.
}
\begin{table}[!ht]
\centering
\caption{\captionukbadjustmentweighting}
\label{tab:ukb-adjustment-weighting}
\begin{tabular}[t]{cccrrr}
\toprule
Adjustment & Weighting & Transition & Coefficient & SE & P.value\\
\midrule
\cellcolor{gray!10}{no} & \cellcolor{gray!10}{no} & \cellcolor{gray!10}{0 $\rightarrow$ 1} & \cellcolor{gray!10}{0.284} & \cellcolor{gray!10}{0.025} & \cellcolor{gray!10}{$<$0.001}\\
\cellcolor{gray!10}{no} & \cellcolor{gray!10}{no} & \cellcolor{gray!10}{1 $\rightarrow$ 2} & \cellcolor{gray!10}{-0.121} & \cellcolor{gray!10}{0.090} & \cellcolor{gray!10}{0.181}\\
\cellcolor{gray!10}{no} & \cellcolor{gray!10}{no} & \cellcolor{gray!10}{2 $\rightarrow$ 3} & \cellcolor{gray!10}{-0.311} & \cellcolor{gray!10}{0.146} & \cellcolor{gray!10}{0.034}\\
\midrule
yes & no & 0 $\rightarrow$ 1 & 0.278 & 0.026 & $<$0.001\\
yes & no & 1 $\rightarrow$ 2 & 0.045 & 0.100 & 0.654\\
yes & no & 2 $\rightarrow$ 3 & -0.159 & 0.178 & 0.371\\
\midrule
\cellcolor{gray!10}{no} & \cellcolor{gray!10}{yes} & \cellcolor{gray!10}{0 $\rightarrow$ 1} & \cellcolor{gray!10}{0.280} & \cellcolor{gray!10}{0.026} & \cellcolor{gray!10}{$<$0.001}\\
\cellcolor{gray!10}{no} & \cellcolor{gray!10}{yes} & \cellcolor{gray!10}{1 $\rightarrow$ 2} & \cellcolor{gray!10}{-0.049} & \cellcolor{gray!10}{0.101} & \cellcolor{gray!10}{0.627}\\
\cellcolor{gray!10}{no} & \cellcolor{gray!10}{yes} & \cellcolor{gray!10}{2 $\rightarrow$ 3} & \cellcolor{gray!10}{-0.134} & \cellcolor{gray!10}{0.179} & \cellcolor{gray!10}{0.454}\\
\midrule
yes & yes & 0 $\rightarrow$ 1 & 0.277 & 0.026 & $<$0.001\\
yes & yes & 1 $\rightarrow$ 2 & 0.028 & 0.101 & 0.784\\
yes & yes & 2 $\rightarrow$ 3 & -0.169 & 0.179 & 0.346\\
\bottomrule
\end{tabular}
\parbox{\linewidth}{\footnotesize
\begin{itemize}[leftmargin=*, noitemsep, label=\textcolor{white}{\textbullet}]
 \item 0: Healthy; 1: Mild CKD; 2: Severe CKD; 3: ESKD.
\end{itemize}
}
\end{table}

\newcommand{\captionukbic}{Comparison of estimation points regarding risk factor effects in UK Biobank.
This table compares effect size estimates (Coef.), standard errors (SE), and P-values of the risk factors rs77924615, sex, diabetes, and smoking status for two scenarios: when the event time point is set to the interval mid-point (left) or to the interval end-point (right).
Both models are fitted to the UK Biobank dataset introduced in Section \ref{ssec:ukb-data}, using \emph{SSTS} PAMs as introduced in Section \ref{ssec:ukb-model-specification}.
}
\begin{table}[!ht]
\centering
\caption{\captionukbic}
\label{tab:ukb-ic}
\fontsize{9}{11}\selectfont
\begin{tabular}[t]{lrrrrrr}
\toprule
\multicolumn{1}{c}{ } & \multicolumn{3}{c}{Mid-Point Estimation} & \multicolumn{3}{c}{End-Point Estimation} \\
\cmidrule(l{3pt}r{3pt}){2-4} \cmidrule(l{3pt}r{3pt}){5-7}
Transition & Coef. & SE & P-value & Coef. & SE & P-value\\
\midrule
\addlinespace[0.3em]
\multicolumn{7}{l}{\textbf{G}}\\
\hspace{1em}\cellcolor{gray!10}{\hspace{1em}0$\rightarrow$1} & \cellcolor{gray!10}{0.278} & \cellcolor{gray!10}{0.026} & \cellcolor{gray!10}{$<$0.001} & \cellcolor{gray!10}{0.278} & \cellcolor{gray!10}{0.026} & \cellcolor{gray!10}{$<$0.001}\\
\hspace{1em}\hspace{1em}1$\rightarrow$2 & 0.045 & 0.100 & 0.654 & 0.005 & 0.099 & 0.959\\
\hspace{1em}\cellcolor{gray!10}{\hspace{1em}2$\rightarrow$3} & \cellcolor{gray!10}{-0.159} & \cellcolor{gray!10}{0.178} & \cellcolor{gray!10}{0.371} & \cellcolor{gray!10}{-0.123} & \cellcolor{gray!10}{0.192} & \cellcolor{gray!10}{0.522}\\
\addlinespace[0.3em]
\multicolumn{7}{l}{\textbf{Sex (Women)}}\\
\hspace{1em}\hspace{1em}0$\rightarrow$1 & 0.055 & 0.027 & 0.044 & 0.059 & 0.027 & 0.029\\
\hspace{1em}\cellcolor{gray!10}{\hspace{1em}1$\rightarrow$2} & \cellcolor{gray!10}{-0.390} & \cellcolor{gray!10}{0.109} & \cellcolor{gray!10}{$<$0.001} & \cellcolor{gray!10}{-0.358} & \cellcolor{gray!10}{0.109} & \cellcolor{gray!10}{0.001}\\
\hspace{1em}\hspace{1em}2$\rightarrow$3 & -0.205 & 0.210 & 0.329 & -0.250 & 0.224 & 0.264\\
\addlinespace[0.3em]
\multicolumn{7}{l}{\textbf{Diabetes}}\\
\hspace{1em}\cellcolor{gray!10}{\hspace{1em}0$\rightarrow$1} & \cellcolor{gray!10}{-0.355} & \cellcolor{gray!10}{0.042} & \cellcolor{gray!10}{$<$0.001} & \cellcolor{gray!10}{-0.361} & \cellcolor{gray!10}{0.042} & \cellcolor{gray!10}{$<$0.001}\\
\hspace{1em}\hspace{1em}1$\rightarrow$2 & -0.326 & 0.123 & 0.008 & -0.284 & 0.124 & 0.022\\
\hspace{1em}\cellcolor{gray!10}{\hspace{1em}2$\rightarrow$3} & \cellcolor{gray!10}{0.036} & \cellcolor{gray!10}{0.241} & \cellcolor{gray!10}{0.880} & \cellcolor{gray!10}{-0.037} & \cellcolor{gray!10}{0.257} & \cellcolor{gray!10}{0.886}\\
\addlinespace[0.3em]
\multicolumn{7}{l}{\textbf{Smoking}}\\
\hspace{1em}\hspace{1em}0$\rightarrow$1 & 0.221 & 0.045 & $<$0.001 & 0.231 & 0.045 & $<$0.001\\
\hspace{1em}\cellcolor{gray!10}{\hspace{1em}1$\rightarrow$2} & \cellcolor{gray!10}{0.172} & \cellcolor{gray!10}{0.157} & \cellcolor{gray!10}{0.273} & \cellcolor{gray!10}{0.113} & \cellcolor{gray!10}{0.162} & \cellcolor{gray!10}{0.485}\\
\hspace{1em}\hspace{1em}2$\rightarrow$3 & 0.358 & 0.293 & 0.222 & 0.222 & 0.325 & 0.495\\
\bottomrule
\end{tabular}
\end{table}

\clearpage

\vskip 0.2in
\bibliographystyle{unsrt}
\bibliography{bibliography}


\end{document}